\PassOptionsToPackage{numbers,sort&compress}{natbib}  
\documentclass[preprint,12pt]{elsarticle}
\usepackage[margin=2.5cm]{geometry}
\usepackage{enumitem}
\newlist{inlinelist}{enumerate*}{1}
\setlist*[inlinelist,1]{%
  label=(\roman*),
}
\usepackage{dirtytalk}
\usepackage{soul}
\usepackage{hhline}
\usepackage{subfig}
\usepackage{multirow}
\usepackage{xcolor}
\usepackage{graphicx}
\usepackage{upgreek}
\usepackage{amssymb}
\usepackage{amsfonts,amsthm,bm,amsmath} 
\usepackage{appendix}
\usepackage{times}
\usepackage{caption}
\usepackage{subfig}
\newcommand{\psubref}[1]{\protect\subref{#1}}
\usepackage{multirow}
\usepackage{xcolor}
\usepackage[linesnumbered,ruled,vlined]{algorithm2e}
\usepackage{tablefootnote}

\SetCommentSty{mycommfont}

\SetKwInput{KwInput}{Input}                
\SetKwInput{KwOutput}{Output}              

\usepackage[colorlinks]{hyperref} 
\hypersetup{ 
    colorlinks=true,       
    linkcolor=red,          
    citecolor=blue,        
    filecolor=magenta,      
    urlcolor=cyan           
}

\newcommand{\fref}[1]{Fig.~\ref{#1}}

\newcommand{\eref}[1]{Eq.~(\ref{#1})}

\newcommand{\sref}[1]{Section~\ref{#1}}
\newcommand{\tref}[1]{Table~\ref{#1}}
\setcitestyle{square}

\journal{arXiv}
\begin{document}

\begin{frontmatter}

\title{On the use of graph neural networks and shape-function-based gradient computation in the deep energy method}
\author[]{Junyan He$^1$}
\author[]{Diab Abueidda$^2$}
\author[]{Seid Koric$^{1,2}$}
\author[]{Iwona Jasiuk$^1$\corref{mycorrespondingauthor}}
\address{$^1$ Department of Mechanical Science and Engineering, University of Illinois at Urbana-Champaign, Champaign, IL, USA \\
$^2$ National Center for Supercomputing Applications, University of Illinois at Urbana-Champaign, Champaign, IL, USA}
\cortext[mycorrespondingauthor]{Corresponding author}
\ead{ijasiuk@illinois.edu}
\begin{abstract}
A graph neural network (GCN) is employed in the deep energy method (DEM) model to solve the momentum balance equation in 3D for the deformation of linear elastic and hyperelastic materials due to its ability to handle irregular domains over the traditional DEM method based on a multilayer perceptron (MLP) network. Its accuracy and solution time are compared to the DEM model based on a MLP network. We demonstrate that the GCN-based model delivers similar accuracy while having a shorter run time through numerical examples. Two different spatial gradient computation techniques, one based on automatic differentiation (AD) and the other based on shape function (SF) gradients, are also accessed. We provide a simple example to demonstrate the strain localization instability associated with the AD-based gradient computation and show that the instability exists in more general cases by four numerical examples. The SF-based gradient computation is shown to be more robust and delivers an accurate solution even at severe deformations. Therefore, the combination of the GCN-based DEM model and SF-based gradient computation is potentially a promising candidate for solving problems involving severe material and geometric nonlinearities.

\end{abstract}

\begin{keyword}
Automatic differentiation \sep Elasticity \sep Hyperelasticity \sep Partial differential equations \sep Physics-informed neural networks
\end{keyword}

\end{frontmatter}

\section{Introduction}
\label{sec:intro}
The deep energy method (DEM) is a physics-informed neural network (PINN) model developed by Nguyen-Thanh et al. \cite{nguyen2020deep}. The method readily applies to engineering systems governed by an energy functional, whose solution coincides with the stationary point of the functional. The DEM model takes a series of points in the simulation domain as inputs and predicts field variables like displacements \cite{abueidda2022deep,nguyen2021parametric,samaniego2020energy} (for mechanics), temperature (for heat transfer), and/or electric potential (for piezoelectricity) \cite{samaniego2020energy} at the same nodal locations. The values of the predicted field variables are then used to numerically compute the energy functional, which is defined as the loss function. Standard optimization packages can be applied to minimize the loss, and the weights and biases of the DEM model are updated through backpropagation. 

Current DEM model implementations are mostly based on a multilayer perceptron (MLP) network \cite{noriega2005multilayer,hastie2009elements}, which consists of multiple layers of fully connected neurons. While shown to be highly effective, it is of interest to study how the underlying neural network (NN) structure affects the performance of the resulting DEM model. Chadha et al. \cite{chadha2022optimizing} studied the effects of changing network architecture (number of neurons and hidden layers) using an MLP-based DEM model. Zhuang et al. \cite{zhuang2021deep} investigated the use of autoencoders in the DEM method. While outside of the DEM framework, graph convolutional networks (GCN) have seen wide applications in the field of engineering mechanics, ranging from mechanics of polycrystals \cite{vlassis2020geometric,frankel2022mesh}, the solution to PDEs \cite{gao2022physics,sanchez2020learning,hernandez2022thermodynamics}, fluid mechanics \cite{ogoke2021graph,chen2021graph,he2022flow}, and crack propagation \cite{schwarzer2019learning}. GCNs are particularly suitable for describing data with structure and a high sense of locality \cite{gao2022physics, vlassis2020geometric}. In DEM, the computational domain is discretized into a series of structured or unstructured nodes, which renders it perfect to be represented by a graph weighted by the Euclidean distance between the nodes. In existing DEM implementations based on MLP networks, most utilized a structured grid to discretized the domain. As shown in \cite{gao2022physics}, PINN models that are based on convolutional neural networks require a fine grid to resolve the solution, and the convolution operator limited it to structured grid. While the GCN-based frameworks can represent unstructured mesh in terms of graphs and apply a graph convolution to it, greatly reducing the burden to generate a structured discretization of the domain. Therefore, in this work, we investigate the use of GCNs as the underlying NN for the DEM method, and study how this affects solution accuracy and time.

Many PINNs, including the DEM model, contain differential operators in the definition of their loss functions. For PINNs that are based on the strong form of the governing differential equation, such as the deep collocation method \cite{raissi2018deep,abueidda2021meshless,guo2021deep,haghighat2021physics}, second-order (e.g., in elasticity, heat transfer and the Navier-Stokes equations) or higher-order (e.g., plate bending) spatial gradients of field variables are required. For PINNs that are based on a variational formulation, such as the deep Ritz method \cite{yu2018deep,liao2019deep}, deep Galerkin method \cite{gao2022physics}, and DEM, first-order spatial gradients of field variables are required. Besides, researchers have used a mixed formulation, combining both the energy method and the strong form, to capture the mechanical response with high solution gradients and stress concentrations \cite{fuhg2022mixed, abueidda2022enhanced, rezaei2022mixed}.

In almost all cases, these spatial gradients are obtained utilizing automatic differentiation (AD) of the NN model \cite{paszke2017automatic,van2018automatic,ketkar2021automatic}. This approach is widely used since AD is already applied in the backpropagation step during NN model training. Most importantly, it alleviates the need to form an element-based discretization of the computational domain as is typically required for the finite element method (FEM), yielding a meshless simulation technique \cite{abueidda2021meshless}. However, the cost of AD can be quite expensive, especially when many nodes are employed in the domain. Therefore, as an alternative, spatial gradients of the field variables can be calculated on a mesh-based discretization of the domain using Sobel filters (finite difference) \cite{zhu2019physics,geneva2020modeling,zhang2020physics,zhang2020physics,wandel2021teaching}. In addition, finite element shape functions (SFs) can be used \cite{chadha2022optimizing,he2022deep,gao2022physics,yao2020fea}. In this case, the gradients are evaluated at the integration points of the finite elements through the shape function gradients, identical to the treatment in classical FEM \cite{logan2016first,fish2007first}. We note that a thorough comparison between the AD- and SF-based gradient computations and their implications on the accuracy and stability of the DEM model has been missing from the literature. Thus, the second objective of this work is to perform such a comparison using case studies. More importantly, we point out that the point-wise AD-based gradient computation is inherently susceptible to instability within the DEM framework, which is a novel finding in the literature.

This paper is organized as follows: \sref{sec:methods} presents an overview of graph convolutional networks, the deep energy method, and a discussion on instability. \sref{sec:results} presents and discusses the results of four numerical examples. \sref{sec:conc} summarizes the outcomes and highlights possible future works. In the following discussion, we denote the DEM model based on MLP networks as MLP-DEM and that based on GCNs as GCN-DEM.

\section{Methods}
\label{sec:methods}
\subsection{Graph convolutional network}
\label{GCN}
\begin{figure}[h!] 
    \centering
     \includegraphics[width=0.65\textwidth]{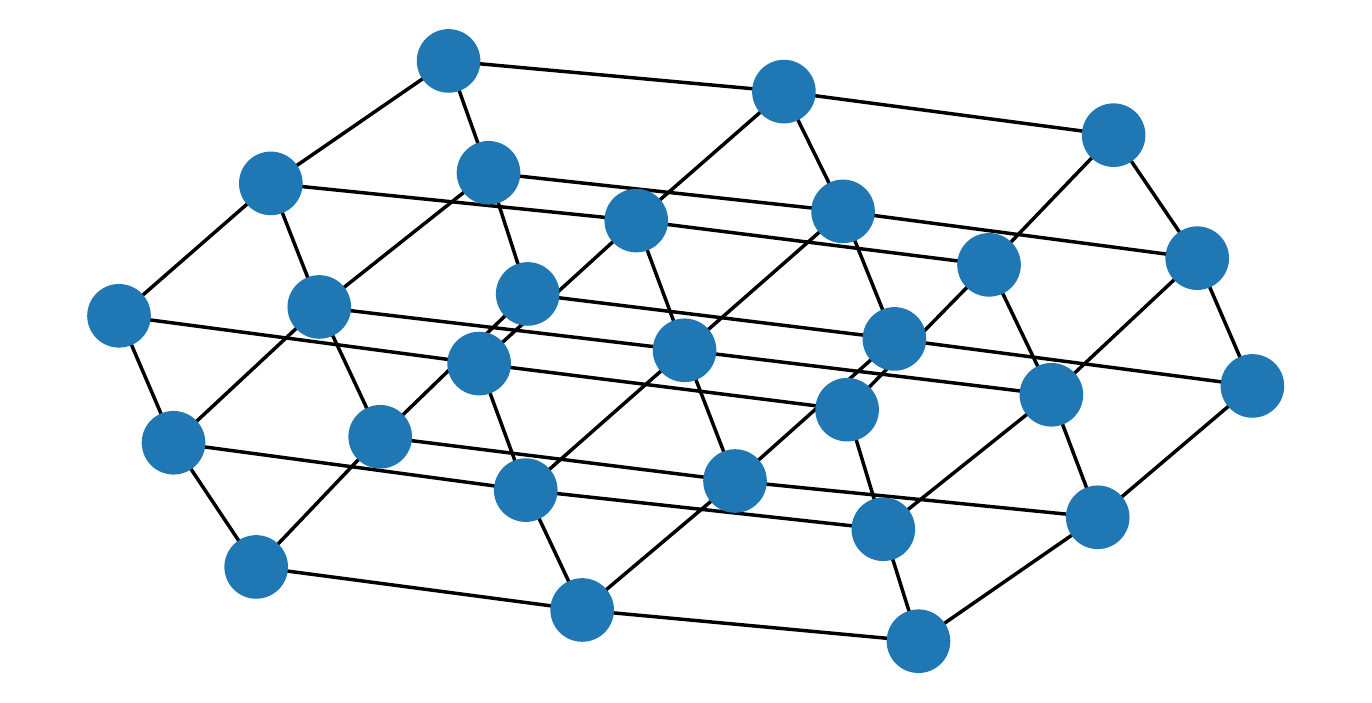}
    \caption{A graph of a cube with 27 nodes and 54 edges. Note that only nodes within a distance of $1/3$ to each other remain connected in the graph.}
    \label{graph}
\end{figure}

A GCN contains a graph that is defined by a set of nodes and the edges between them. For computational mechanics problems, where geometries are typically discretized into meshes, the nodes of domain discretization is a natural choice for constructing a graph. The pair-wise Euclidean distance between the nodes becomes the weight of each edge in the graph. The graph's locality, or sparseness, is ensured by removing edges whose weights exceed a threshold radius $r$. Therefore, only nodes within a distance $r$ from each other remain connected in the graph. An example graph for a 1-by-1-by-1 cube discretized by 27 nodes, forming a 3-by-3-by-3 grid is shown in \fref{graph}; a threshold radius $r=1/3$ was used to generate the graph. In this work, the GCN consists of an input graph of nodal coordinates $\bm{X}$, which is mapped to an output graph of nodal displacement vectors $\bm{u}$. The threshold distance is set to $r = L_x / ( N_x - 1 )$, where $L_x$ and $N_x$ denote the length and number of nodes along the X-axis.

The graph convolution is a key operation in GCN, whose implementation in PyTorch Geometric \cite{Fey/Lenssen/2019} is based on the Chebyshev spectral graph convolution operator \cite{defferrard2016convolutional}. The message passing function is defined as \cite{defferrard2016convolutional}:
\begin{equation}
    \mathbf{X}^{i} = f_{\rm{act}} \left( \sum_{k=1}^{K} \mathbf{Z}^{k} \cdot \mathbf{\Theta}^{k} \right),
\end{equation}
where $i$ denotes the $i^{th}$ layer, K is the polynomial order, $\bm{\Theta}$ denotes the trainable parameters, and the $k^{th}$ basis vectors $\bm{Z}^{(k)}$ is defined recursively as \cite{defferrard2016convolutional,gao2022physics}:
\begin{equation}
\begin{aligned}
    \bm{Z}^{i-1,1} = \bm{X}^{i-1}, \\
    \bm{Z}^{ i-1 , 2 } = \bm{\hat{L}} \cdot \bm{X}^{i-1}, \\
    \bm{Z}^{ i-1 , k } = 2 \cdot \bm{\hat{L}} \cdot
    \bm{Z}^{ i-1 , k-1 } - \mathbf{Z}^{ i-1 , k-2 }, 
\end{aligned}
\end{equation}
and:  
\begin{equation}
    \begin{aligned}
    \bm{\hat{L}} = \bm{L} - \bm{I}, \\
    \bm{L} = \bm{I} - \bm{D}^{-\frac{1}{2}} \bm{A} \bm{D}^{-\frac{1}{2}}.
    \end{aligned}
\end{equation}
$\bm{I}$, $\bm{A}$, and $\bm{D}$ denote the identity, adjacency, and degree matrix of the graph, respectively. The hyperbolic tangent (tanh) function is chosen to be the nonlinear activation function for the Chebyshev convolution layers, and the maximum polynomial degree K=1 is used in this work. Training of the GCN refers to the iterative process where the parameters $\bm{\Theta}$ of each layer are updated by gradient descent \cite{pattanayak2017pro} using the L-BFGS algorithm \cite{zhu1997algorithm}.

\subsection{The deep energy method based on graph convolution network}
\label{GCN-DEM}
In this section, we describe the deep energy method based on graph convolution network (GCN-DEM) in the context of the elasticity equation in structural mechanics. This work considers two material models: linear elastic material in small deformation and Neo-Hookean material in finite deformation. In the absence of any body and inertial forces, the equilibrium equations and boundary conditions (Dirichlet and Neumann) under small deformation can be stated in terms of the Cauchy stress tensor $\bm{\sigma}$ as
\begin{equation}
\begin{aligned}
    \nabla_{\bm{x}} \cdot \bm{\sigma} = \bm{0}, \;\; \forall \bm{X} \in \Omega,\\
    \bm{ u } = \Bar{\bm{u}}, \;\; \forall \bm{X} \in \partial \Omega_u,\\
    \bm{\sigma} \cdot \bm{ n } = \Bar{\bm{t}} , \;\; \forall \bm{X} \in \partial \Omega_t,
\end{aligned}
\end{equation}
where $\bm{n}$, $\Bar{\bm{u}}$ and $\Bar{\bm{t}}$ denote the outward boundary normal, prescribed displacement, and prescribed traction, respectively. $\nabla_{\bm{x}}$ denotes the gradient operator in the current configuration. In the small deformation setting, the strain tensor is given by:
\begin{equation}
    \bm{\epsilon} = \frac{1}{2} ( \nabla_{\bm{x}} \bm{u} + \nabla_{\bm{x}} \bm{u}^T ).
    \label{strain}
\end{equation}
For linear elastic materials, the stress can be computed from the constitutive law as:
\begin{equation}
    \bm{\sigma} = \frac{E}{1+\nu} \bm{\epsilon} + \frac{E\nu}{ (1+\nu)(1-2\nu) } tr(\bm{\epsilon}) \bm{I},
    \label{stress}
\end{equation}
where $E$ and $\nu$ are Young's modulus and Poisson's ratio. The strain energy density $\Psi_{LE}$ in the linear elastic case is given by:
\begin{equation}
    \Psi_{LE} = \frac{1}{2} \bm{\sigma} : \bm{\epsilon}.
\end{equation}

In the finite deformation setting, the equilibrium equations and boundary conditions (Dirichlet and Neumann) can be stated in terms of the first Piola-Kirchhoff stress tensor $\bm{P}$ as:
\begin{equation}
\begin{aligned}
    \nabla_{\bm{X}} \cdot \bm{P} = \bm{0}, \;\; \forall \bm{X} \in \Omega,\\
    \bm{ u } = \Bar{\bm{u}}, \;\; \forall \bm{X} \in \partial \Omega_u,\\
    \bm{P} \cdot \bm{ N } = \Bar{\bm{t}} , \;\; \forall \bm{X} \in \partial \Omega_t,
\end{aligned}
\end{equation}
where $\nabla_{\bm{X}}$ and $\bm{N}$ denote the gradient operator and outward normal in the reference configuration, respectively. $\bm{P}$ is power conjugate to the deformation gradient tensor:
\begin{equation}
    \bm{F} = \nabla_{\bm{X}} \bm{ u } + \bm{I},
\end{equation}
and can be computed from the hyperelastic strain energy density $\Psi_{NH}$ as:
\begin{equation}
    \bm{P} = \frac{\partial \Psi_{NH}}{ \partial \bm{F}}.
\end{equation}
For Neo-Hookean material, $\Psi_{NH}$ is given by \cite{Abaqus2021}:
\begin{equation}
    \Psi_{NH} = C_{10} [ tr( \Bar{\bm{F}} \cdot \Bar{\bm{F}}^T ) - 3 ] + \frac{1}{D_1} [ {\rm{det}}(\bm{F}) - 1 ]^2,
\end{equation}
where:
\begin{equation}
    \Bar{\bm{F}} = {\rm{det}}(\bm{F})^{-\frac{1}{3}} \bm{F},
\end{equation}
is the deviatoric part of the deformation gradient, and $C_{10}$ and $D_1$ are material constants.

In either case, the DEM model seeks the solution to the equilibrium equations via the principle of minimum potential energy (PMPE). For a body in static equilibrium with no applied body forces, the potential energy of the system reads:
\begin{equation}
    \psi(\bm{u}) = \int_{\Omega} \Psi \, dV - \int_{\partial \Omega_t} \Bar{\bm{t}} \cdot \bm{u} \, dA.
    \label{PE}
\end{equation}
The loss function ($\mathcal{L}$) in GCN-DEM is defined identically as the potential of the system:
\begin{equation}
    \mathcal{L}( \bm{u} ) = \psi(\bm{u}).
    \label{loss}
\end{equation}
The Neumann boundary conditions are enforced by the boundary integral part of \eref{PE}, and Dirichlet boundary conditions are enforced directly as in the work of He et al. \cite{he2022deep} to avoid modification of the loss function. The solution to the elasticity problem, as given by the GCN-DEM model, is defined as the stationary point of the potential energy functional:
\begin{equation}
    \bm{u}^* = {\rm{arg}}\min_{ \bm{u} } \, \mathcal{L}( \bm{u} ).
\end{equation}

\subsection{Spatial gradient computation and instability}
\label{grad_cal}
A central component in calculating the loss function value is the evaluation of the spatial gradients of $\bm{u}$. In many previous studies \cite{nguyen2020deep,samaniego2020energy,abueidda2021meshless,abueidda2022deep}, the spatial gradients of $\bm{u}$ are evaluated at discrete points inside the domain through AD of the underlying NN model:
\begin{equation}
    \frac{ \partial \bm{u}(\bm{X}) }{ \partial \bm{X} } = \frac{ \partial \bm{u} }{ \partial \bm{z}^n } \cdot \frac{ \partial \bm{z}^n }{ \partial \bm{z}^{n-1} } \dots \cdot \frac{ \partial \bm{z}^2 }{ \partial \bm{z}^{1} } \cdot \frac{ \partial \bm{z}^1 }{ \partial \bm{X} },
\end{equation}
where $\bm{z}^i$ denotes the output of the $i^{th}$ layer of the NN model. The first and last layers are the input and output layer, respectively. After calculating gradients at the nodes $\bm{X}$ in the domain, Simpson's rule or trapezoidal rule is typically used to perform the integration. The AD-based approach is straightforward to implement, as AD is already used in the training of the NN model during backpropagation. This point-based method also alleviates the need to form a mesh of the computational domain like in FEM.

The studies by Chadha et al. \cite{chadha2022optimizing} and He et al. \cite{he2022deep} demonstrated gradient computation and numerical integration through finite element SFs and Gauss quadrature. In this case, the spatial gradients are given by:
\begin{equation}
    \frac{ \partial \bm{u}(\bm{X}) }{ \partial \bm{X} } \approx \frac{\partial \bm{\phi} }{\partial \bm{\xi}} \bm{J}^{-1} \cdot \bm{u},
\end{equation}
where $\bm{\phi}$ , $\bm{\xi}$, and $\bm{u}$ denote the finite element SFs, natural coordinates, and displacement vector (evaluated at discrete nodes), respectively. $\bm{J} = \frac{\partial \bm{X} }{\partial \bm{\xi}}$ denotes the Jacobian matrix of the isoparametric mapping of the finite element. Instead of evaluating the gradients at the nodes, the gradients are evaluated at the quadrature integration points of the 'elements', reminiscent to classical FEM procedures. Therefore, the SF-based approach requires the formation of isoparametric elements from the nodes in the computational domain.

\begin{figure}[h!] 
    \centering
     \subfloat[]{
         \includegraphics[width=0.45\textwidth]{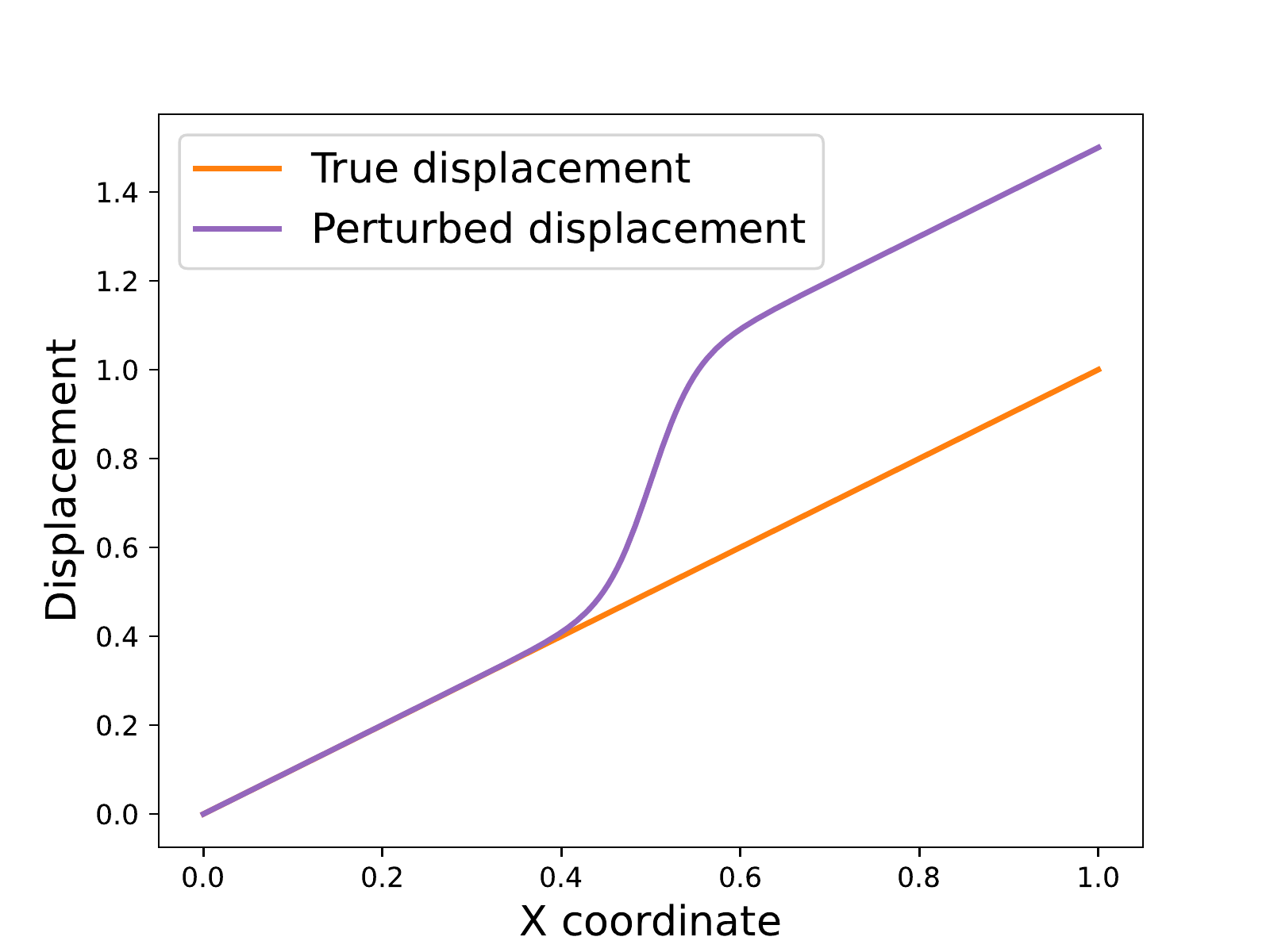}
         \label{theory_disp}
     }
     \subfloat[]{
         \includegraphics[width=0.45\textwidth]{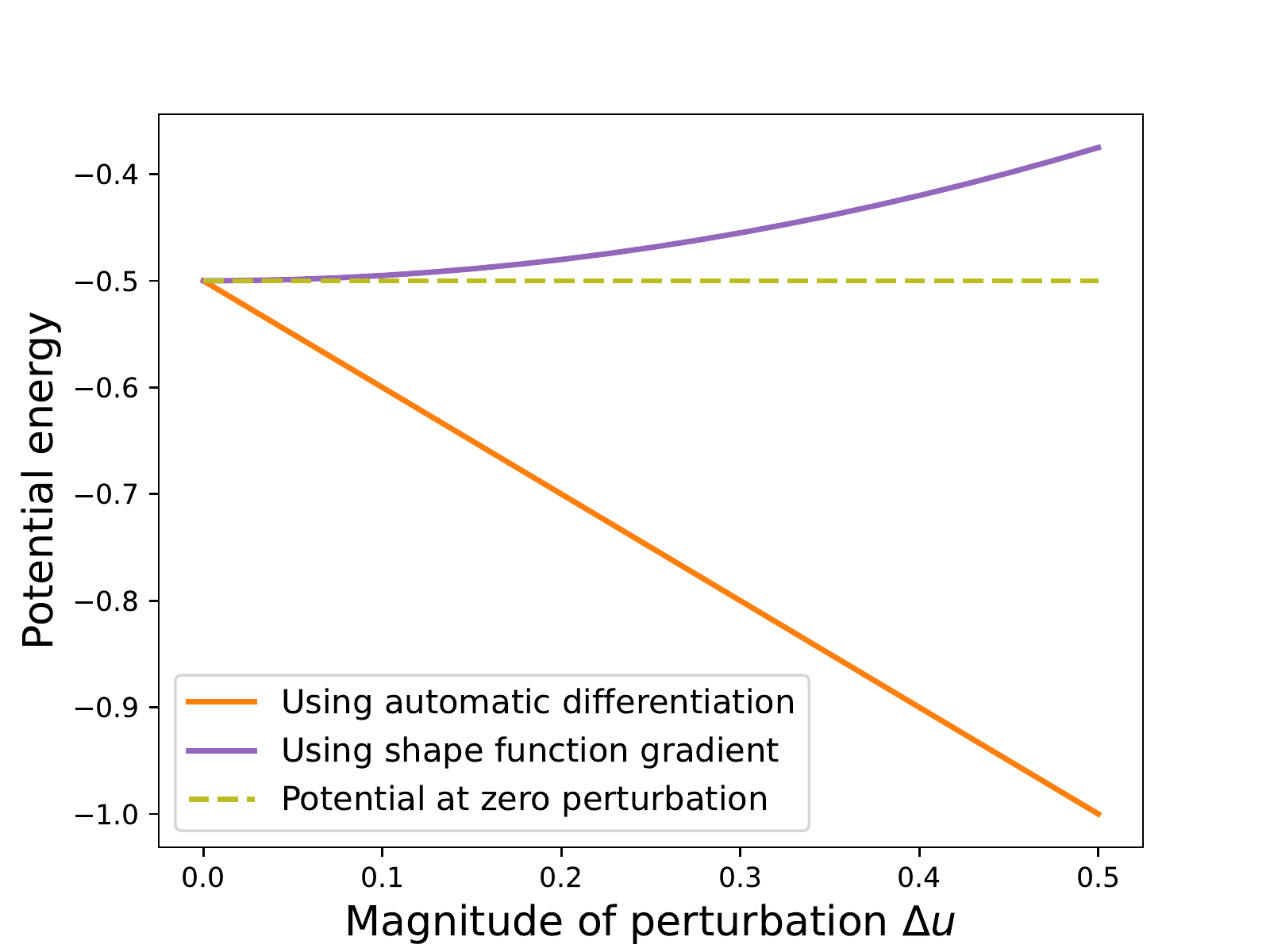}
         \label{theory_energy}
     }
    \caption{1D bar example: \psubref{theory_disp} True and perturbed displacements. $\Delta u$ is taken to be 0.5 in the plot. \psubref{theory_energy} Comparison of system potential energy at different $\Delta u$}
\end{figure}

We shall demonstrate with a simple 1D example that, despite being straightforward to implement, the AD-based spatial gradient computation can be susceptible to instability that leads to divergence of the DEM method. Consider a 1D bar whose dimensions and properties are all unity, subject to a unit tensile force at its right end. The displacement is simply $u(x) = x$. Let the bar be discretized into two nodes located at $x=0$ and $x=1$, and let $u_p (x)$ be a perturbation to the displacement field given by:
\begin{equation}
    u_p (x) = \frac{\Delta u}{2} ( {\rm{tanh}}[ 20 ( x - 0.5 ) ] + 1 ),
\end{equation}
where $\Delta u$ characterizes the magnitude of the perturbation\footnote{The factor 20 is chosen randomly and does not affect the argument we want to make.}. This perturbation can be thought of as a strain localization at the center of the bar. The true displacement $u$ and the perturbed displacement $\Tilde{u} = u + u_p$ are plotted in \fref{theory_disp}. 
First, we use AD (in this case, it corresponds to analytical differentiation) and trapezoidal rule to evaluate \eref{PE} using the two nodes. Then, we use SF to approximate the displacement gradient and use one-point Gauss quadrature to evaluate \eref{PE}. We repeat the process for multiple values of $\Delta u$ and plot the results in \fref{theory_energy}. At zero perturbation, both methods yield the true potential energy of the system, which is -0.5. However, as the perturbation increases, the integrated potential calculated from 
AD keeps decreasing, thereby violating PMPE (the unperturbed equilibrium state has minimum potential). While that calculated from the SF gradient remains larger than the true minimum and does not violate PMPE. This behavior is expected, as the tanh function has vanishing gradient away from its center. Therefore the point-wise AD fails to detect strain localization that occurs in between the nodes, while the boundary integral part of \eref{PE} keeps increasing with increased perturbation. On the contrary, strain localization can always be detected by an element-based SF gradient, as the gradient is computed from the difference of the displacements of the two nodes. Therefore, we argue that the SF-based gradient computation is more stable than its AD-based counterpart with respect to strain localization.

The inability of AD to detect and penalize strain localization between nodes can adversely affect the solution of energy-based methods, as the loss function has a lower value than the true equilibrium solution when strain localization occurs, thus driving the solution to the incorrect direction of increasing strain localization.  Although the argument is presented using a simple example, it can be shown easily that the argument generalizes to arbitrary geometries with arbitrary spacing between the nodes. In \sref{sec:results}, we demonstrate that this type of instability is indeed observed when AD-based gradient computation is used, especially when a large load is applied.

\section{Results and discussion}
\label{sec:results}
In this section, we critically compare GCN-DEM and MLP-DEM using four case studies. In addition, we compare the solution stability of AD- and SF-based gradient computations. In all cases except \sref{external_example}, the geometry considered is a cantilever beam of dimensions 4-by-1-by-1 units, subject to downward traction at its right surface. In the first two cases, 3700 nodes were placed in the domain, forming a 37-by-10-by-10 grid. Although a structured hexahedral grid was used in all the examples, we highlight that the GCN-based DEM can be easily extended to unstructured tetrahedral (or triangular for 2D) meshes as in the work of Fuhg et al. \cite{fuhg2022mixed}. In the first case, we consider a linear elastic material with $E=1000$ and $\nu=0.3$, subjected to sequentially increasing loads. In the second case, we consider a Neo-Hookean material with $C_{10}=192.31$ and $D_1=0.0024$, subjected to the same loads as in case 1. Then, we demonstrate that instability can occur in other AD-based DEM implementations, not limited to the ones presented in our work. In the last case, we investigate how grid refinement can remedy the instabilities caused by AD-based gradient computation. In all cases except \sref{external_example}, we compared the displacement results with those obtained from FEM using Abaqus/Standard \cite{Abaqus2021}. The GCN-DEM and MLP-DEM models were implemented in PyTorch (version 1.11.0) \cite{NEURIPS2019_9015}. The MLP-DEM implementation was adopted from Abueidda et al. \cite{abueidda2022deep}. All training of the NNs was done on an Intel i7-11800H processor. For a fair and consistent comparison, the GNN in GCN-DEM and the MLP model in MLP-DEM share the same network structure: they have 6 layers (including input and output). The number of neurons in each layer is 3, 16, 32, 64, 32 , 16, and 3, respectively. The hyperbolic tangent function was used as activation function for all layers except the output, which has linear activation. The L-BFGS optimizer \cite{zhu1997algorithm} with a fixed learning rate of 0.01 is used to train the models. Training process is stopped when a maximum of 20 training iterations is reached, or when the relative change in loss function value is less than $5\times10^{-5}$. To test the robustness of both methods, we applied the full magnitude of the external load in a single load step, which is in contrast to FEM, where large loads are applied gradually throughout several load steps for better convergence. This approach puts the stability and robustness of the methods at severe deformation to test.

\subsection{Linear elastic material}
\label{LE}
\begin{figure}[h!] 
    \centering
     \subfloat[]{
         \includegraphics[width=0.33\textwidth]{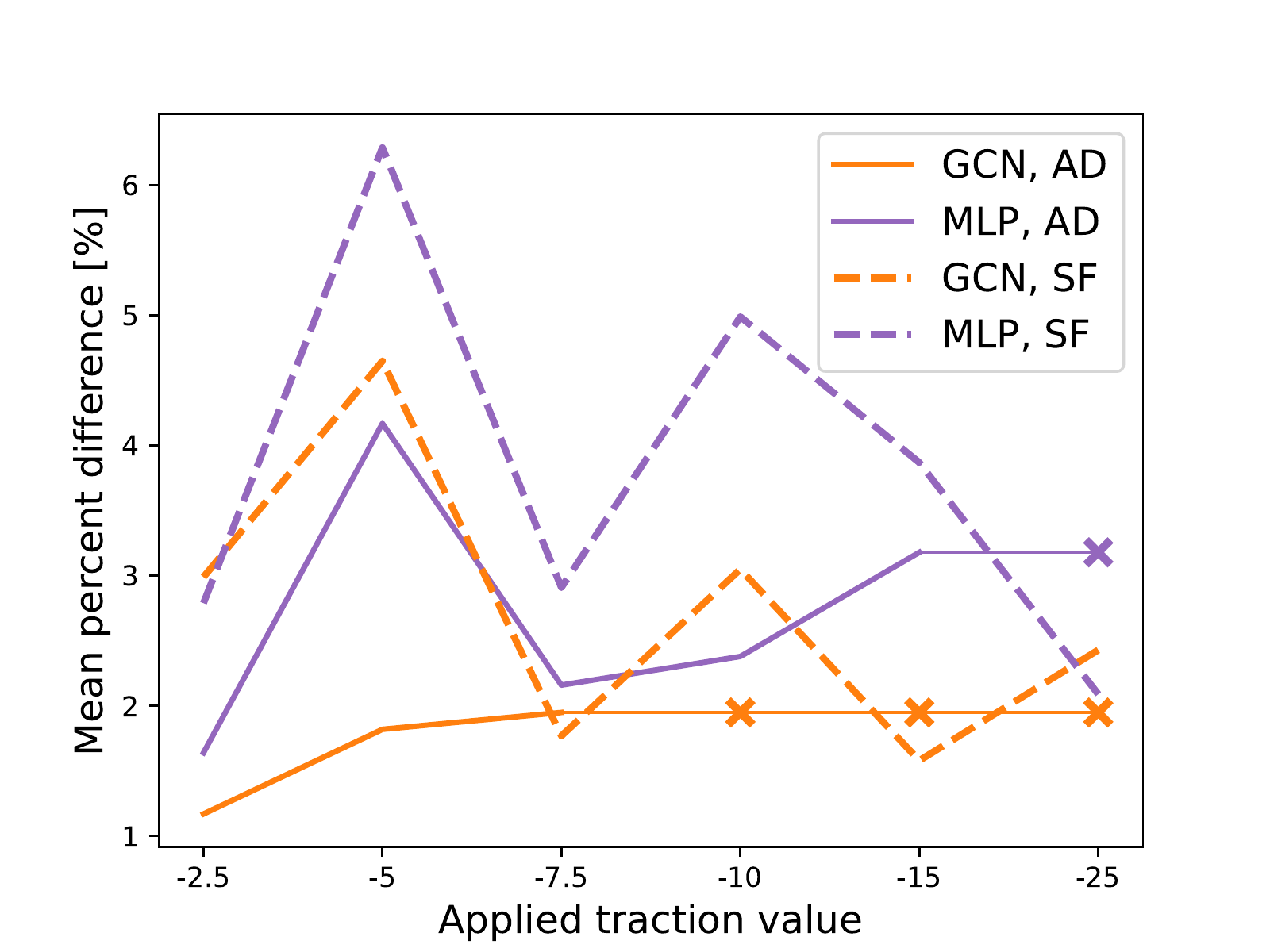}
         \label{fig:LE_RD}
     }
     \subfloat[]{
         \includegraphics[width=0.33\textwidth]{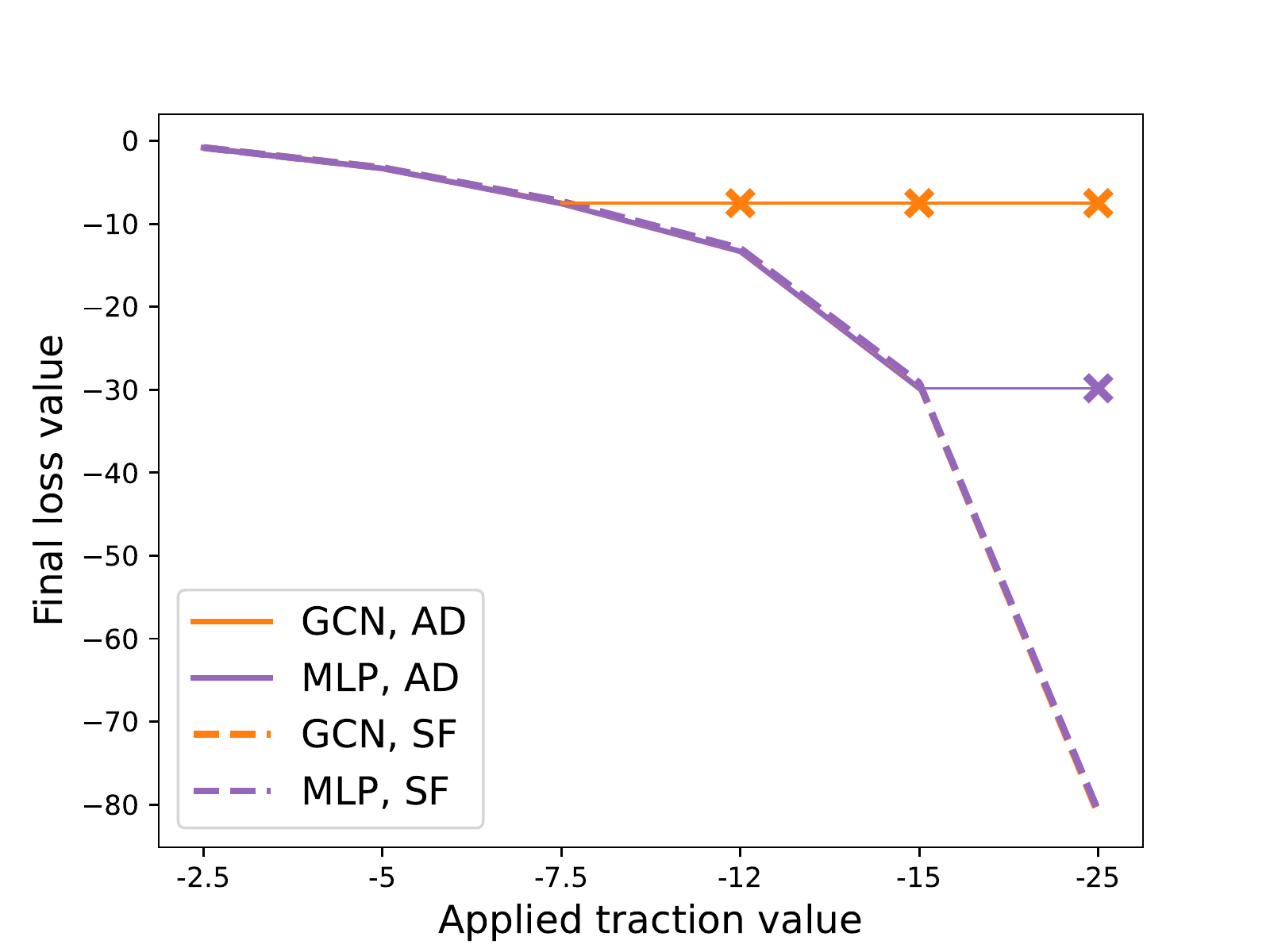}
         \label{fig:LE_loss}
     }
     \subfloat[]{
         \includegraphics[width=0.33\textwidth]{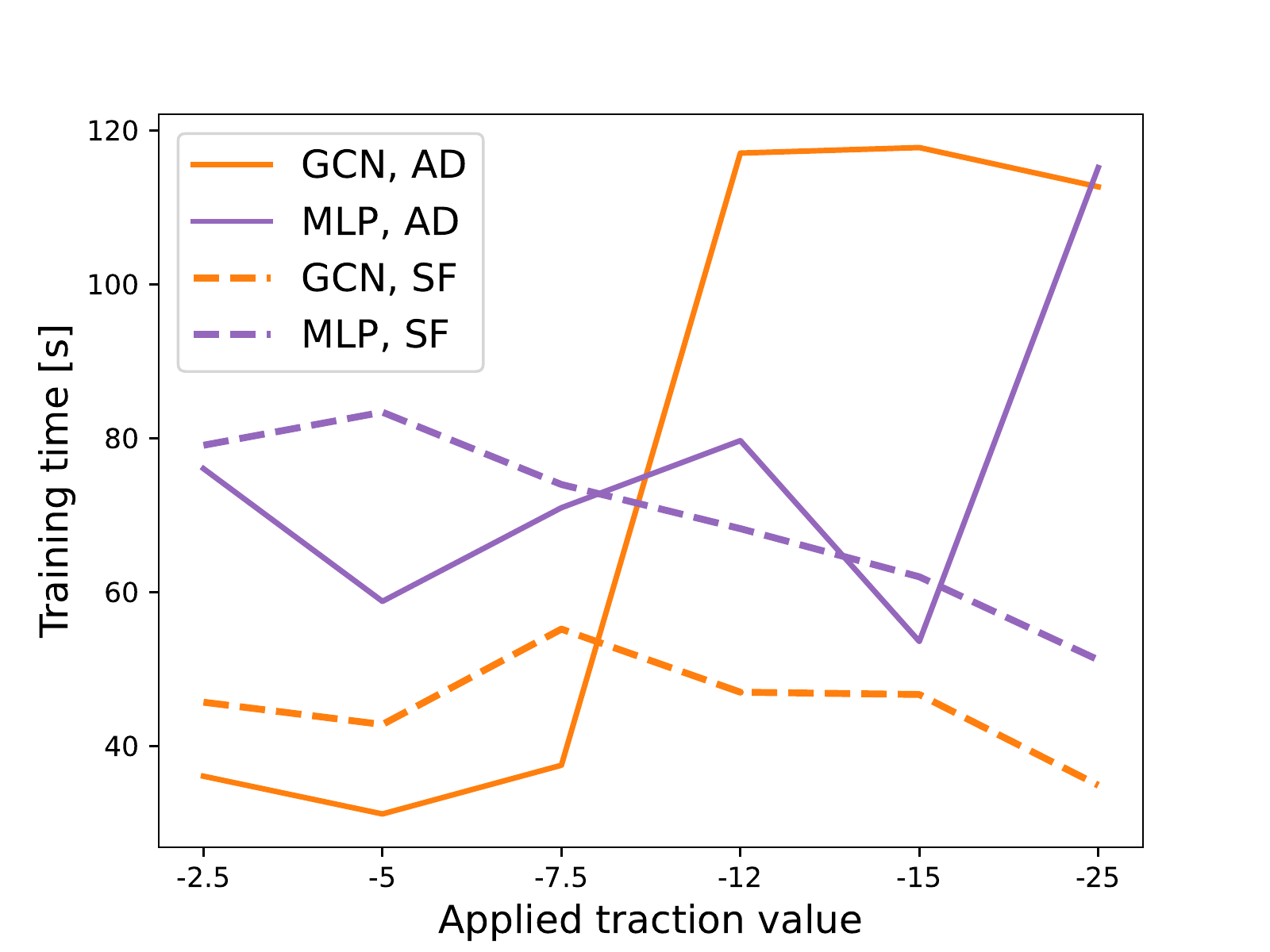}
         \label{fig:LE_time}
     }
    \caption{Comparing GCN-DEM and MLP-DEM, AD-based and SF-based gradient computation: \psubref{fig:LE_RD} Relative difference in displacements. \psubref{fig:LE_loss} Final loss value.
    \psubref{fig:LE_time} Training time for simulation. In \psubref{fig:LE_RD} and \psubref{fig:LE_loss}, simulations that failed to converge are marked with a $\bm{\times}$.}
    \label{LE_lineplot}
\end{figure}

\begin{figure}[h!]
\newcommand\x{0.2}
    \centering
    \begin{tabular}{ c c c c c c }
    \begin{minipage}[c]{\x\textwidth}
       \centering 
        \subfloat[GCN, RD$_x$=0.42\%]{\includegraphics[trim={22cm 7cm 16cm 12cm},clip,width=\textwidth]{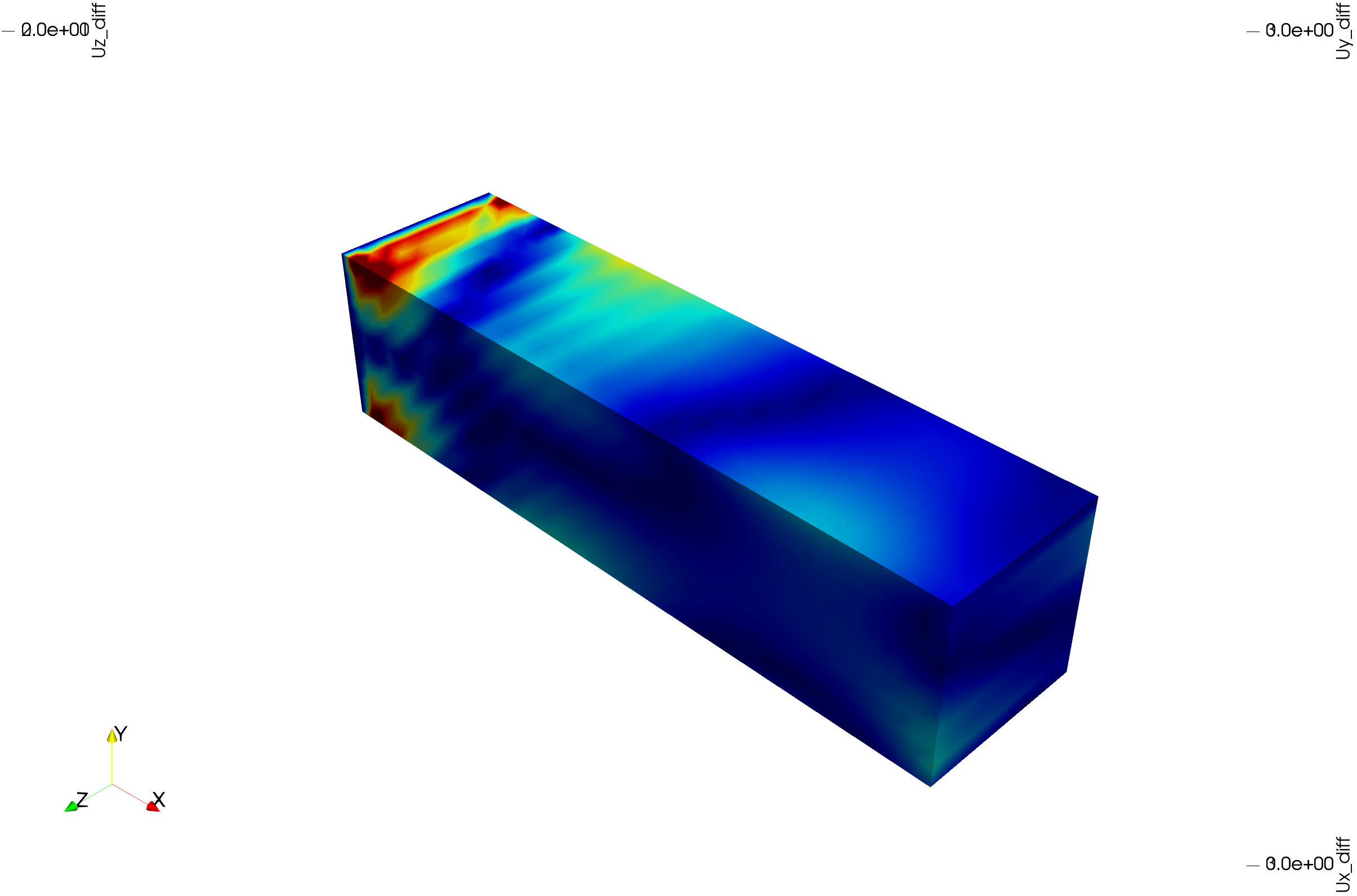}
        \label{fig:gax}}
    \end{minipage} &
    \multirow{2}{*}{
    \begin{minipage}[c]{0.05\textwidth}
       \centering 
        \includegraphics[trim={72cm 0cm 1cm 25cm},clip,width=\textwidth]{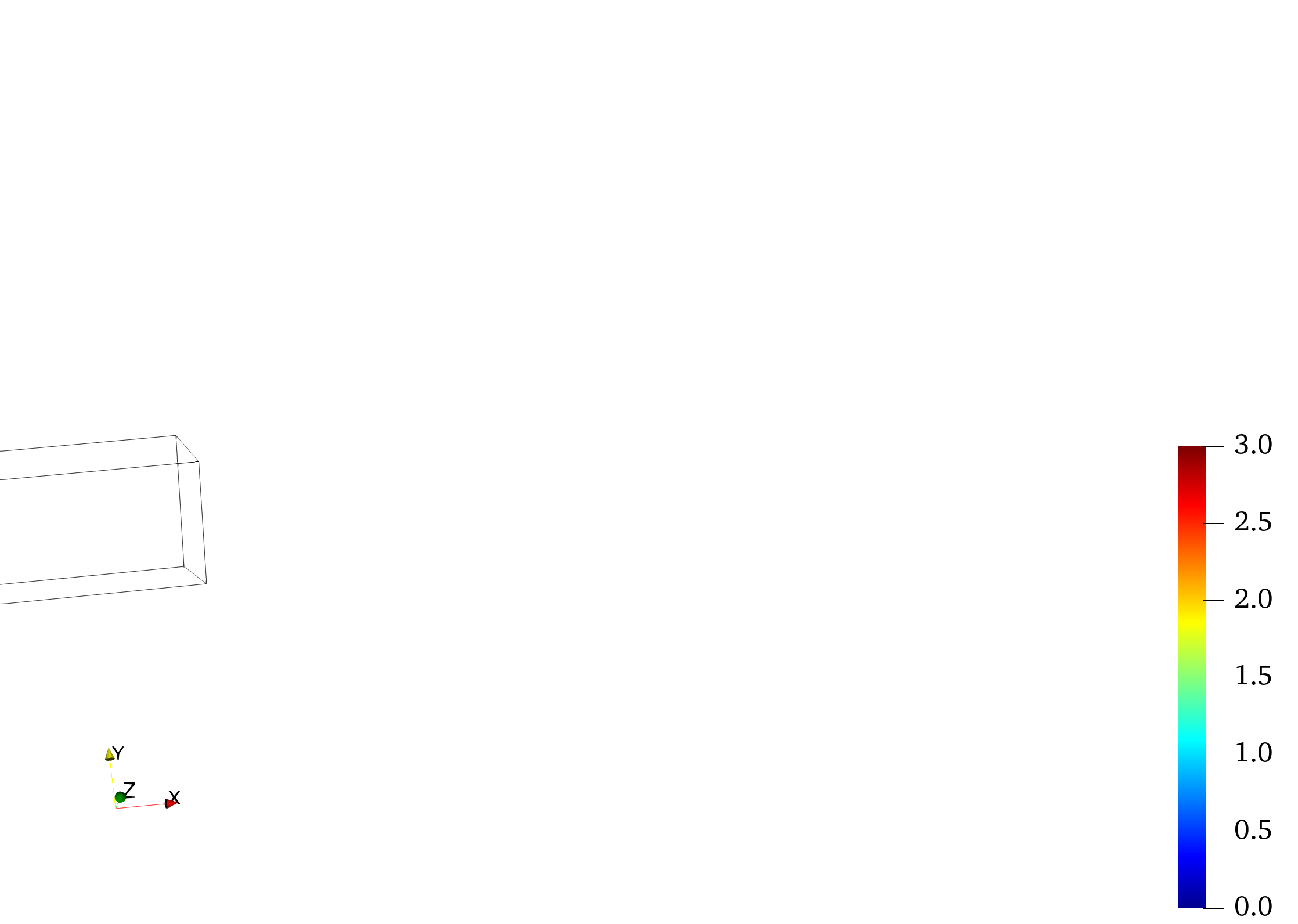}
    \end{minipage}
    } &
    \begin{minipage}[c]{\x\textwidth}
       \centering 
        \subfloat[GCN, RD$_y$=0.15\%]{\includegraphics[trim={22cm 7cm 16cm 12cm},clip,width=\textwidth]{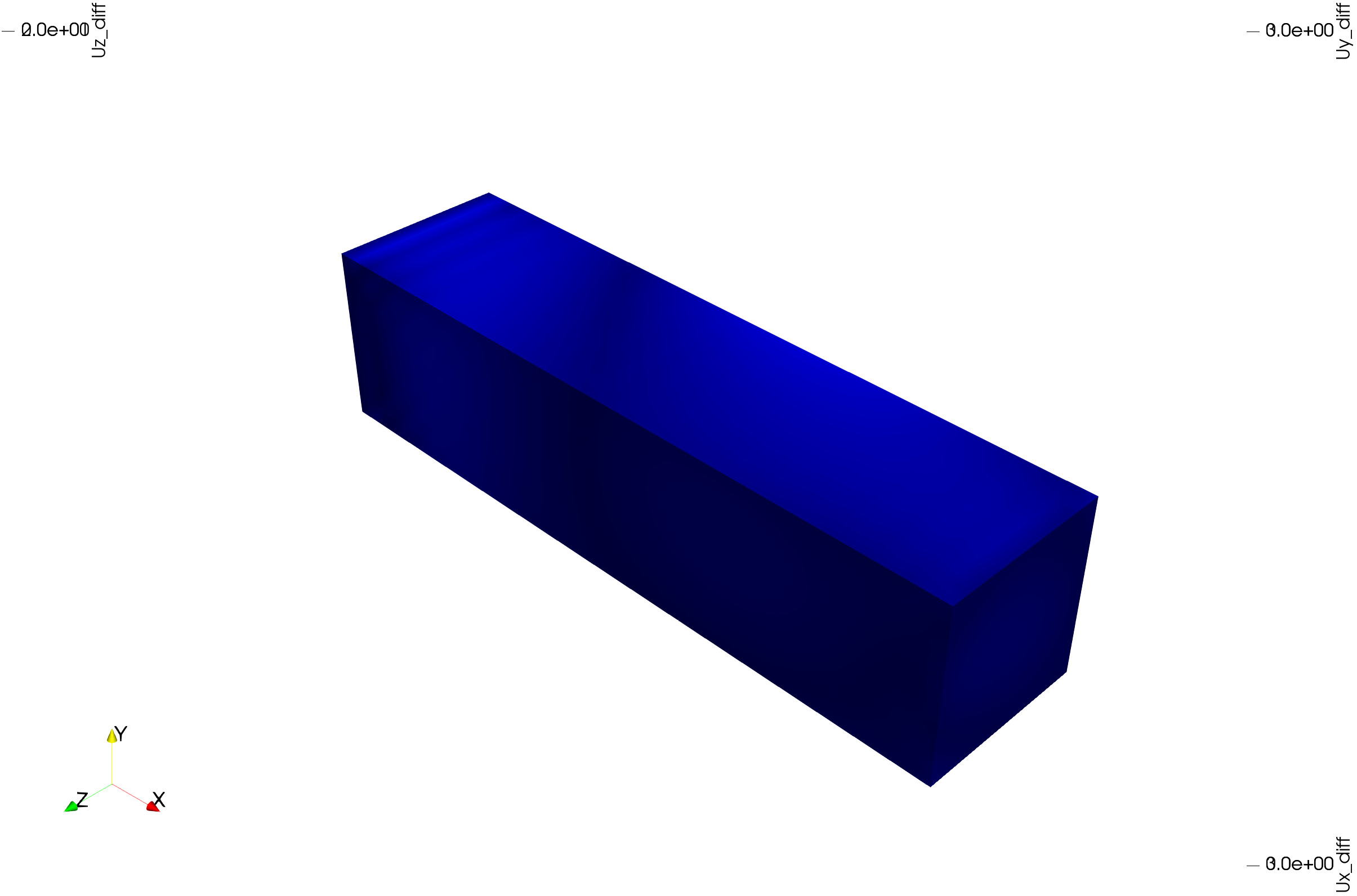}
        \label{fig:gay}}
    \end{minipage} &
    \multirow{2}{*}{
    \begin{minipage}[c]{0.05\textwidth}
       \centering 
        \includegraphics[trim={72cm 0cm 1cm 25cm},clip,width=\textwidth]{Figures/CB1.png}
    \end{minipage}
    } &
    \begin{minipage}[c]{\x\textwidth}
       \centering 
        \subfloat[GCN, RD$_z$=5.30\%]{\includegraphics[trim={22cm 7cm 16cm 12cm},clip,width=\textwidth]{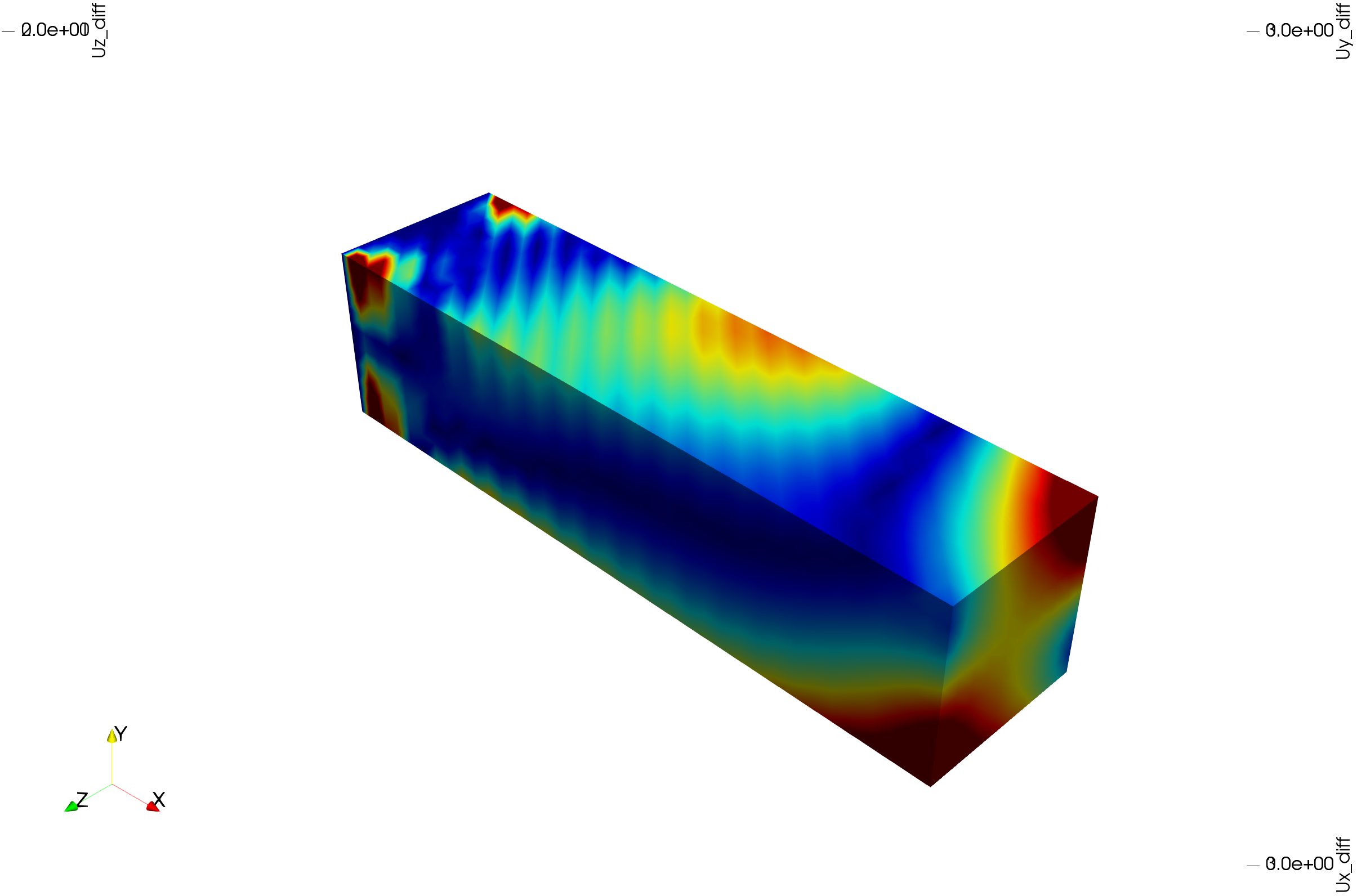}
        \label{fig:gaz}}
    \end{minipage} &
    \multirow{2}{*}{
    \begin{minipage}[c]{0.05\textwidth}
       \centering 
        \includegraphics[trim={71cm 0cm 1.5cm 24cm},clip,width=\textwidth]{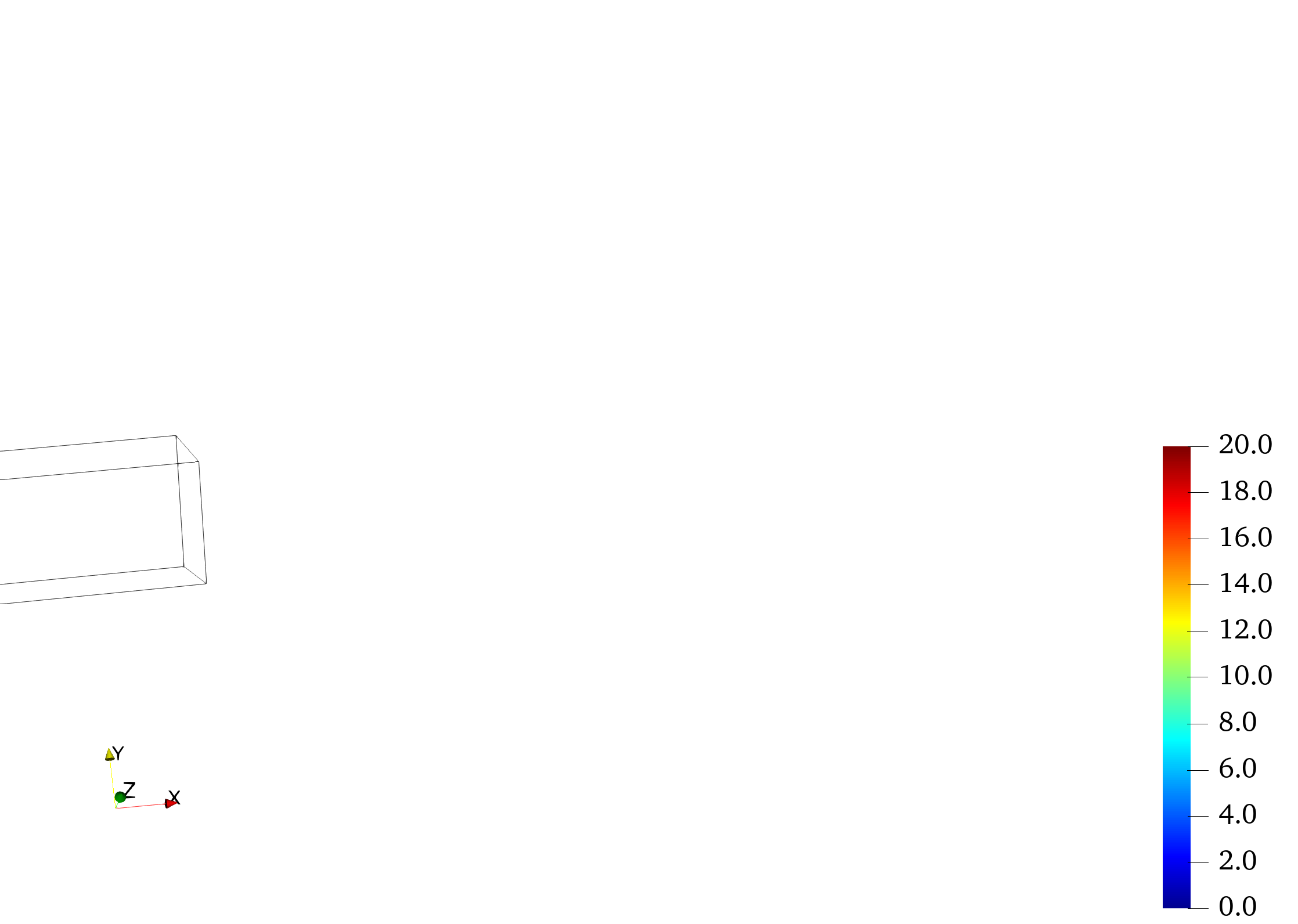}
    \end{minipage}
    } \\
    \begin{minipage}[c]{\x\textwidth}
       \centering 
        \subfloat[MLP, RD$_x$=0.50\%]{\includegraphics[trim={22cm 7cm 16cm 12cm},clip,width=\textwidth]{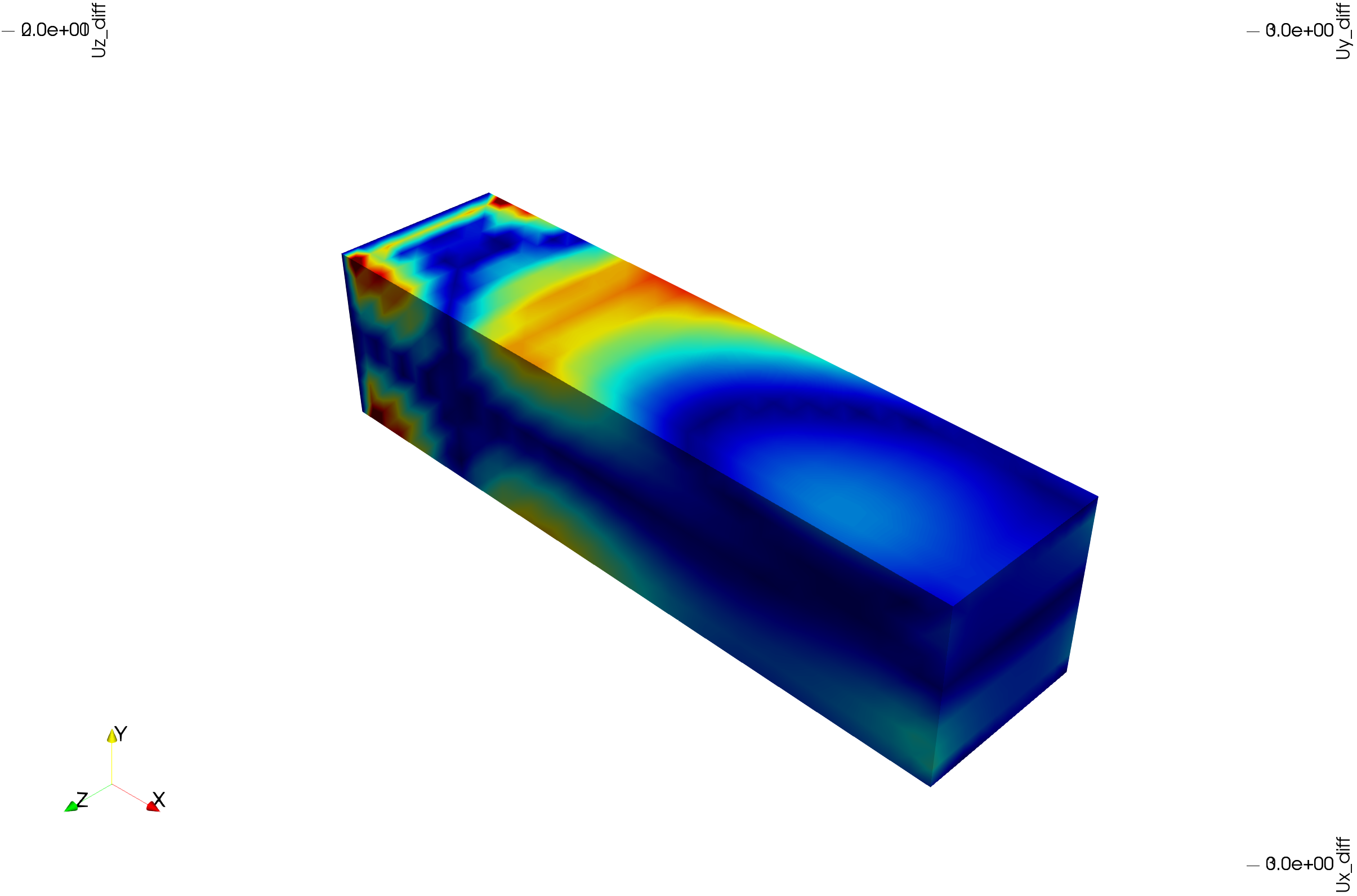}
        \label{fig:dax}}
    \end{minipage} & &
    \begin{minipage}[c]{\x\textwidth}
       \centering 
        \subfloat[MLP, RD$_y$=0.49\%]{\includegraphics[trim={22cm 7cm 16cm 12cm},clip,width=\textwidth]{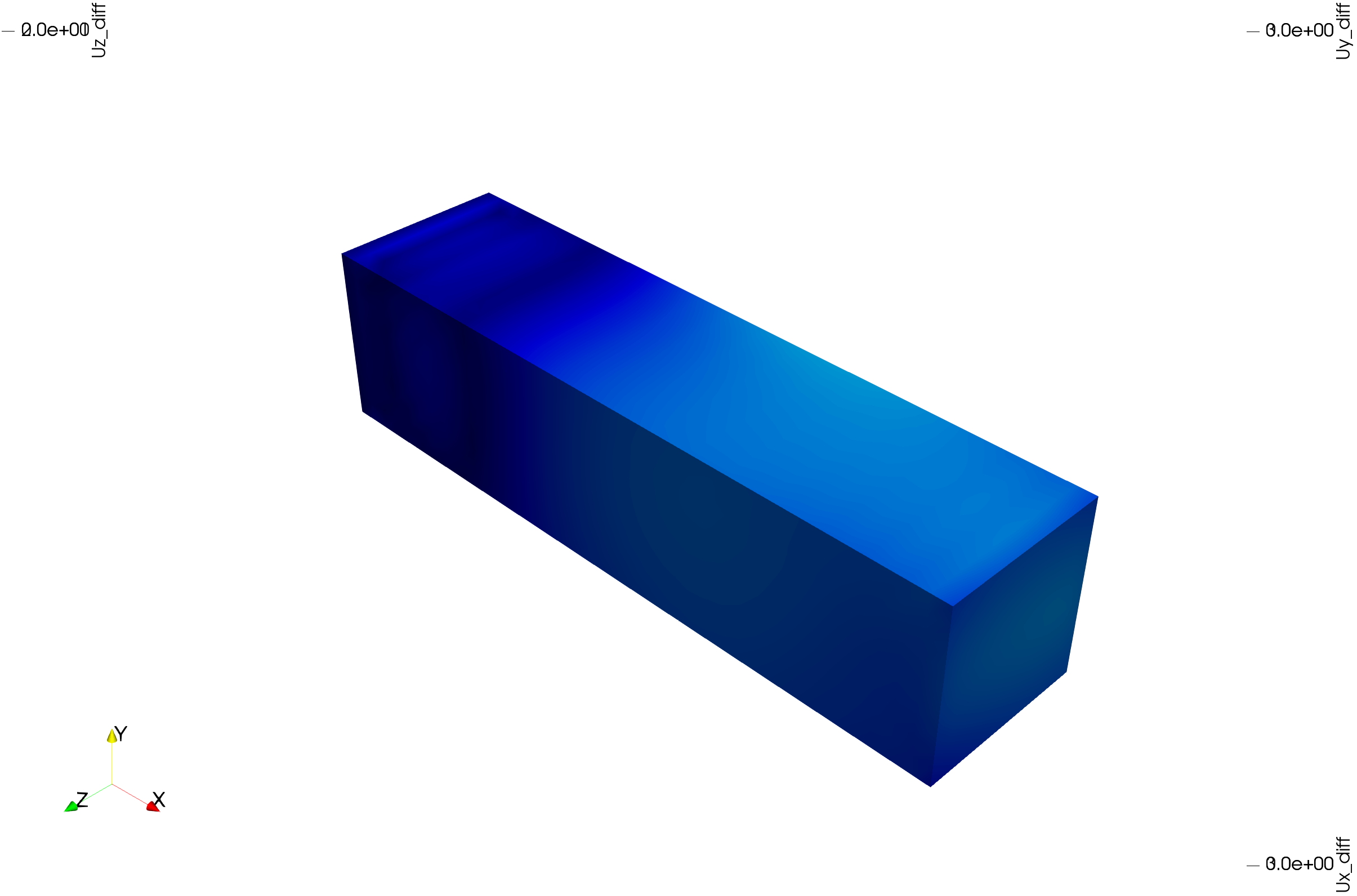}
        \label{fig:day}}
    \end{minipage} & &
    \begin{minipage}[c]{\x\textwidth}
       \centering 
        \subfloat[MLP, RD$_z$=5.50\%]{\includegraphics[trim={22cm 7cm 16cm 12cm},clip,width=\textwidth]{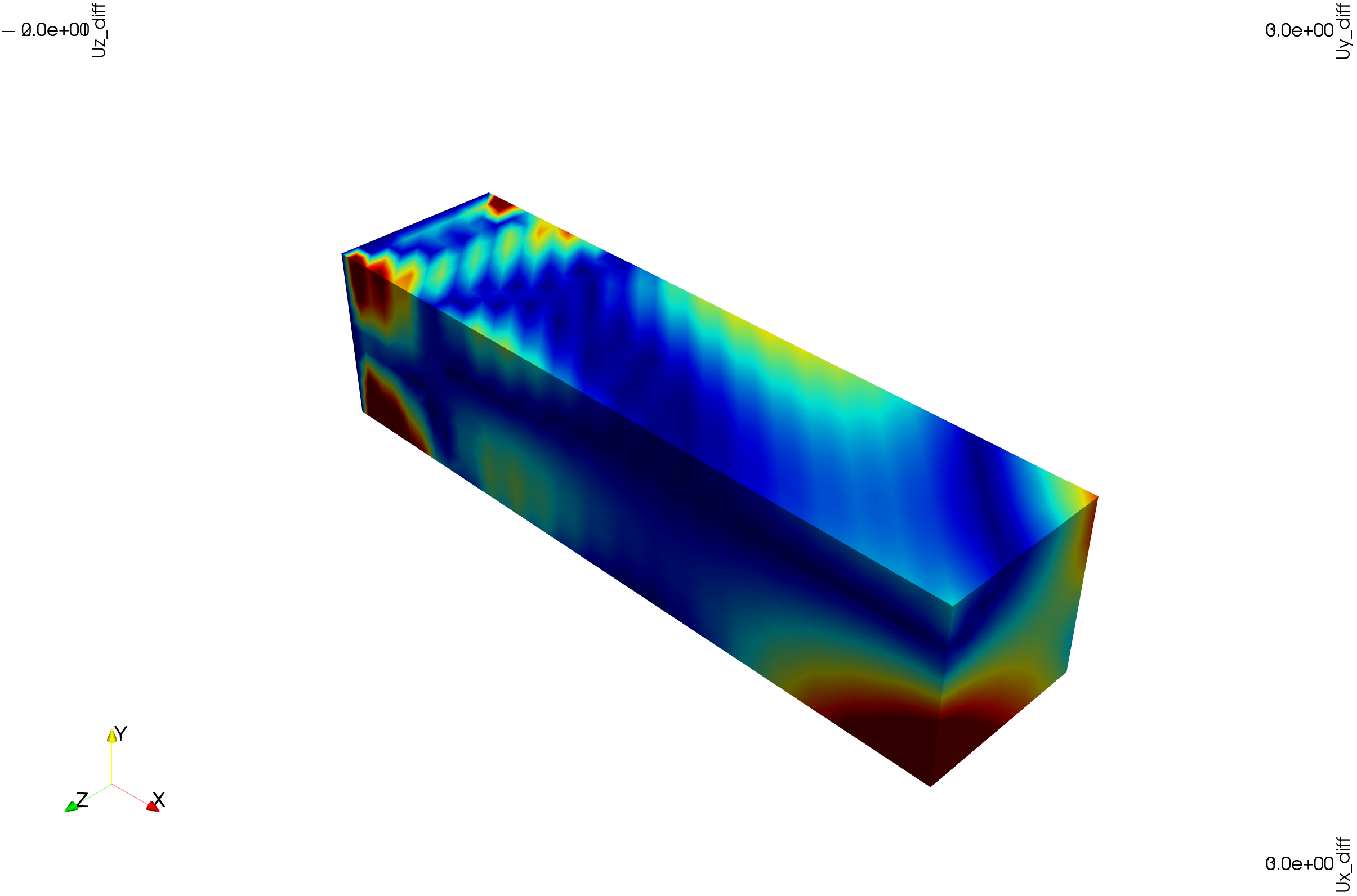}
        \label{fig:daz}}
    \end{minipage} & \\
    \end{tabular}
    \caption{Comparison of relative displacement difference between GCN-DEM and MLP-DEM, using AD for gradient computation, linear elastic material. The mean relative difference for each displacement component is reported in the caption.}
    \label{LE_AD}
\end{figure}

\begin{figure}[h!]
\newcommand\x{0.2}
    \centering
    \begin{tabular}{ c c c c c c }
    \begin{minipage}[c]{\x\textwidth}
       \centering 
        \subfloat[GCN, RD$_x$=0.99\%]{\includegraphics[trim={22cm 7cm 16cm 12cm},clip,width=\textwidth]{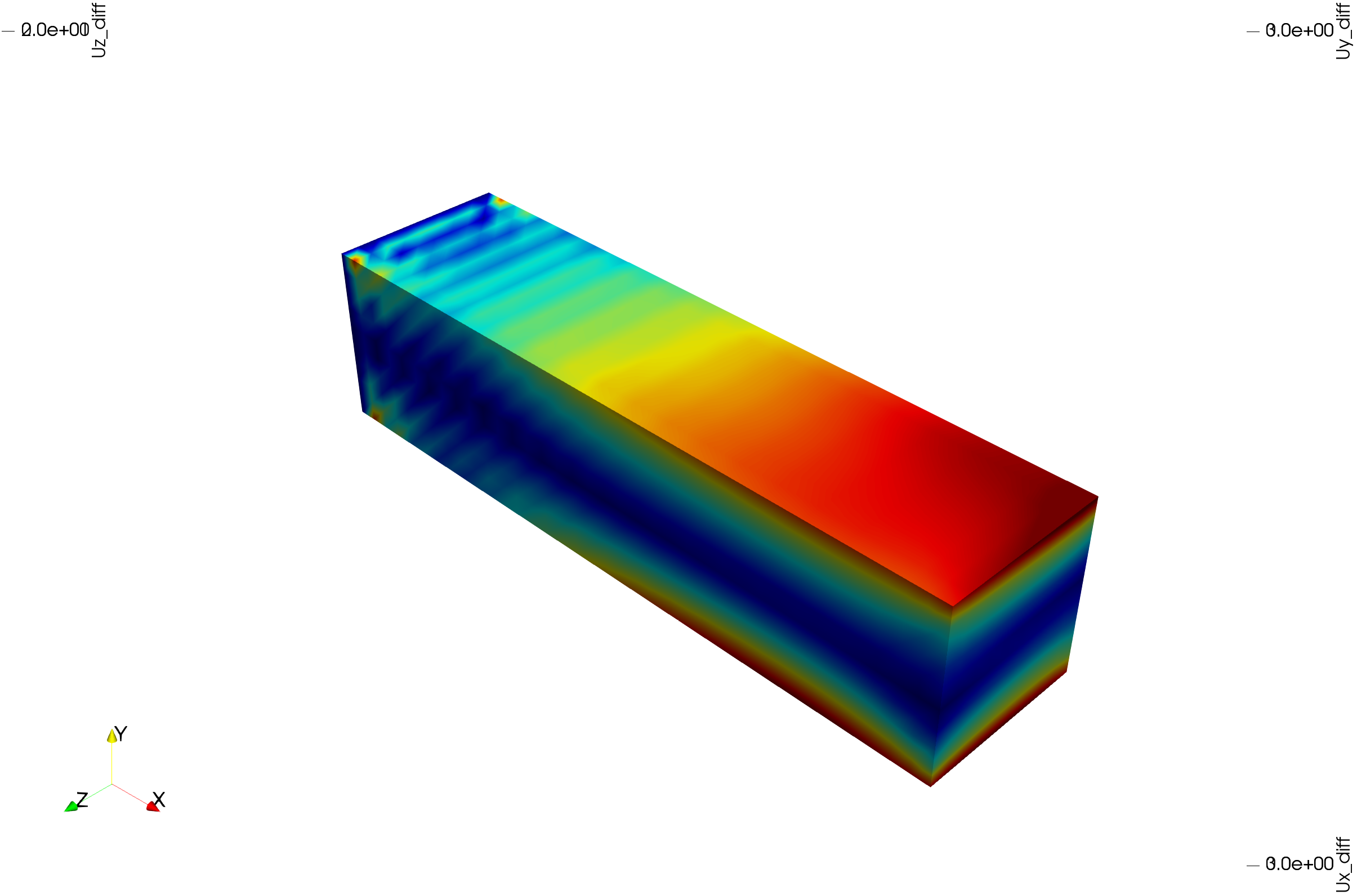}
        \label{fig:gax}}
    \end{minipage} &
    \multirow{2}{*}{
    \begin{minipage}[c]{0.05\textwidth}
       \centering 
        \includegraphics[trim={72cm 0cm 1cm 25cm},clip,width=\textwidth]{Figures/CB1.png}
    \end{minipage}
    } &
    \begin{minipage}[c]{\x\textwidth}
       \centering 
        \subfloat[GCN, RD$_y$=1.06\%]{\includegraphics[trim={22cm 7cm 16cm 12cm},clip,width=\textwidth]{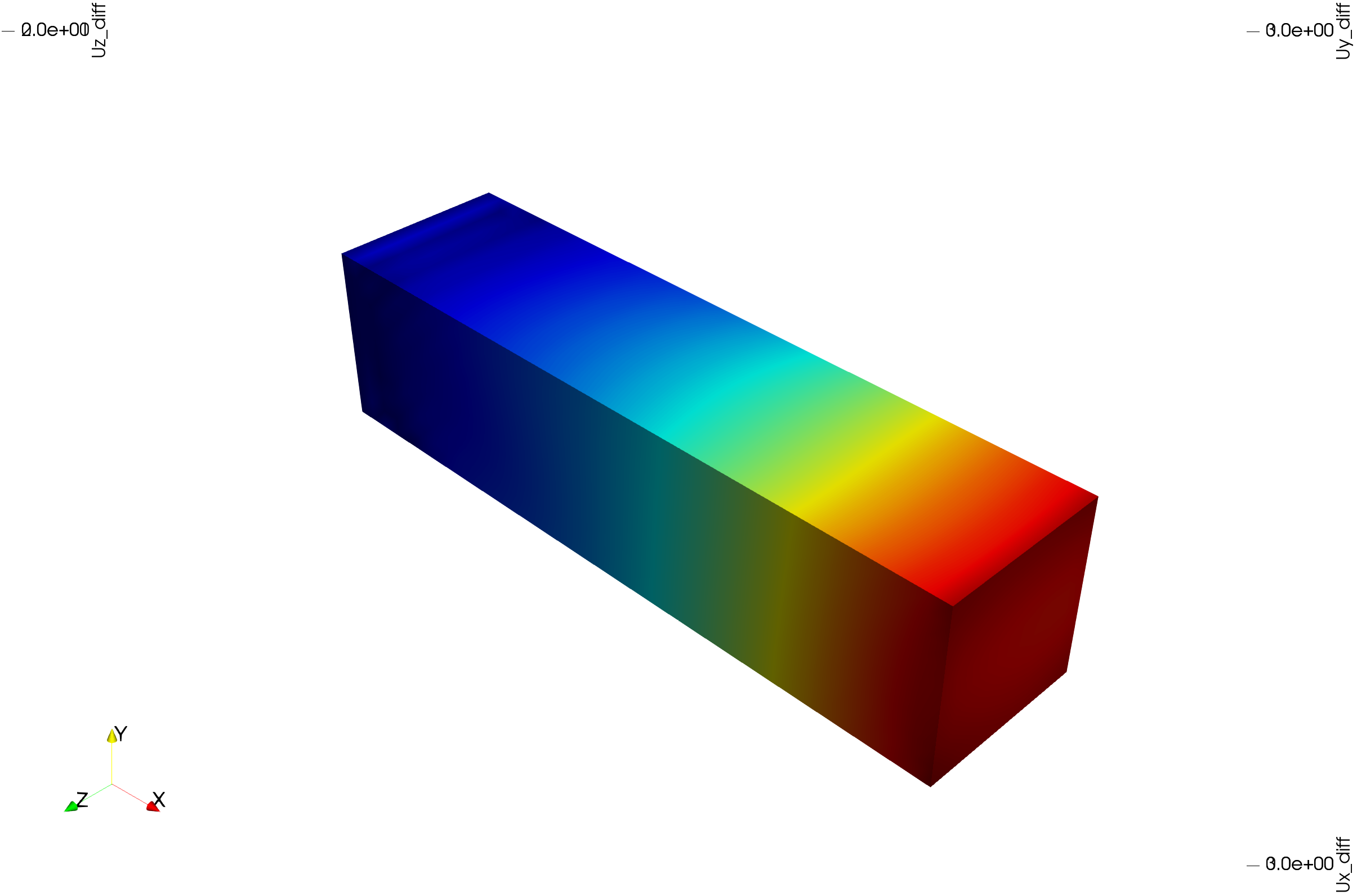}
        \label{fig:gay}}
    \end{minipage} &
    \multirow{2}{*}{
    \begin{minipage}[c]{0.05\textwidth}
       \centering 
        \includegraphics[trim={72cm 0cm 1cm 25cm},clip,width=\textwidth]{Figures/CB1.png}
    \end{minipage}
    } &
    \begin{minipage}[c]{\x\textwidth}
       \centering 
        \subfloat[GCN, RD$_z$=3.27\%]{\includegraphics[trim={22cm 7cm 16cm 12cm},clip,width=\textwidth]{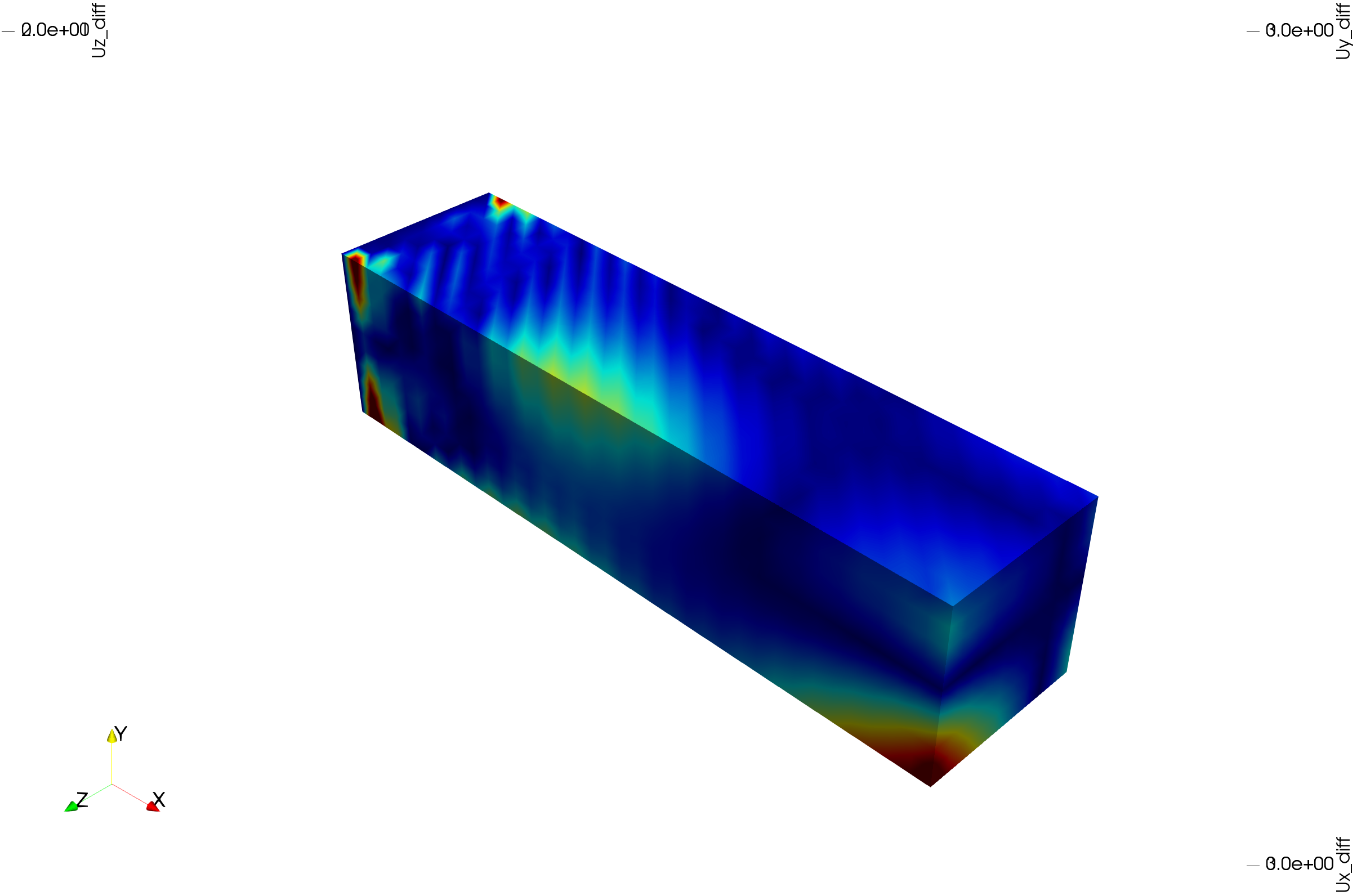}
        \label{fig:gaz}}
    \end{minipage} &
    \multirow{2}{*}{
    \begin{minipage}[c]{0.05\textwidth}
       \centering 
        \includegraphics[trim={71cm 0cm 1.5cm 24cm},clip,width=\textwidth]{Figures/CB2.png}
    \end{minipage}
    } \\
    \begin{minipage}[c]{\x\textwidth}
       \centering 
        \subfloat[MLP, RD$_x$=1.18\%]{\includegraphics[trim={22cm 7cm 16cm 12cm},clip,width=\textwidth]{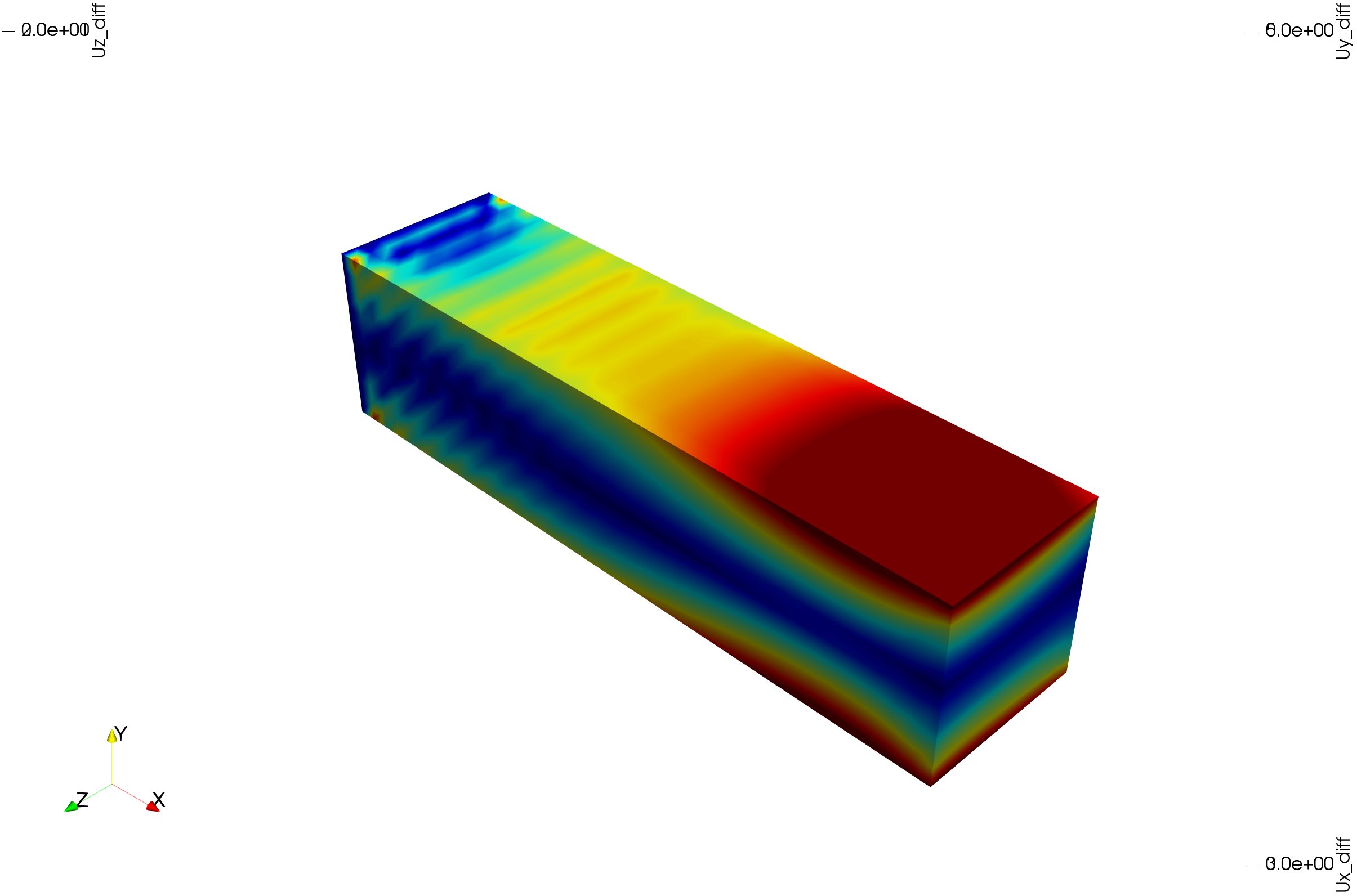}
        \label{fig:dax}}
    \end{minipage} & &
    \begin{minipage}[c]{\x\textwidth}
       \centering 
        \subfloat[MLP, RD$_y$=1.25\%]{\includegraphics[trim={22cm 7cm 16cm 12cm},clip,width=\textwidth]{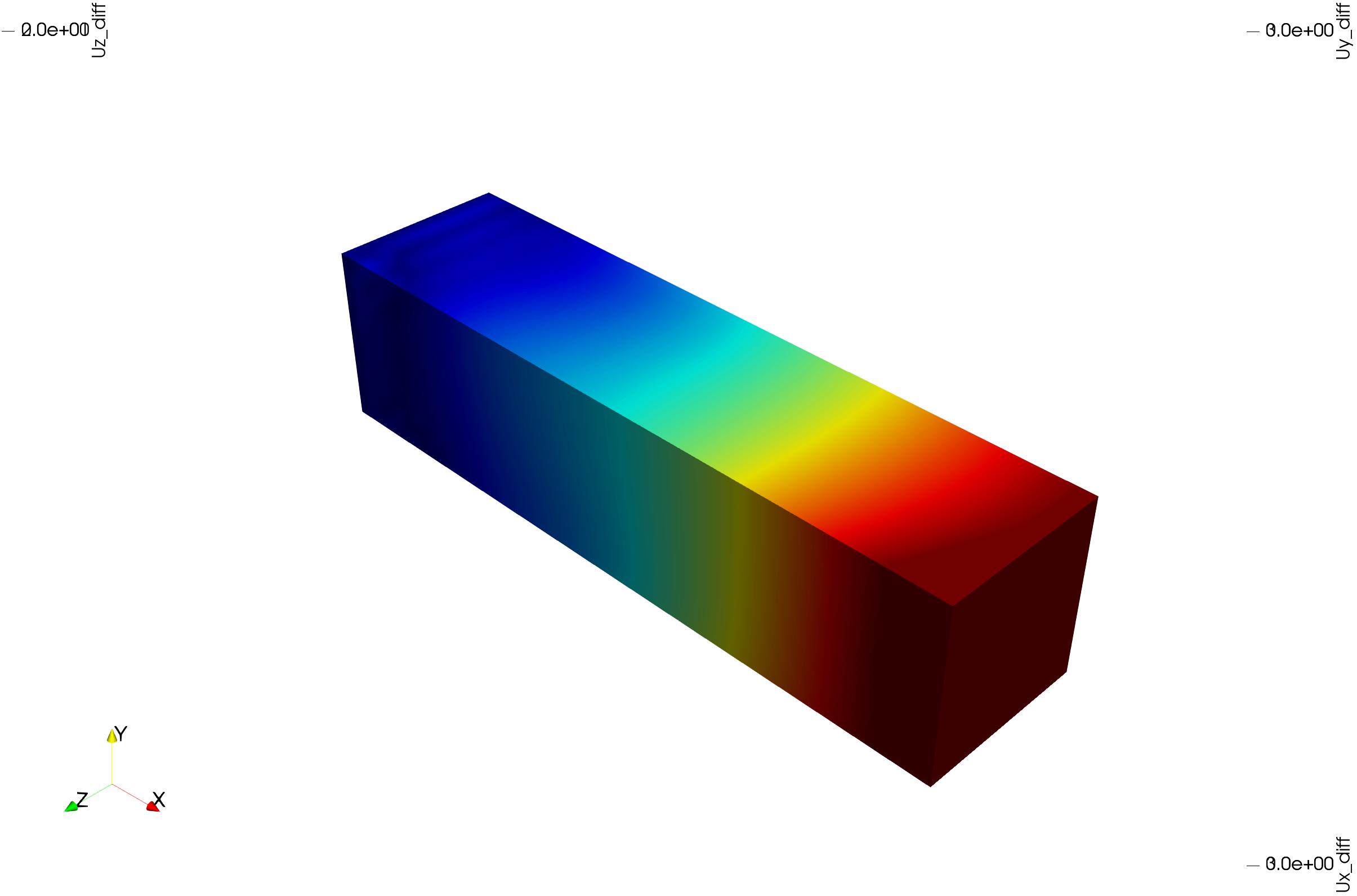}
        \label{fig:day}}
    \end{minipage} & &
    \begin{minipage}[c]{\x\textwidth}
       \centering 
        \subfloat[MLP, RD$_z$=6.30\%]{\includegraphics[trim={22cm 7cm 16cm 12cm},clip,width=\textwidth]{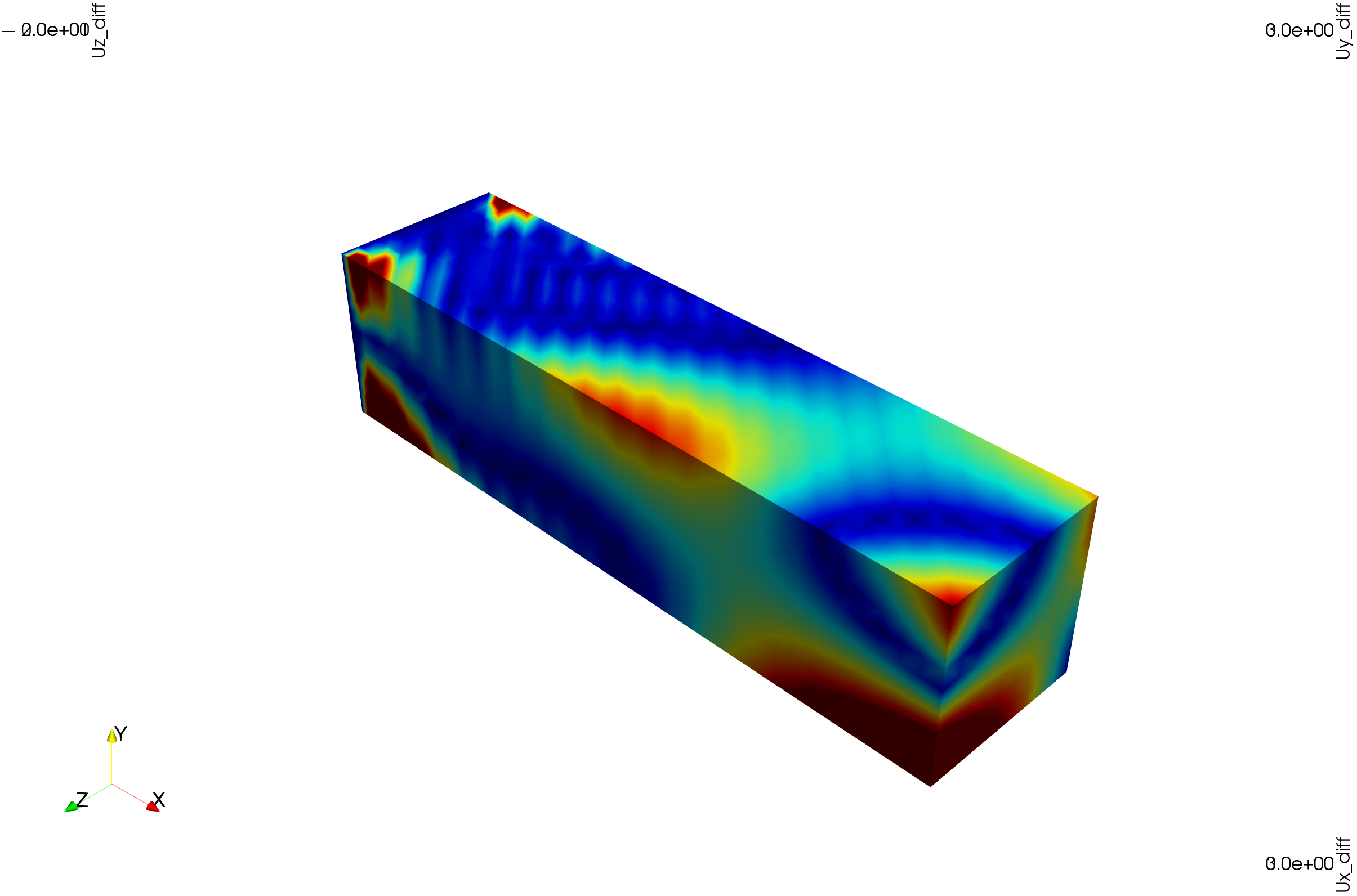}
        \label{fig:daz}}
    \end{minipage} & \\
    \end{tabular}
    \caption{Comparison of relative displacement difference between GCN-DEM and MLP-DEM, using FE shape functions for gradient computation, linear elastic material. The mean relative difference for each displacement component is reported in the caption.}
    \label{LE_SF}
\end{figure}

\begin{figure}[h!] 
    \centering
     \subfloat[GCN-DEM, t = -25]{
         \includegraphics[trim={0cm 0cm 1cm 14cm},clip,width=0.45\textwidth]{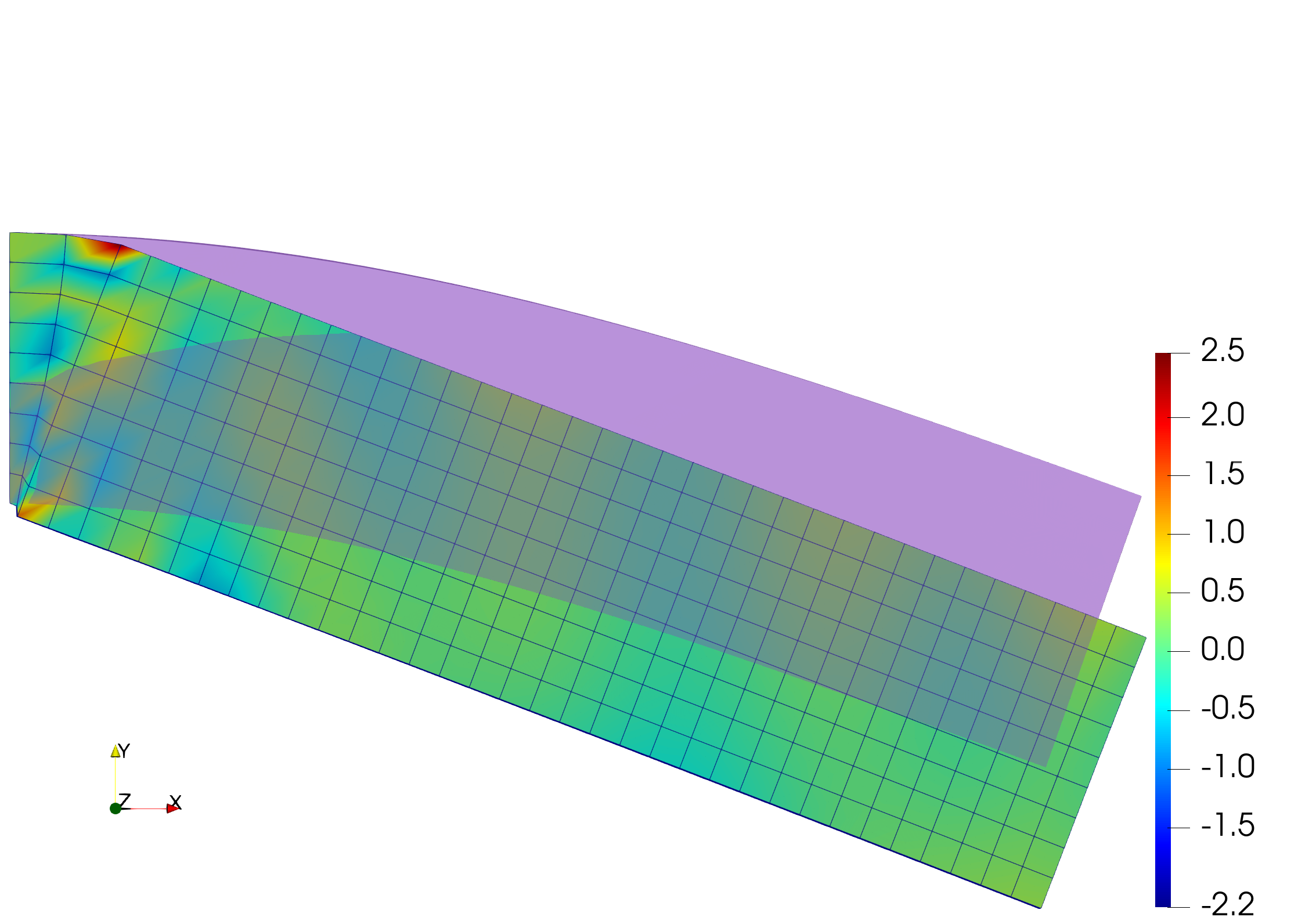}
         \label{fig:LE_def_G}
     }
     \subfloat[MLP-DEM, t = -25]{
         \includegraphics[trim={0cm 0cm 1cm 14cm},clip,width=0.45\textwidth]{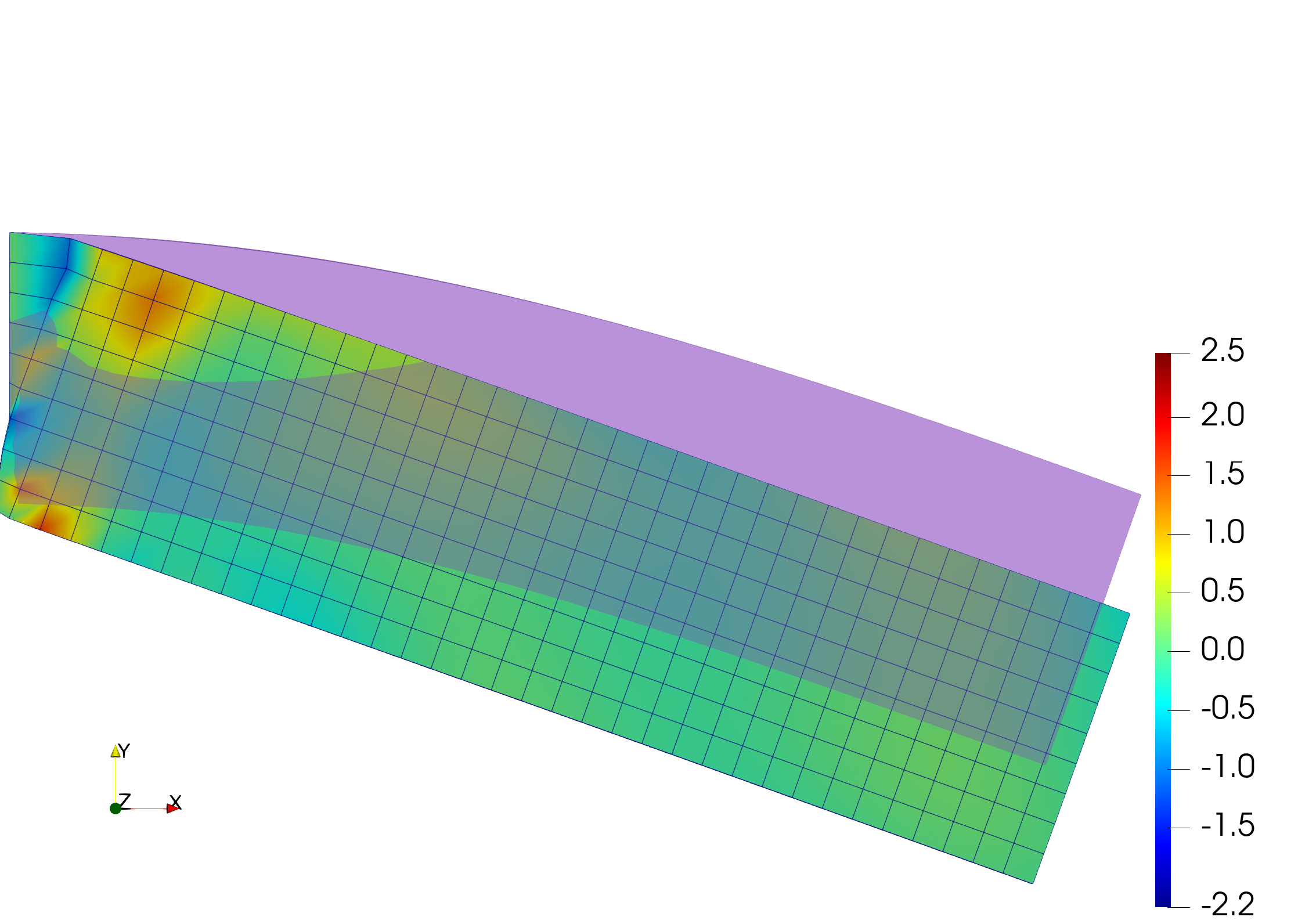}
         \label{fig:LE_def_D}
     }
    \caption{Comparing deformed shapes computed from GCN-DEM and MLP-DEM: \psubref{fig:LE_def_G} Using GCN-DEM. \psubref{fig:LE_def_D} Using MLP-DEM. The deformed shape computed by SF-based gradient computation is shown in translucent purple with a scaling factor of 0.15. $\epsilon_{11}$ contour is overlaid onto the deformed shape obtained from AD-based method with a scaling factor of 0.015. Note that for the plot to look reasonable, the scaling factor in this figure is \emph{ten} times smaller for the AD cases. }
    \label{LE_def}
\end{figure}

In this case, we gradually increased the magnitude of the applied traction in six different simulations. The applied loads were: $t = $ -2.5, -5, -7.5, -10, -15, and -25, respectively. To quantify the model accuracy, we compare the GCN-/MLP-DEM solutions to FEM solutions generated using identical node layouts. The relative difference in displacement is computed as:
\begin{equation}
    RD_i = \frac{ |\bm{u}^i_{NN} - \bm{u}^i_{FE}| }{ {\rm{max}}( |\bm{u}^i_{FE}| ) } \times 100\% , \;\;\; i = x,y,z.
\end{equation}
The mean relative difference (averaged over all nodes and all three displacement components), final loss value, and train time are presented in \tref{tab:elastic} and plotted in \fref{LE_lineplot}. 
Specifically, we highlighted the cases that failed to converge in red. To give a more direct visualization of the distribution of the displacement error, contour plots of the displacement error for different methods at the case $t=-7.5$ are presented in \fref{LE_AD} and \fref{LE_SF}. The deformed shapes at $t=-25$ are presented in \fref{LE_def} to highlight the occurrence of instability for the AD-based gradient computation.
\begin{table}[h!]
    \caption{Performance comparison of GCN-DEM and MLP-DEM, linear elastic model}
    \small
    \centering
    \begin{tabular}{cccccccc}
     Method & \vline & t = -2.5  & t = -5 & t = -7.5 & t = -10 & t = -15 & t = -25 \\
    
    \hline
     & \vline & \multicolumn{6}{c}{Mean percent difference (\%)} \\
    GCN-DEM (AD) & \vline  & 1.17 & 1.82 & 1.95 & \textcolor{red}{964.60} & \textcolor{red}{772.20} & \textcolor{red}{721.00}\\
    MLP-DEM (AD) & \vline  & 1.64 & 4.17 & 2.16 & 2.38 & 3.18 & \textcolor{red}{604.60}\\
    GCN-DEM (SF) & \vline  & 2.99 & 4.65 & 1.77 & 3.05 & 1.58 & 2.43\\
    MLP-DEM (SF) & \vline  & 2.79 & 6.29 & 2.91 & 4.99 & 3.87 & 2.09\\
    \hline
     & \vline & \multicolumn{6}{c}{Final loss function value} \\
    GCN-DEM (AD) & \vline  & -0.83 & -3.33 & -7.51 & \textcolor{red}{-592.05} & \textcolor{red}{-1014.66} & \textcolor{red}{-2146.11}\\
    MLP-DEM (AD) & \vline  & -0.83 & -3.33 & -7.56 & -13.32 & -29.83 & \textcolor{red}{-1981.89}\\
    GCN-DEM (SF) & \vline  & -0.81 & -3.24 & -7.31 & -12.99 & -29.26 & -81.13\\
    MLP-DEM (SF) & \vline  & -0.81 & -3.25 & -7.29 & -12.96 & -29.17 & -80.80\\
    \hline
     & \vline & \multicolumn{6}{c}{Train time [s]} \\
    GCN-DEM (AD) & \vline  & 36.08 & 31.17 & 37.50 & \textcolor{red}{117.08} & \textcolor{red}{117.80} & \textcolor{red}{112.70}\\
    MLP-DEM (AD) & \vline  & 76.06 & 58.81 & 70.98 & 79.70 & 53.60 & \textcolor{red}{115.20}\\
    GCN-DEM (SF) & \vline  & 45.70 & 42.80 & 55.20 & 47.00 & 46.70 & 34.90\\
    MLP-DEM (SF) & \vline  & 79.10 & 83.40 & 74.00 & 68.26 & 62.00 & 51.20\\

    \end{tabular}
    \label{tab:elastic}
\end{table}

From \tref{tab:elastic}, we immediately see that the AD-based gradient computation leads to instability in several cases, especially at loads greater than -10. In contrast, the SF-based gradient computation leads to stable results in all cases. Inspecting the final loss value for the cases that failed to converge, we see that they are much smaller than the converged value for the same loading, which is in complete agreement with our analysis presented in \sref{grad_cal}, and serves as a direct proof that this is the cause of instability. Even for the cases where both AD- and SF-based formulation converged, we notice that the AD-based methods always yield a lower loss value than their SF-based counterparts, hinting that there exists a threshold load magnitude beyond which divergence may occur. Further, when comparing the first three columns of \tref{tab:elastic}, we see that at small loads, the AD-based formulation outperforms the SF-based formulation, yielding a shorter solution time and higher accuracy. When the gradient computation method is fixed, we see that GCN-DEM outperforms MLP-DEM in 4 out of the 6 cases, again showing higher accuracy with a shorter run time. 

\fref{LE_AD} and \fref{LE_SF} provide much more information regarding the performances of GCN-DEM and MLP-DEM. Since the loading is in the XY plane and the Z planes are unconstrained except at the root, stresses in the Z direction are expected to be small and only account for a small percentage of the overall system strain energy $\Psi$. In this sense, it is reasonable to expect that the relative error in the Z direction to be higher than those in the X- and Y-directions, as it has a smaller weight in the loss function. Comparing the two rows in \fref{LE_AD}, we see that GCN-DEM outperforms MLP-DEM in this case, yielding lower relative error in all three displacement components. When inspecting the SF-based cases in \fref{LE_SF}, we again notice the same trend, where GCN-DEM provides better results in all displacement components. Comparing \fref{LE_AD} and \fref{LE_SF}, it is visually apparent that SF-based formulation leads to a higher error than the AD-based formulation in this case. However, we remark that the errors remain below 1.5\% for the SF cases in the X- and Y-components.

These observations indicate that when the applied load is small, the GCN-DEM method based on AD gradient computation tends to provide the best performance in terms of accuracy and run time. When the load magnitude increases, the GCN-DEM method based on SF gradient computation offers a much more robust solution in general.

\fref{LE_def} provides a closer look into the strain localization instability that occurs with AD-based gradient computations. From the grids that are overlaid onto the deformed shape of the AD-based simulations, we see that severe deformation happened at the root of the beam, while to the right of the strain localization, negligible deformation is seen. From a continuum mechanics standpoint, such severe localization must be accompanied by high local strain values, which are grossly neglected by the point-based AD gradient computation. The observed displacement field of the AD simulations also resembles the 1D example presented in \sref{grad_cal}, where large displacement jumps occur in between two nodes of the domain. This behavior further strengthens our argument and shows that the inability to detect strain localization between the nodes is the root cause of instability.

\subsection{Neo-Hookean material}
\label{NH}
\begin{figure}[h!]
\newcommand\x{0.2}
    \centering
    \begin{tabular}{ c c c c c c }
    \begin{minipage}[c]{\x\textwidth}
       \centering 
        \subfloat[GCN, RD$_x$=0.77\%]{\includegraphics[trim={22cm 7cm 16cm 12cm},clip,width=\textwidth]{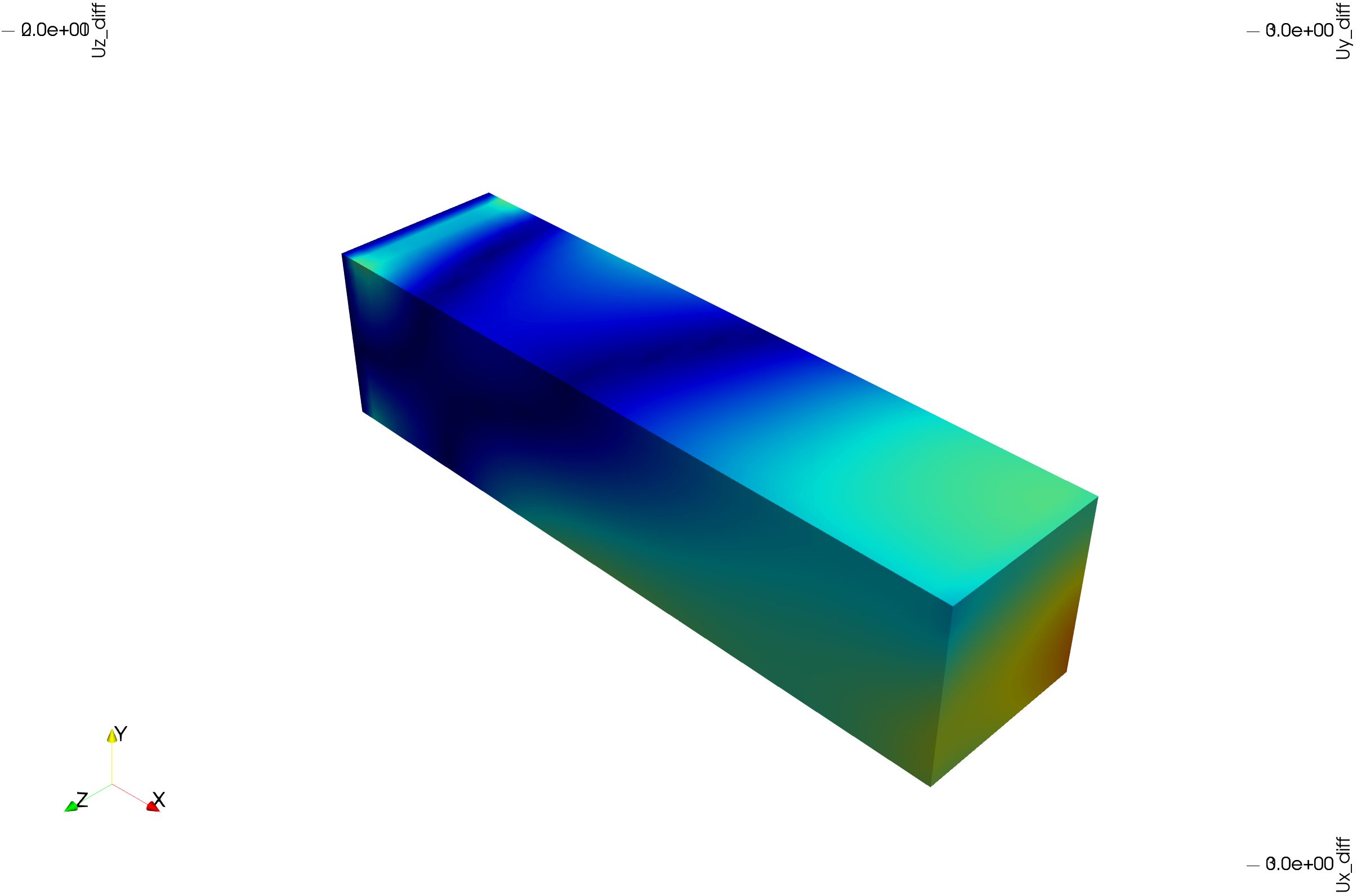}
        \label{fig:gax}}
    \end{minipage} &
    \multirow{2}{*}{
    \begin{minipage}[c]{0.05\textwidth}
       \centering 
        \includegraphics[trim={72cm 0cm 1cm 25cm},clip,width=\textwidth]{Figures/CB1.png}
    \end{minipage}
    } &
    \begin{minipage}[c]{\x\textwidth}
       \centering 
        \subfloat[GCN, RD$_y$=0.54\%]{\includegraphics[trim={22cm 7cm 16cm 12cm},clip,width=\textwidth]{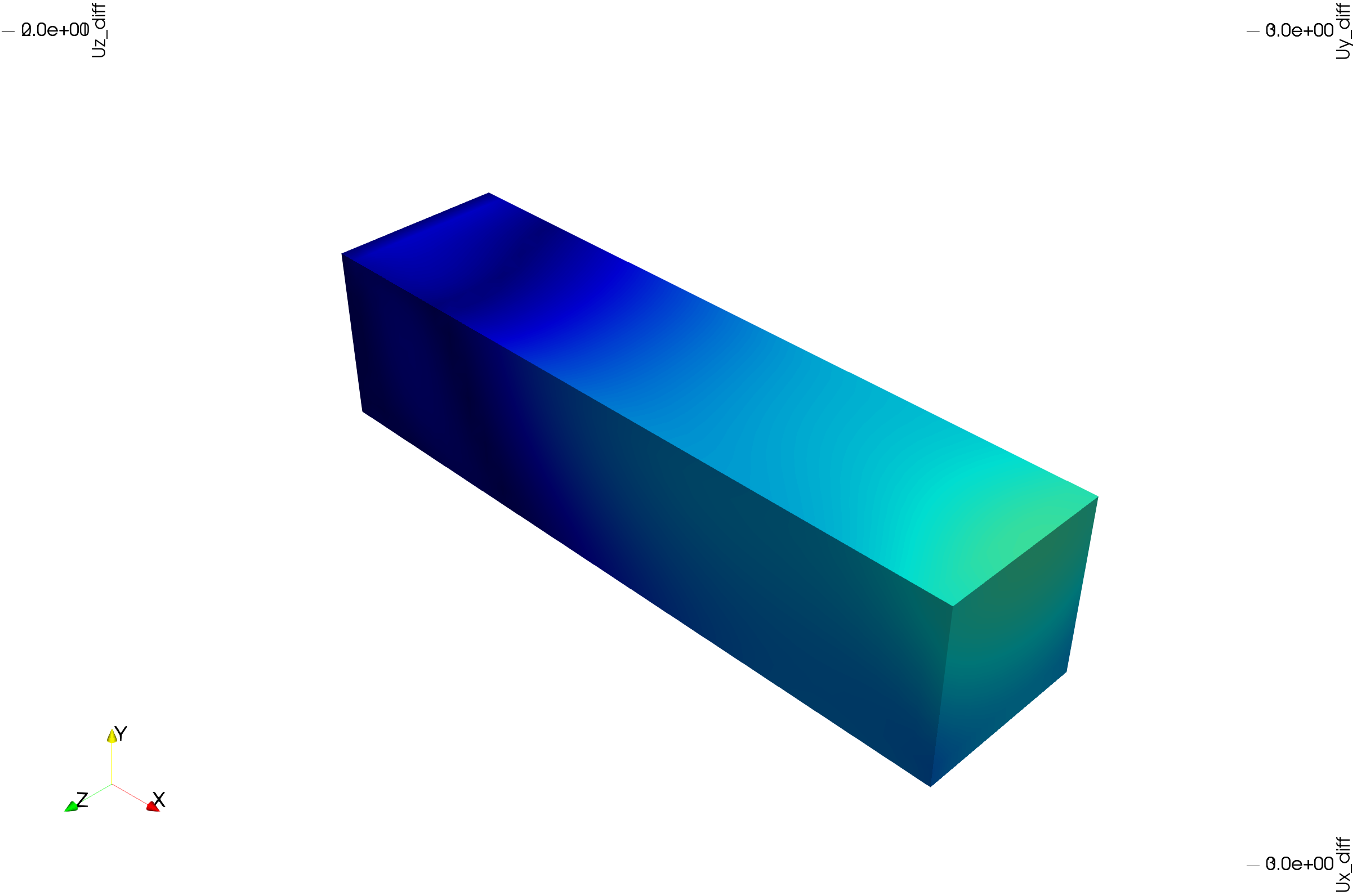}
        \label{fig:gay}}
    \end{minipage} &
    \multirow{2}{*}{
    \begin{minipage}[c]{0.05\textwidth}
       \centering 
        \includegraphics[trim={72cm 0cm 1cm 25cm},clip,width=\textwidth]{Figures/CB1.png}
    \end{minipage}
    } &
    \begin{minipage}[c]{\x\textwidth}
       \centering 
        \subfloat[GCN, RD$_z$=3.25\%]{\includegraphics[trim={22cm 7cm 16cm 12cm},clip,width=\textwidth]{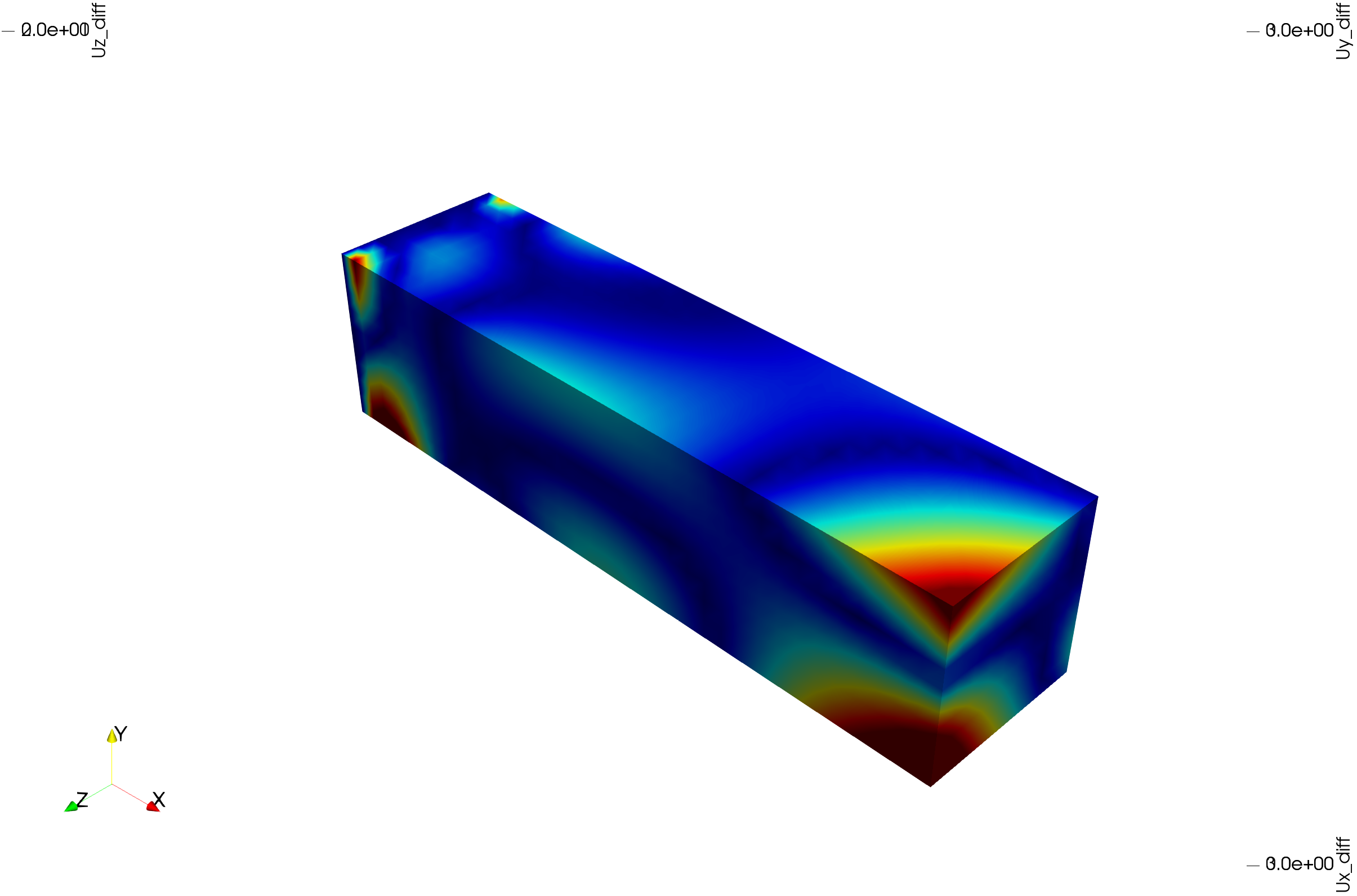}
        \label{fig:gaz}}
    \end{minipage} &
    \multirow{2}{*}{
    \begin{minipage}[c]{0.05\textwidth}
       \centering 
        \includegraphics[trim={71cm 0cm 1.5cm 24cm},clip,width=\textwidth]{Figures/CB2.png}
    \end{minipage}
    } \\
    \begin{minipage}[c]{\x\textwidth}
       \centering 
        \subfloat[MLP, RD$_x$=0.56\%]{\includegraphics[trim={22cm 7cm 16cm 12cm},clip,width=\textwidth]{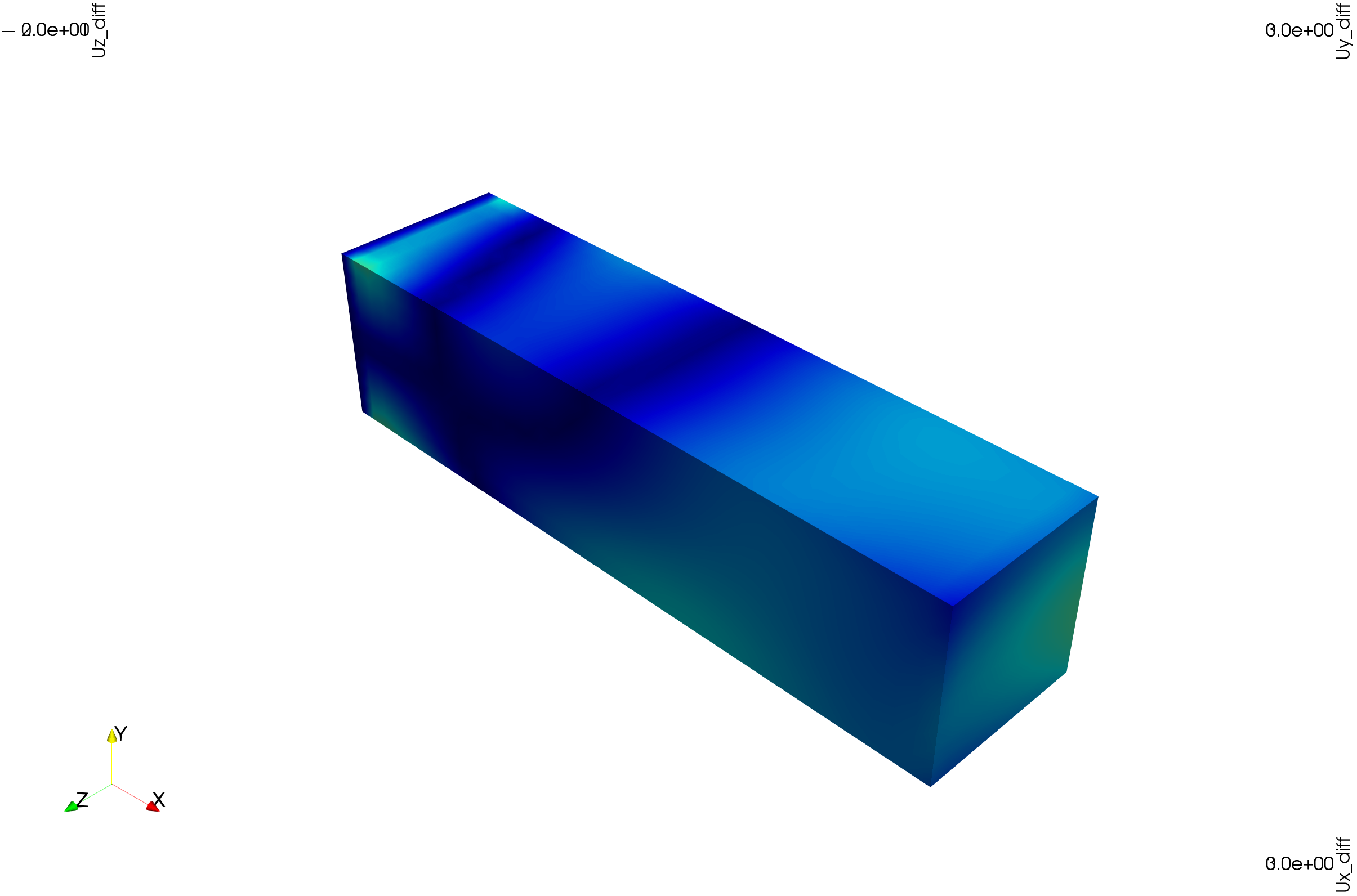}
        \label{fig:dax}}
    \end{minipage} & &
    \begin{minipage}[c]{\x\textwidth}
       \centering 
        \subfloat[MLP, RD$_y$=0.40\%]{\includegraphics[trim={22cm 7cm 16cm 12cm},clip,width=\textwidth]{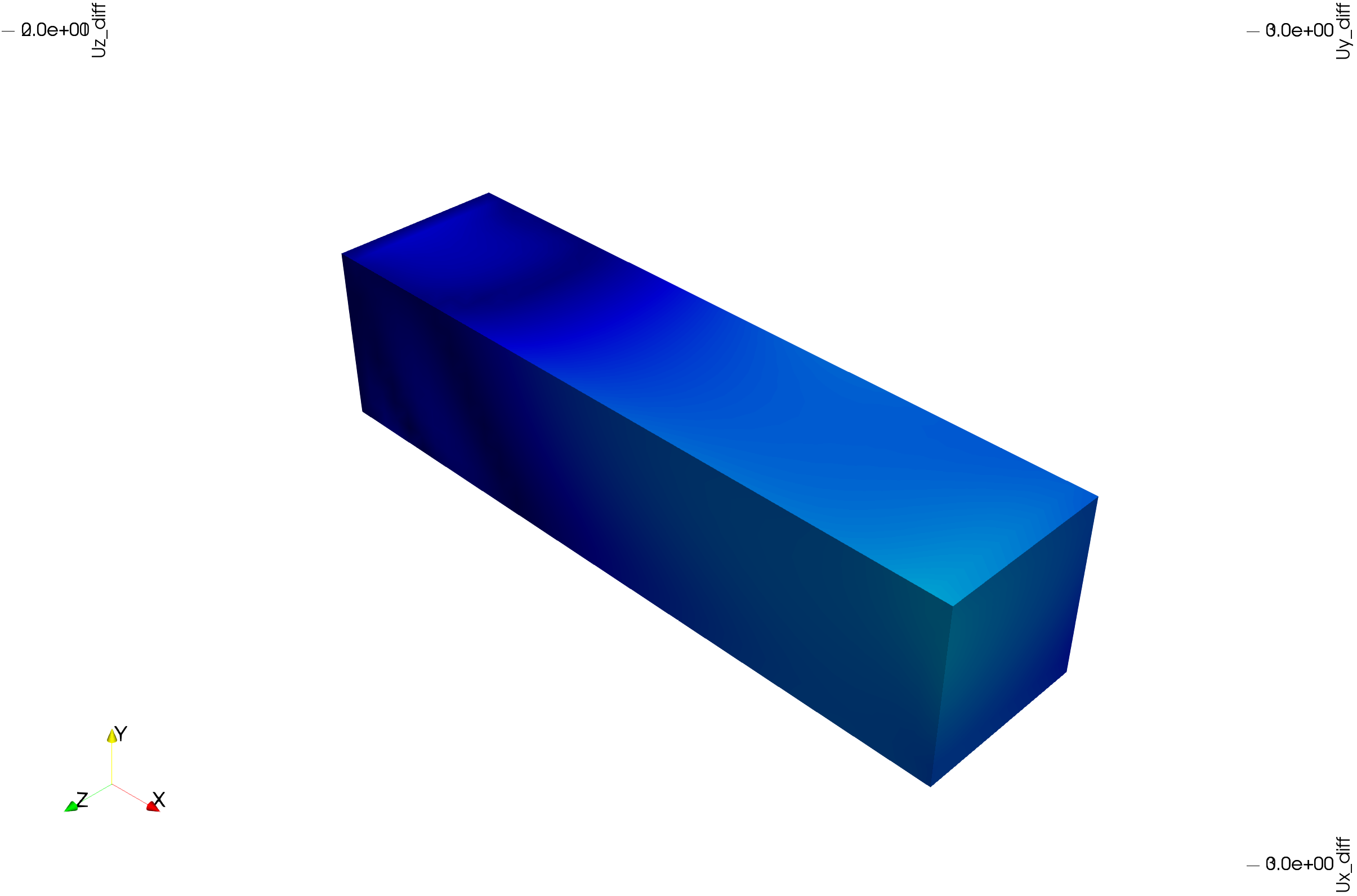}
        \label{fig:day}}
    \end{minipage} & &
    \begin{minipage}[c]{\x\textwidth}
       \centering 
        \subfloat[MLP, RD$_z$=10.33\%]{\includegraphics[trim={22cm 7cm 16cm 12cm},clip,width=\textwidth]{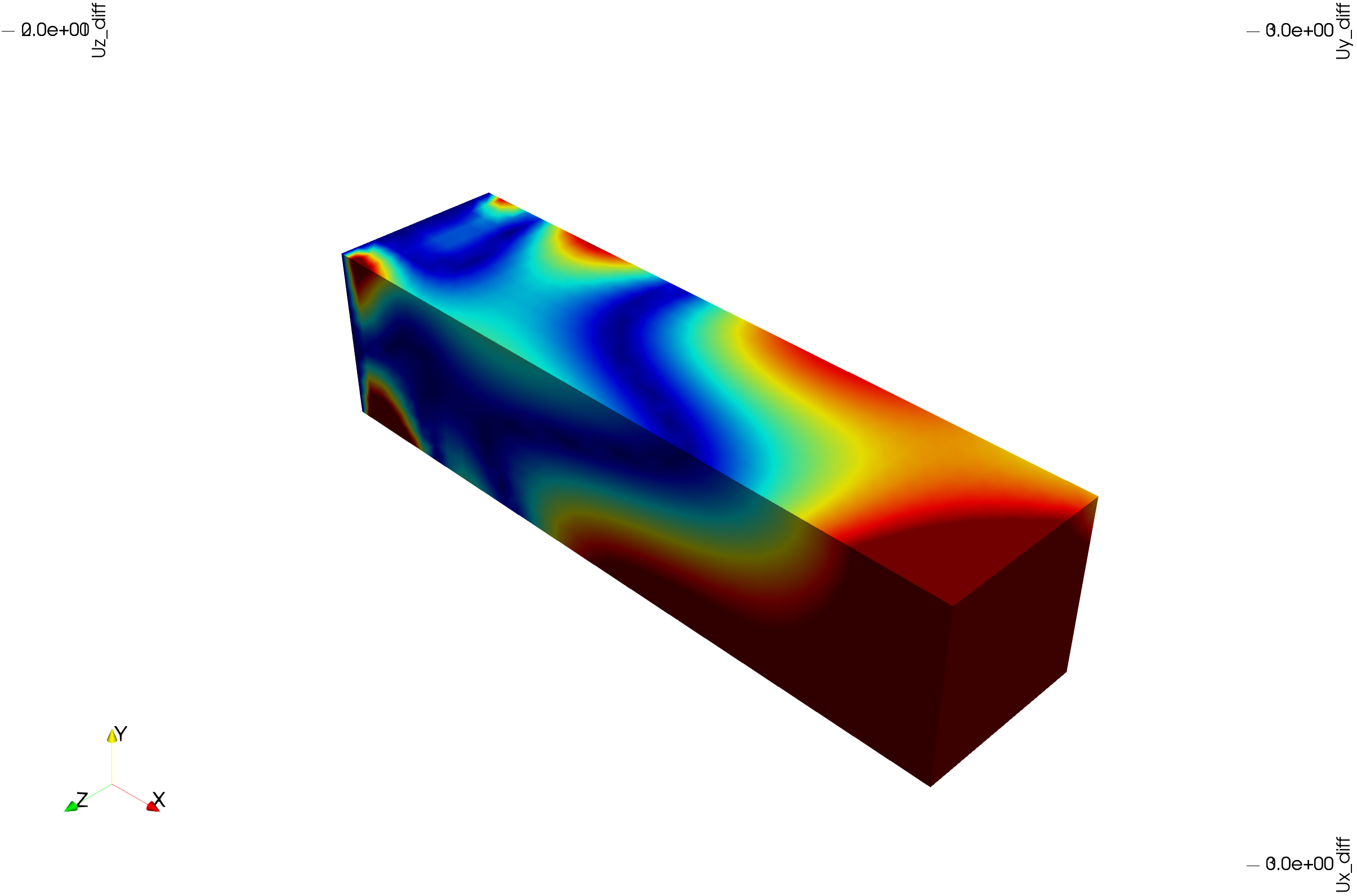}
        \label{fig:daz}}
    \end{minipage} & \\
    \end{tabular}
    \caption{Comparison of relative displacement difference between GCN-DEM and MLP-DEM, using AD for gradient computation, Neo-Hookean material. The mean relative difference for each displacement component is reported in the caption.}
    \label{NH_AD}
\end{figure}

\begin{figure}[h!]
\newcommand\x{0.2}
    \centering
    \begin{tabular}{ c c c c c c }
    \begin{minipage}[c]{\x\textwidth}
       \centering 
        \subfloat[GCN, RD$_x$=0.53\%]{\includegraphics[trim={22cm 7cm 16cm 12cm},clip,width=\textwidth]{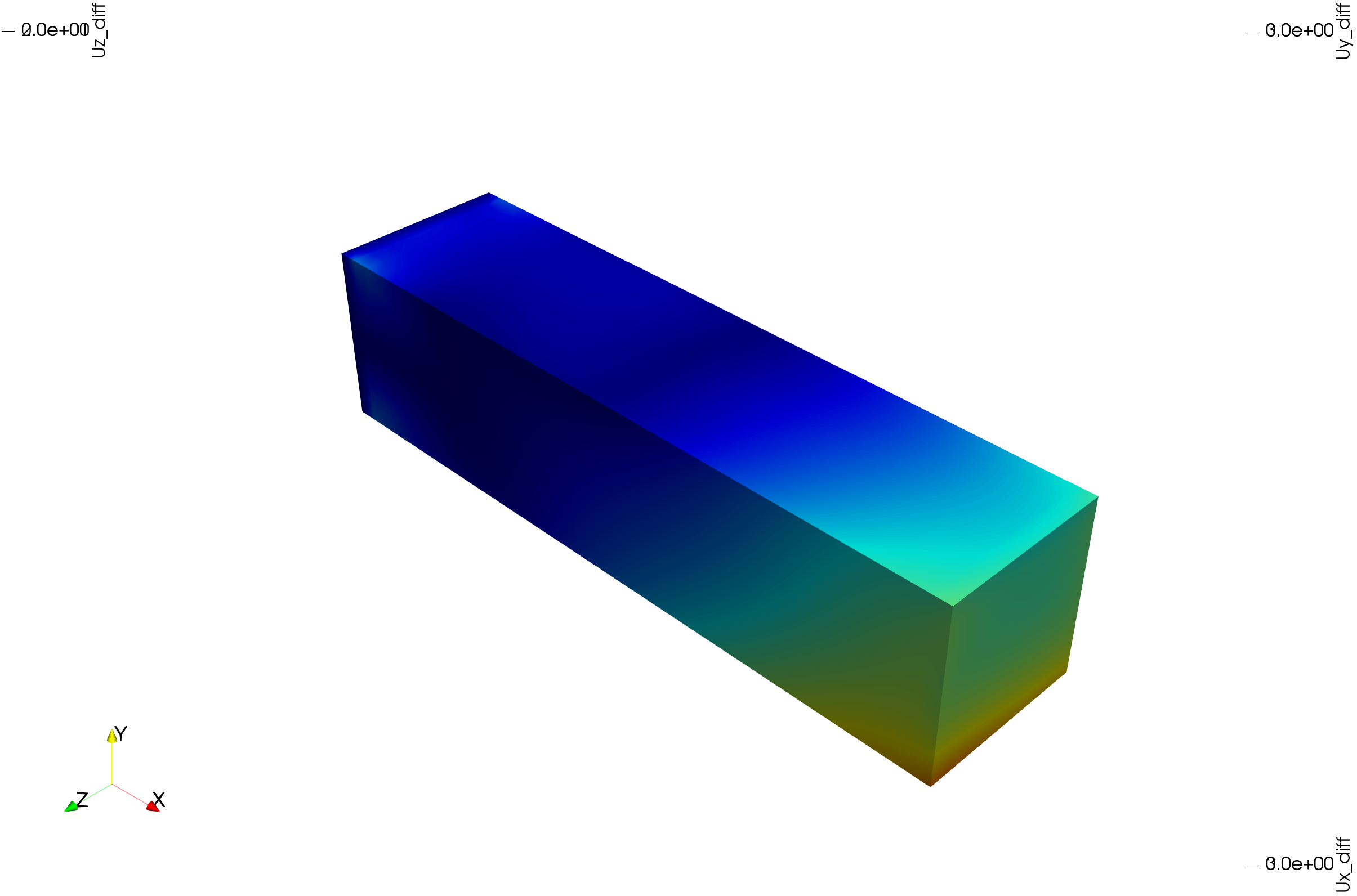}
        \label{fig:gax}}
    \end{minipage} &
    \multirow{2}{*}{
    \begin{minipage}[c]{0.05\textwidth}
       \centering 
        \includegraphics[trim={72cm 0cm 1cm 25cm},clip,width=\textwidth]{Figures/CB1.png}
    \end{minipage}
    } &
    \begin{minipage}[c]{\x\textwidth}
       \centering 
        \subfloat[GCN, RD$_y$=0.47\%]{\includegraphics[trim={22cm 7cm 16cm 12cm},clip,width=\textwidth]{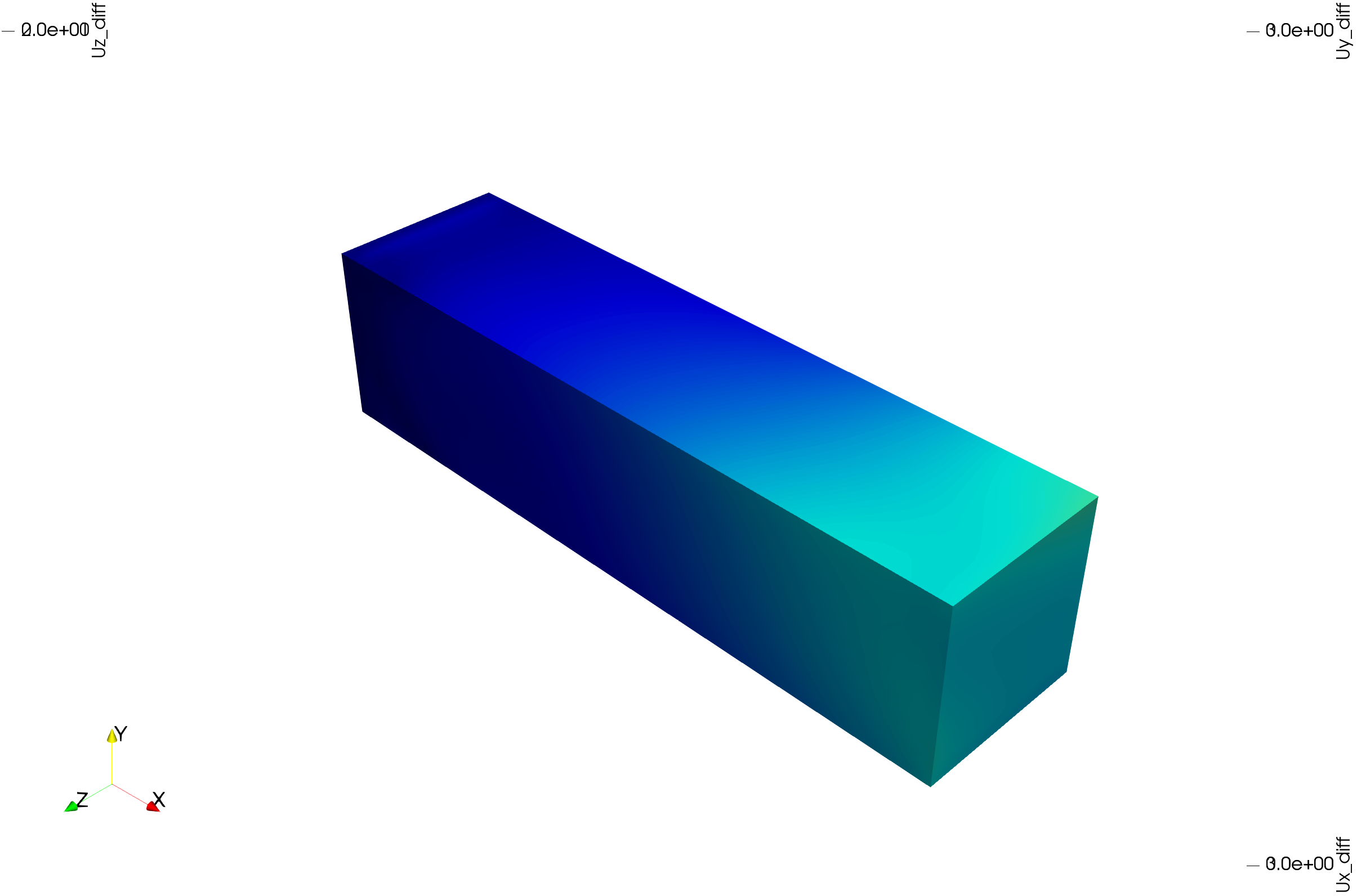}
        \label{fig:gay}}
    \end{minipage} &
    \multirow{2}{*}{
    \begin{minipage}[c]{0.05\textwidth}
       \centering 
        \includegraphics[trim={72cm 0cm 1cm 25cm},clip,width=\textwidth]{Figures/CB1.png}
    \end{minipage}
    } &
    \begin{minipage}[c]{\x\textwidth}
       \centering 
        \subfloat[GCN, RD$_z$=3.28\%]{\includegraphics[trim={22cm 7cm 16cm 12cm},clip,width=\textwidth]{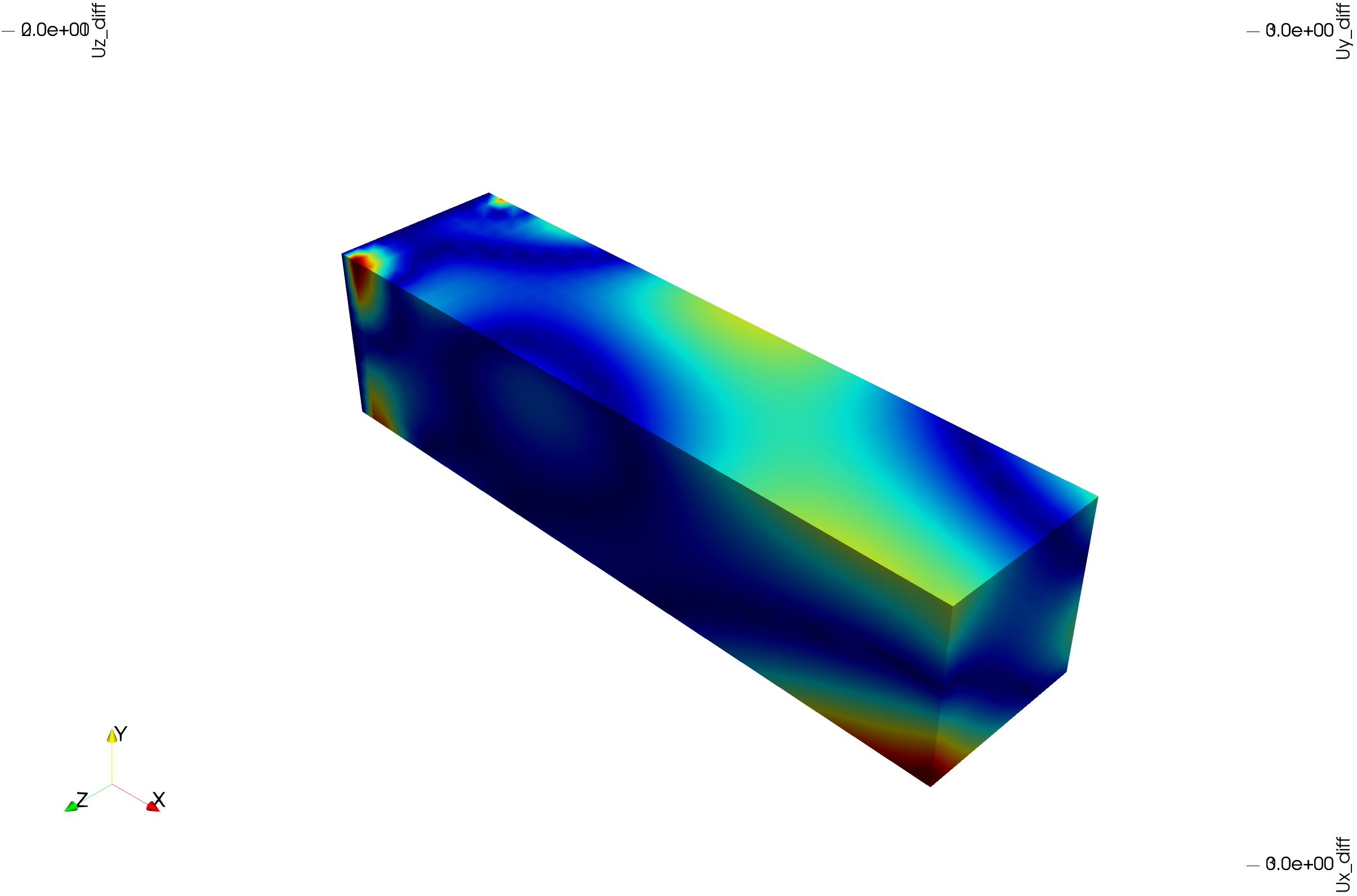}
        \label{fig:gaz}}
    \end{minipage} &
    \multirow{2}{*}{
    \begin{minipage}[c]{0.05\textwidth}
       \centering 
        \includegraphics[trim={71cm 0cm 1.5cm 24cm},clip,width=\textwidth]{Figures/CB2.png}
    \end{minipage}
    } \\
    \begin{minipage}[c]{\x\textwidth}
       \centering 
        \subfloat[MLP, RD$_x$=0.52\%]{\includegraphics[trim={22cm 7cm 16cm 12cm},clip,width=\textwidth]{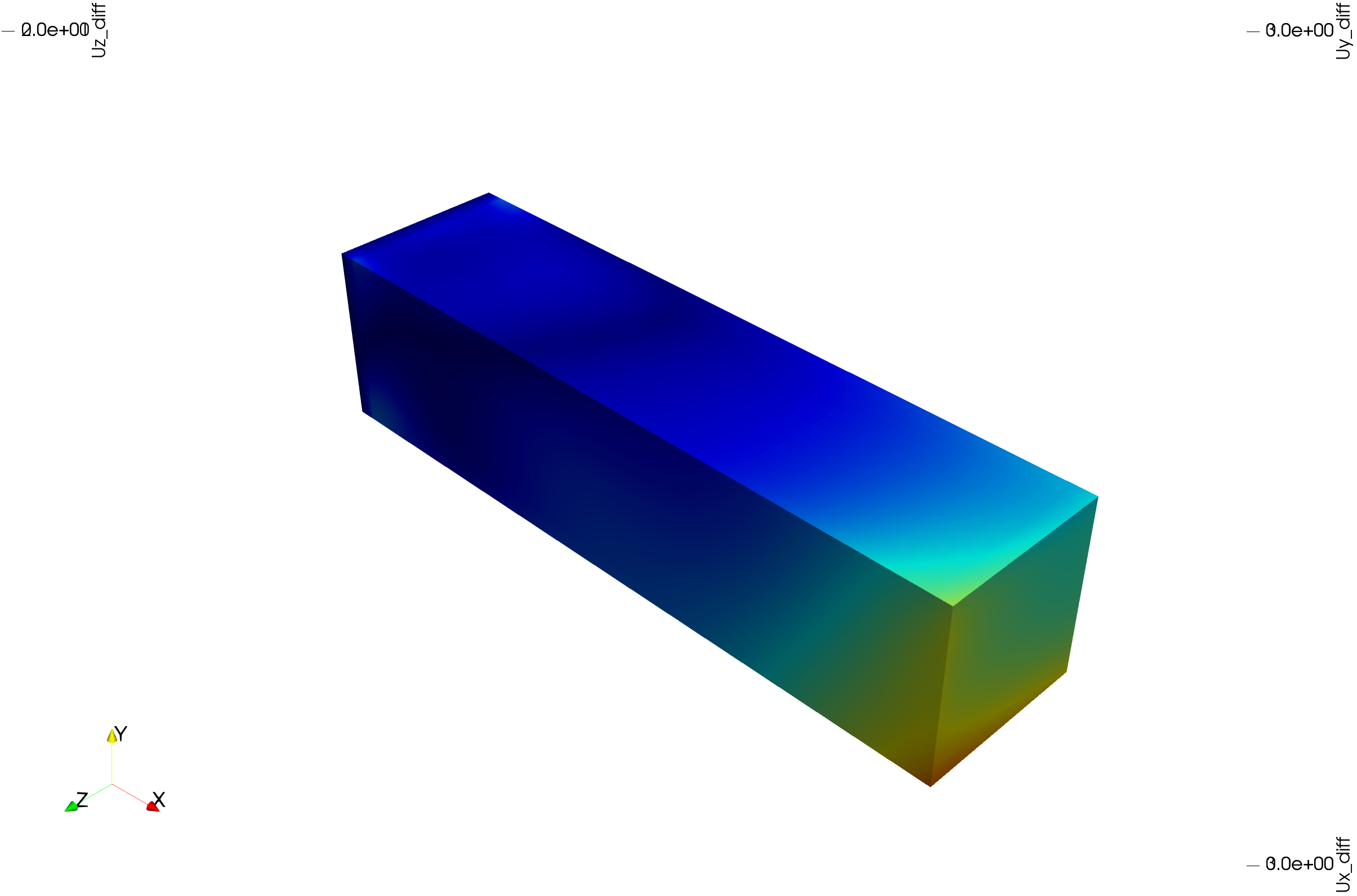}
        \label{fig:dax}}
    \end{minipage} & &
    \begin{minipage}[c]{\x\textwidth}
       \centering 
        \subfloat[MLP, RD$_y$=0.42\%]{\includegraphics[trim={22cm 7cm 16cm 12cm},clip,width=\textwidth]{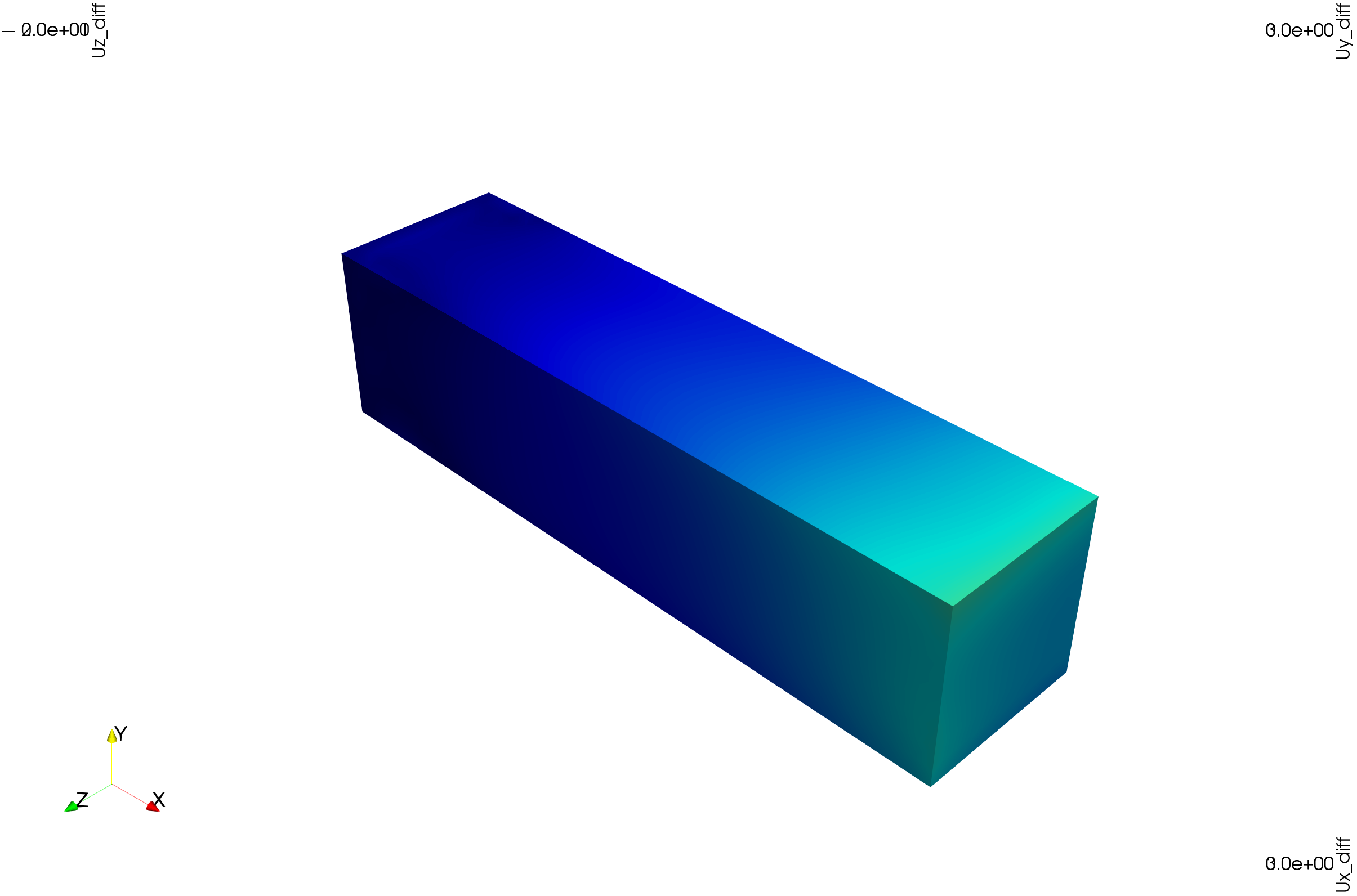}
        \label{fig:day}}
    \end{minipage} & &
    \begin{minipage}[c]{\x\textwidth}
       \centering 
        \subfloat[MLP, RD$_z$=7.71\%]{\includegraphics[trim={22cm 7cm 16cm 12cm},clip,width=\textwidth]{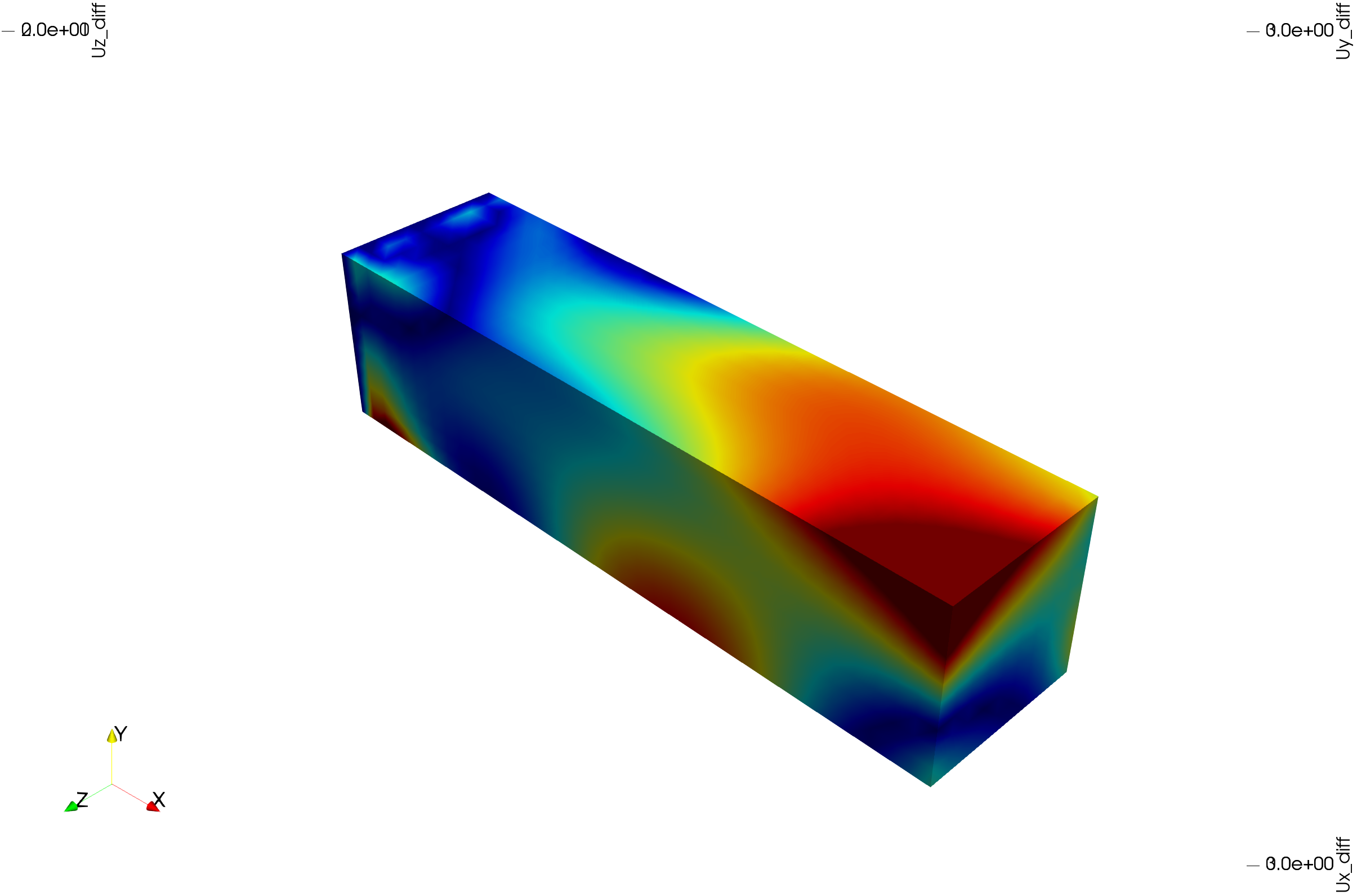}
        \label{fig:daz}}
    \end{minipage} & \\
    \end{tabular}
    \caption{Comparison of relative displacement difference between GCN-DEM and MLP-DEM, using FE shape functions for gradient computation, Neo-Hookean material. The mean relative difference for each displacement component is reported in the caption.}
    \label{NH_SF}
\end{figure}

\begin{figure}[h!] 
    \centering
     \subfloat[GCN-DEM, t = -25]{
         \includegraphics[trim={0cm 0cm 1cm 14cm},clip,width=0.45\textwidth]{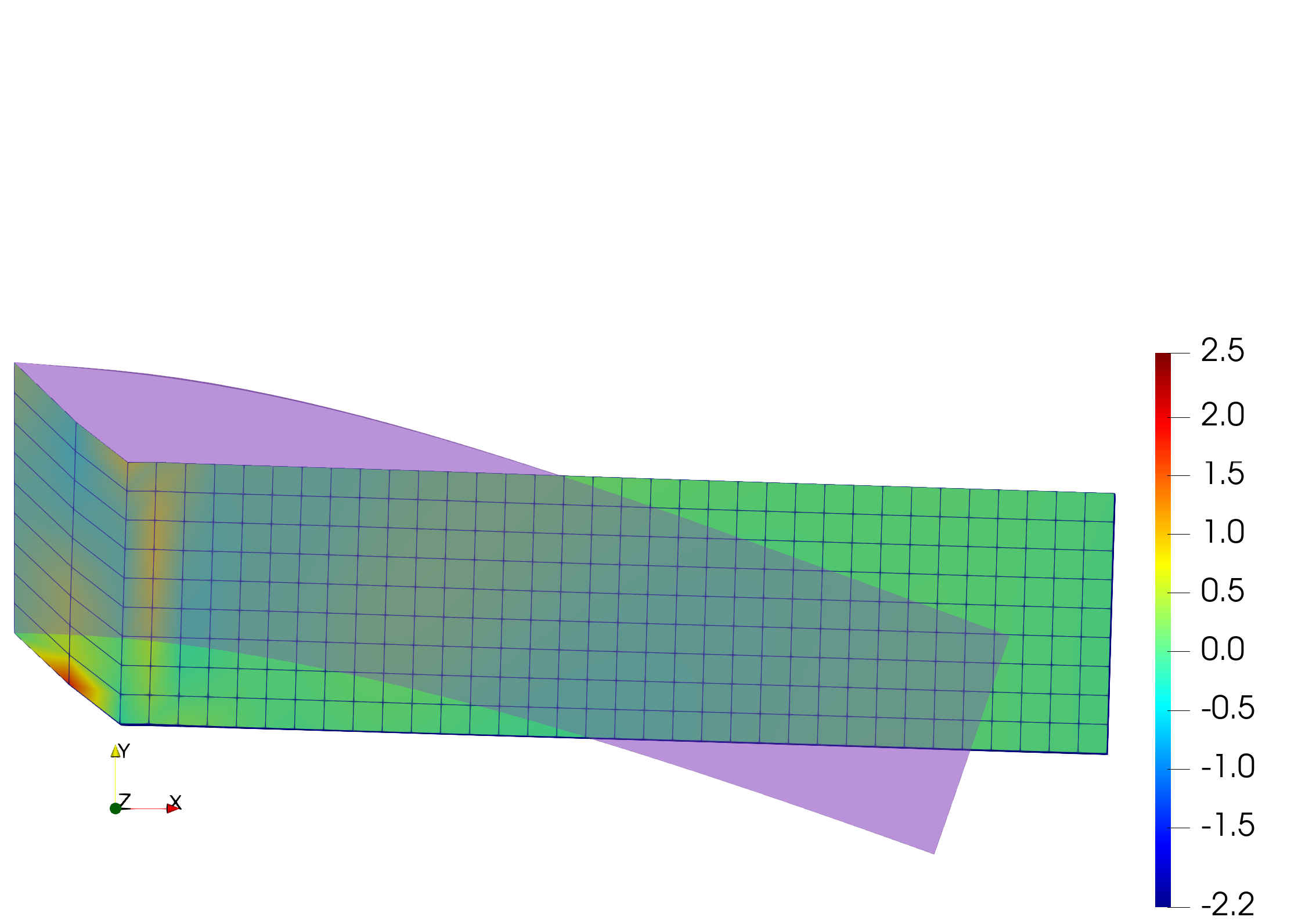}
         \label{fig:NH_def_G}
     }
     \subfloat[MLP-DEM, t = -25]{
         \includegraphics[trim={0cm 0cm 1cm 14cm},clip,width=0.45\textwidth]{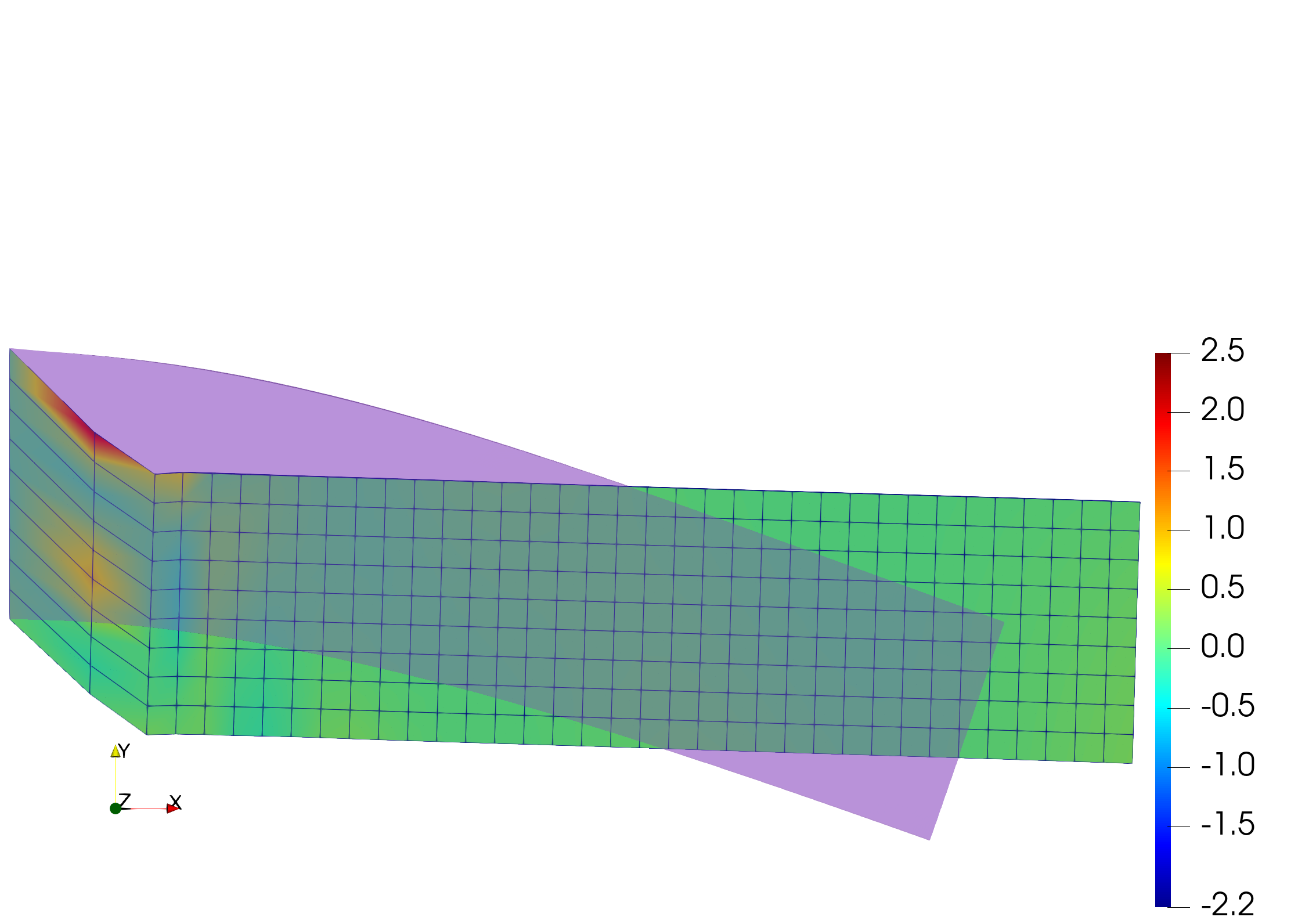}
         \label{fig:NH_def_D}
     }
    \caption{Comparing deformed shapes computed from GCN-DEM and MLP-DEM: \psubref{fig:NH_def_G} Using GCN-DEM. \psubref{fig:NH_def_D} Using MLP-DEM. The deformed shape computed by SF-based gradient computation is shown in translucent purple with a scaling factor of 0.3. $\epsilon_{11}$ contour is overlaid onto the deformed shape obtained from AD-based method with a scaling factor of 0.03. Note that the scaling factor in this figure is \emph{ten} times smaller for the AD cases. }
    \label{NH_def}
\end{figure}

In the previous case, linear elastic material under a small strain formulation provides neither geometric nor material nonlinearity to the system. Therefore, we devote this section to study how GCN-DEM and MLP-DEM compare when both sources of nonlinearity are present. Same loads were applied as in \sref{LE}, and the Neo-Hookean model in Abaqus/Standard was used to generate FEM comparisons. The mean relative difference, final loss value, and train time are presented in \tref{tab:hyper} and plotted in \fref{LE_lineplot} for graphical visualization. 
Cases that failed to converge are highlighted in red. Contour plots of the displacement error at the case $t=-10$ are presented in \fref{NH_AD} and \fref{NH_SF}. The deformed shapes at $t=-25$ are presented in \fref{NH_def}.

\begin{table}[h!]
    \caption{Performance comparison of GCN-DEM and MLP-DEM, Neo-Hookean model}
    \small
    \centering
    \begin{tabular}{cccccccc}
     Method & \vline & t = -2.5  & t = -5 & t = -7.5 & t = -10 & t = -15 & t = -25 \\
    \hline
     & \vline & \multicolumn{6}{c}{Mean percent difference (\%)} \\
    GCN-DEM (AD) & \vline  & 8.06 & 6.03 & 6.26 & 1.52 & \textcolor{red}{254.30} & \textcolor{red}{292.60}\\
    MLP-DEM (AD) & \vline  & 14.17 & 5.56 & 5.23 & 3.76 & \textcolor{red}{361.10} & \textcolor{red}{811.70}\\
    GCN-DEM (SF) & \vline  & 2.72 & 2.96 & 3.57 & 1.42 & 2.76 & 1.33\\
    MLP-DEM (SF) & \vline  & 4.10 & 3.11 & 3.03 & 2.88 & 1.13 & 2.39\\
    \hline
     & \vline & \multicolumn{6}{c}{Final loss function value} \\
    GCN-DEM (AD) & \vline  & -0.79 & -3.15 & -6.80 & -11.57 & \textcolor{red}{-186.68} & \textcolor{red}{-343.66}\\
    MLP-DEM (AD) & \vline  & -0.71 & -3.08 & -6.80 & -11.55 & \textcolor{red}{-177.65} & \textcolor{red}{-413.19}\\
    GCN-DEM (SF) & \vline  & -0.80 & -3.11 & -6.70 & -11.29 & -22.71 & -50.79\\
    MLP-DEM (SF) & \vline  & -0.79 & -3.10 & -6.69 & -11.30 & -22.69 & -50.71\\
    \hline
     & \vline & \multicolumn{6}{c}{Train time [s]} \\
    GCN-DEM (AD) & \vline  & 20.85 & 22.62 & 28.40 & 48.37 & \textcolor{red}{130.10} & \textcolor{red}{127.80}\\
    MLP-DEM (AD) & \vline  & 21.26 & 35.89 & 63.22 & 59.05 & \textcolor{red}{133.10} & \textcolor{red}{141.40}\\
    GCN-DEM (SF) & \vline  & 92.73 & 85.43 & 101.30 & 72.35 & 74.19 & 62.31\\
    MLP-DEM (SF) & \vline  & 147.10 & 140.00 & 116.90 & 141.10 & 136.90 & 98.80\\

    \end{tabular}
    \label{tab:hyper}
\end{table}

\begin{figure}[h!] 
    \centering
     \subfloat[]{
         \includegraphics[width=0.33\textwidth]{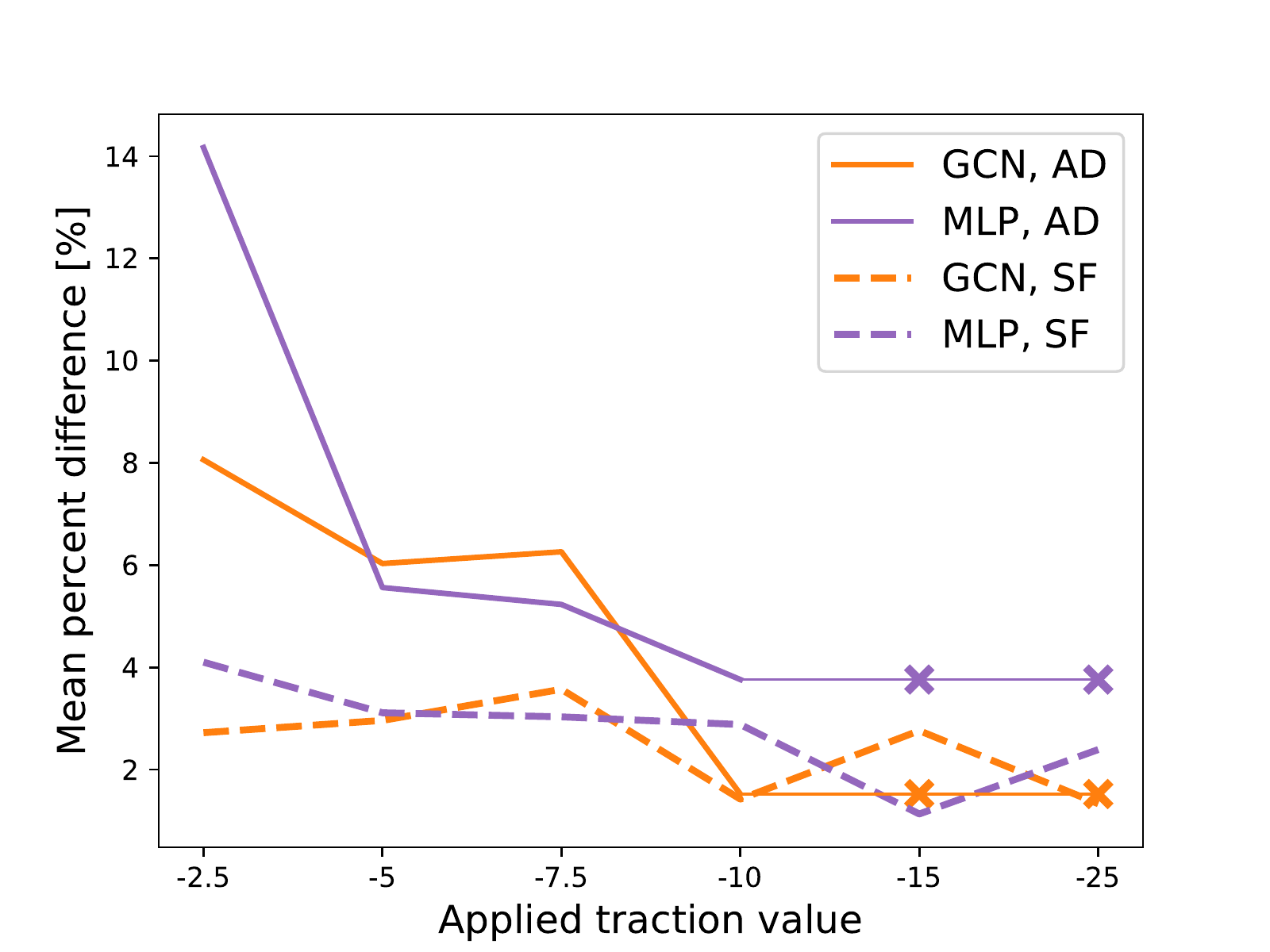}
         \label{fig:NH_RD}
     }
     \subfloat[]{
         \includegraphics[width=0.33\textwidth]{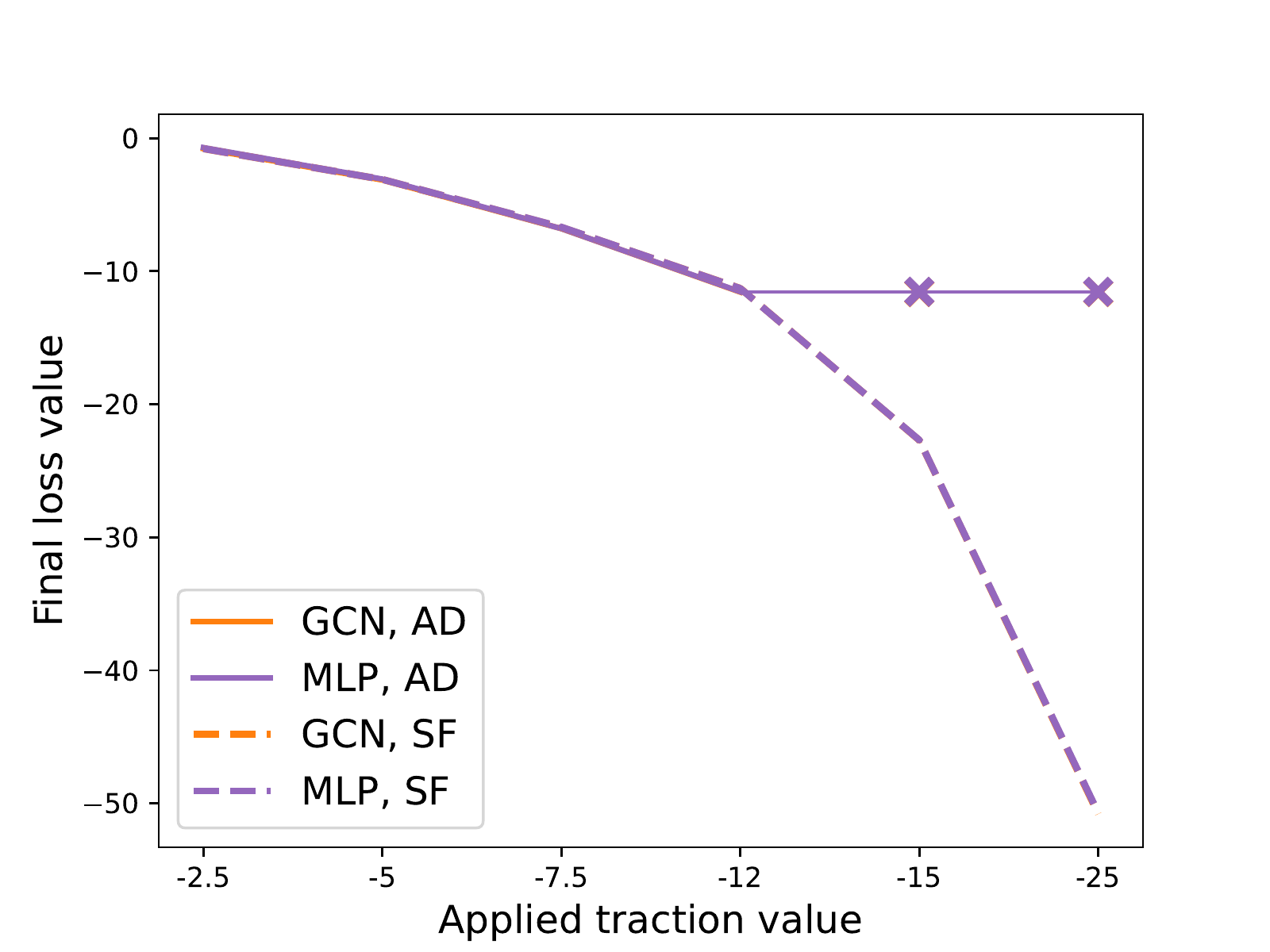}
         \label{fig:NH_loss}
     }
     \subfloat[]{
         \includegraphics[width=0.33\textwidth]{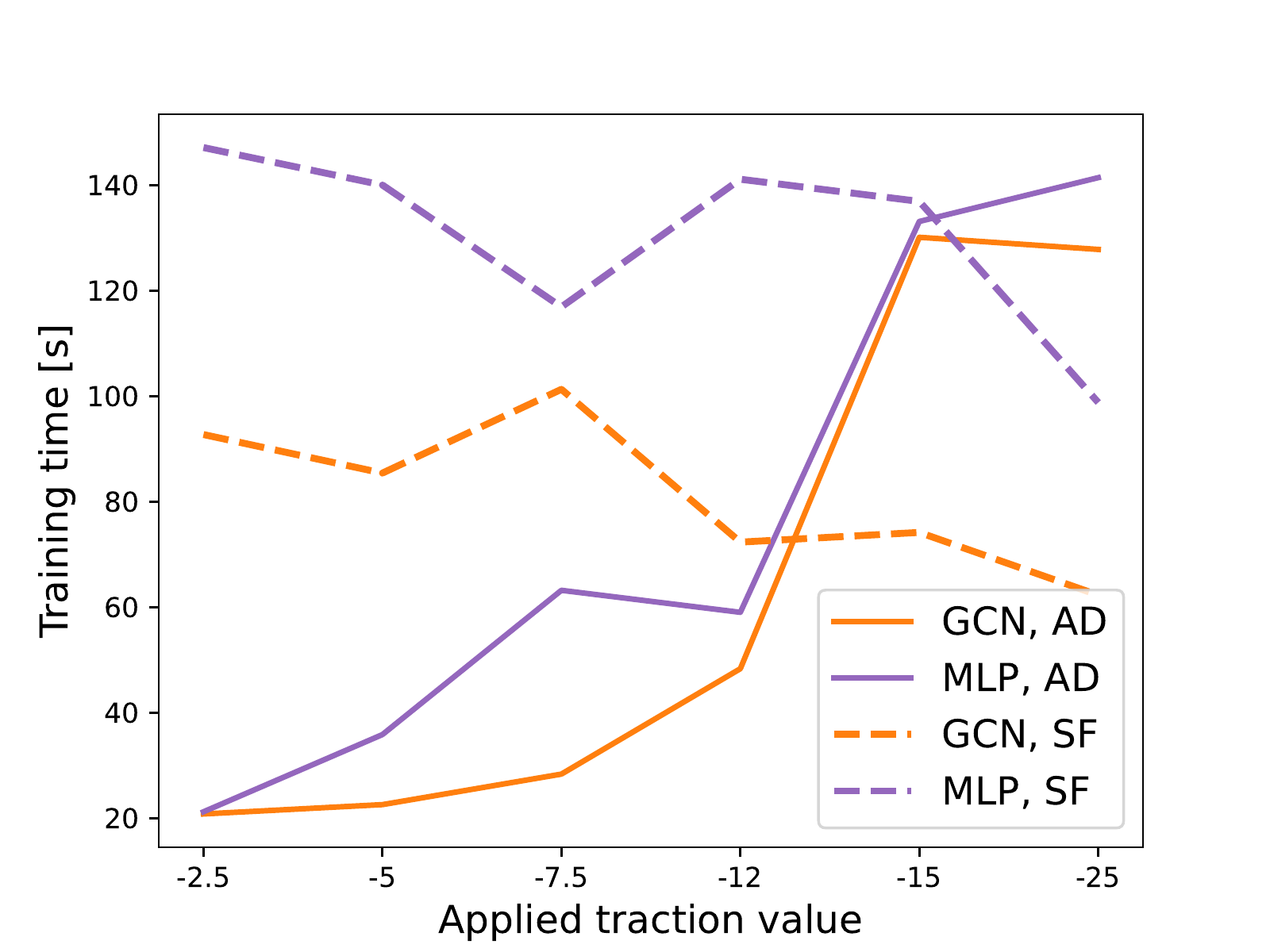}
         \label{fig:NH_time}
     }
    \caption{Comparing GCN-DEM and MLP-DEM, AD-based and SF-based gradient computation: \psubref{fig:NH_RD} Relative difference in displacements. \psubref{fig:NH_loss} Final loss value.
    \psubref{fig:NH_time} Training time for simulation. In \psubref{fig:NH_RD} and \psubref{fig:NH_loss}, simulations that failed to converge are marked with a $\bm{\times}$.}
    \label{NH_lineplot}
\end{figure}

\tref{tab:hyper} shows very similar trends compared to \tref{tab:elastic}. AD-based formulations become unstable when load magnitude is greater than 10, while SF-based formulations remain stable for all loads tested. Comparing the four cases where AD-based formulations were stable, we see that in the hyperelastic case, GCN-DEM and MLP-DEM provide very similar accuracy, each having higher accuracy in 2 of the 4 cases. While the accuracy is similar, GCN-DEM still provides a shorter run time than MLP-DEM. When using SF-based gradient computation, GCN-DEM generally delivers a more accurate solution than MLP-DEM, with a run time that is always shorter than MLP-DEM. Interestingly, AD-based formulation no longer holds an advantage on accuracy against SF-based formulation in small loads as in \sref{LE} and instead delivers worse performance than SF-based formulation in all cases.

Comparing AD-based GCN-DEM and MLP-DEM in \fref{NH_AD}, we see that MLP-DEM outperforms GCN-DEM in the X- and Y-components, which are the primary deformation modes. For the SF-based formulation in \fref{NH_SF}, we see that the performance of GCN-DEM and MLP-DEM is again very comparable. The AD-based and SF-based formulations provide very similar levels of accuracy in the hyperelastic case, which is distinct from the observations made in \sref{LE}. \fref{NH_def} demonstrates similar strain localization at the root of the beam for AD-based simulations, indicating that this phenomenon is due to the gradient computation method and is independent of the material model. AD-based gradient computation again failed to capture the severe strain localization at the beam's root.

To conclude, in the nonlinear hyperelastic case, the solution accuracy of GCN-DEM and MLP-DEM is very comparable. However, GCN-DEM is more computationally effective and trains in a shorter time. SF-based gradient computation still holds a massive advantage over AD-based methods in terms of stability and robustness in severe deformation.

\subsection{Occurrence of instability in the MLP-DEM implementation by Nguyen-Thanh et al.}
\label{external_example}
\begin{figure}[h!] 
    \centering
     \subfloat[t = -1]{
         \includegraphics[trim={0cm 0cm 1cm 20cm},clip,width=0.33\textwidth]{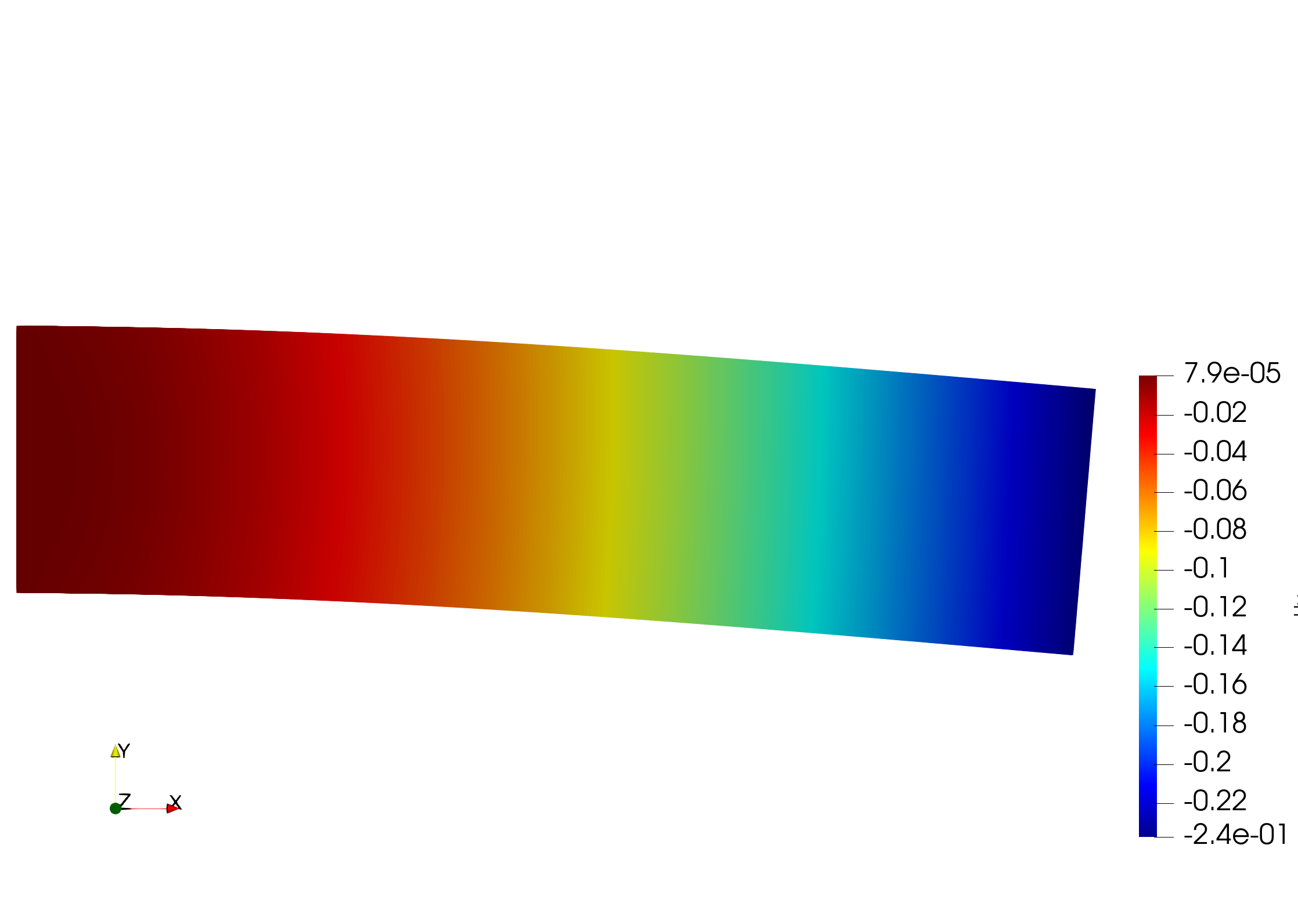}
         \label{fig:L1}
     }
     \subfloat[t = -2.5]{
         \includegraphics[trim={0cm 0cm 1cm 20cm},clip,width=0.33\textwidth]{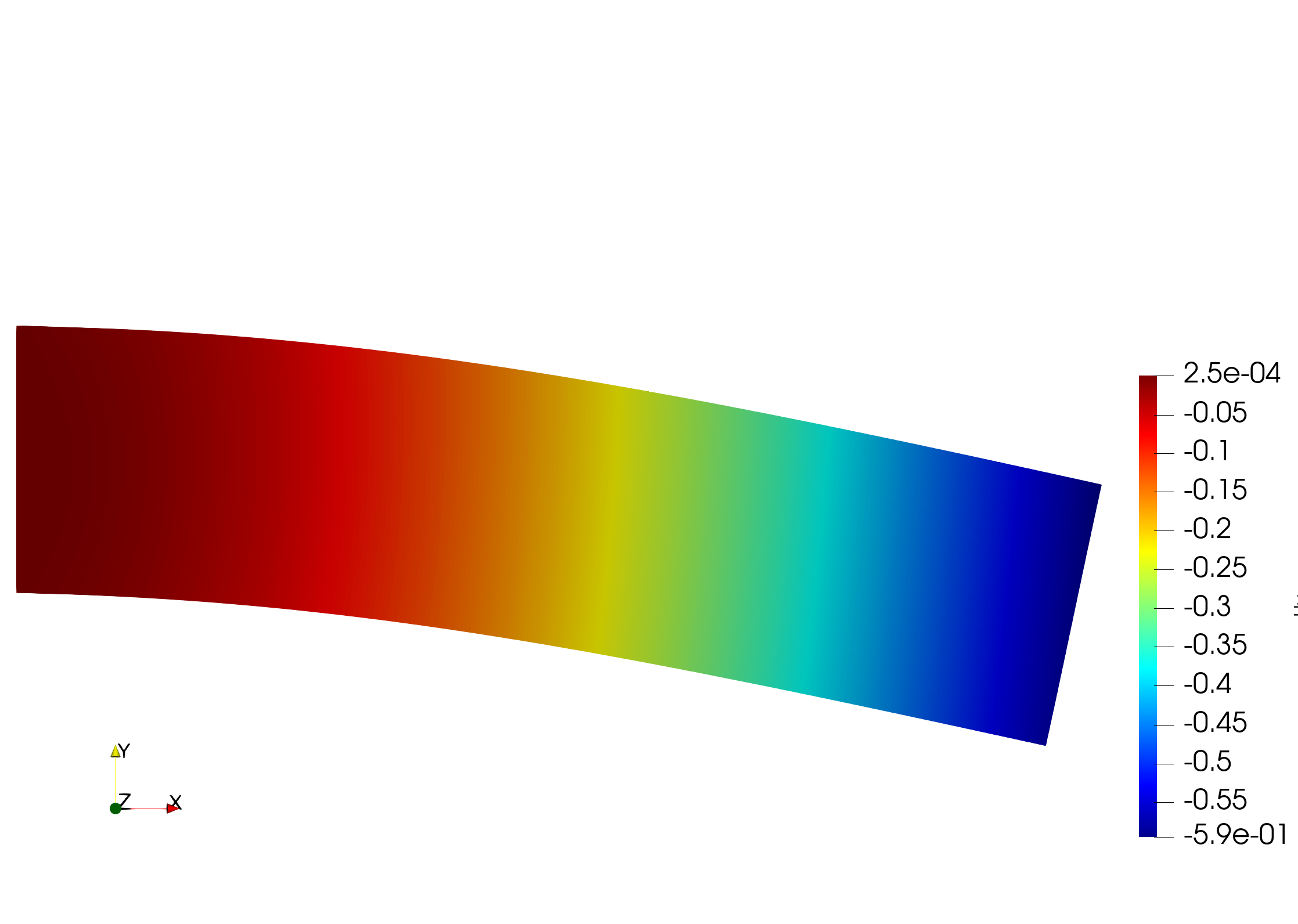}
         \label{fig:L2}
     }
     \subfloat[t = -5]{
         \includegraphics[trim={0cm 0cm 1cm 20cm},clip,width=0.33\textwidth]{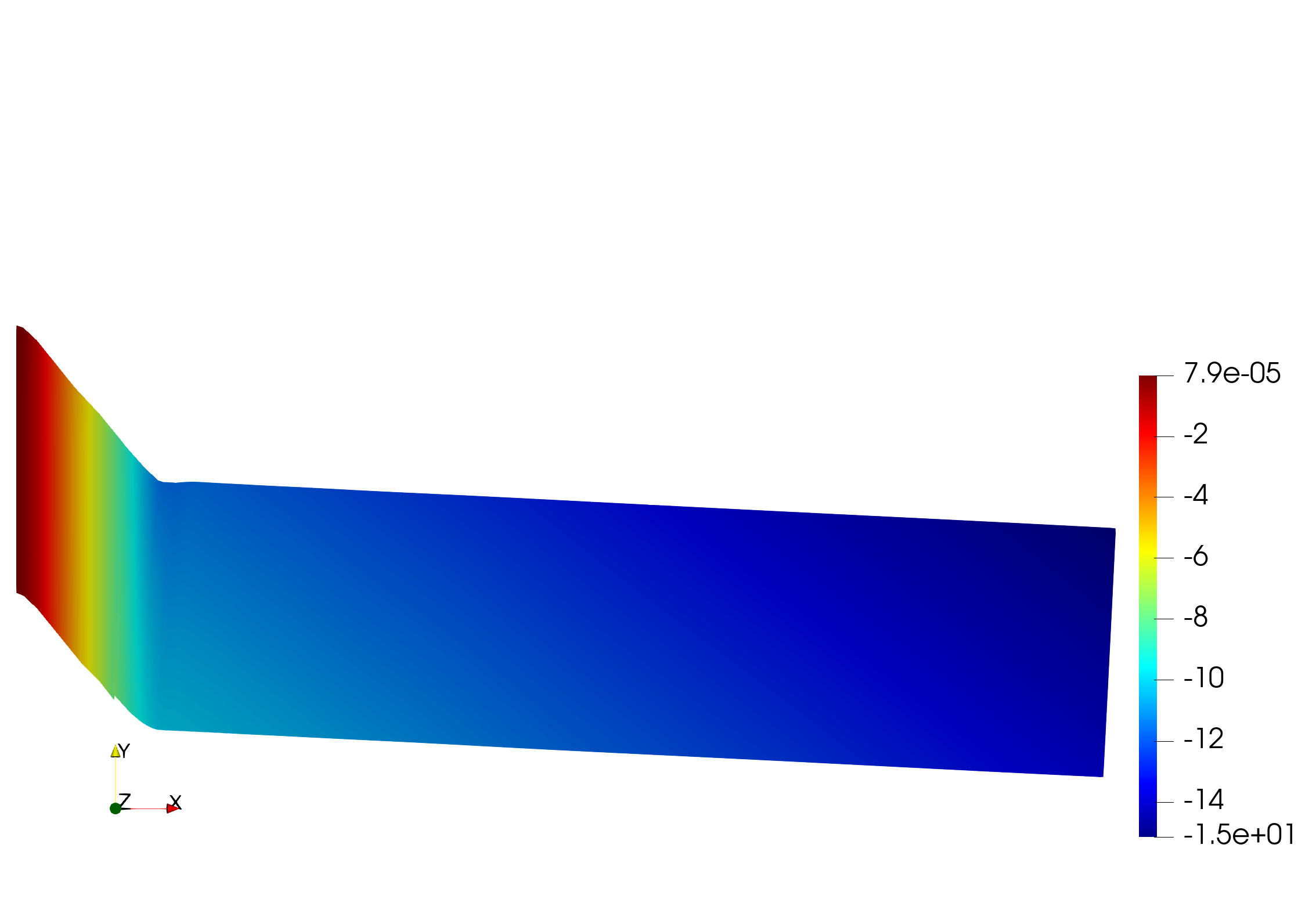}
         \label{fig:L3}
     }
    \caption{MLP-DEM simulations of a Neo-Hookean cantilever beam subject to increasing downward load. A scaling factor of 1 is used to plot the deformed shape in \psubref{fig:L1} and \psubref{fig:L2}, while a scaling factor of 0.05 is used in \psubref{fig:L3}. Note the occurrence of strain localization instability in the largest load case.}
    \label{extEx_contour}
\end{figure}

In this example, we demonstrate that the strain localization instability is not limited to our own implementation of GCN-DEM and MLP-DEM, but also exists in other AD-based MLP-DEM implementations. For this purpose, we used the MLP-DEM code developed in the work of Nguyen-Thanh et al. \cite{nguyen2020deep}. The model was implemented in PyTorch, and AD was used to evaluate the displacement gradients at the nodes. The trapezoidal rule was used to perform the domain integration. For simplicity, we used the 2D Neo-Hookean cantilever beam example, which has a dimension of 4-by-1 units. Materials properties are as defined in the example, and no modifications were made. To match similar gird spacing as used in our examples, we changed the grid dimension to 37-by-10. Three different downward loads were applied at the right edge; they are -1, -2.5, and -5. A total of 200 iterations were conducted during the training of the MLP-DEM model. Contour plots of the Y component of the displacements are shown in \fref{extEx_contour}.

From \fref{extEx_contour}, we see the occurrence of strain localization instability when downward traction of -5 was applied. Comparing \fref{fig:L3} and \fref{NH_def}, we notice very similar strain localization near the root of the beam. This result demonstrates that strain localization instability is not limited to our implementation of the GCN-DEM and MLP-DEM models, but also exists in other implementations that use AD for displacement gradient computations. However, we do point out that both frameworks tested in this work were implemented in PyTorch, and the existence of similar AD-induced instability is not checked in models implemented in TensorFlow \cite{tensorflow2015-whitepaper}, another commonly used package for building NNs.

\subsection{Effect of grid refinement}
\label{refinement}
\begin{figure}[h!] 
    \centering
     \subfloat[37$\times$10 grid]{
         \includegraphics[trim={0cm 0cm 1cm 20cm},clip,width=0.33\textwidth]{Figures/L3.png}
         \label{fig:L3M1}
     }
     \subfloat[74$\times$20 grid]{
         \includegraphics[trim={0cm 0cm 1cm 20cm},clip,width=0.33\textwidth]{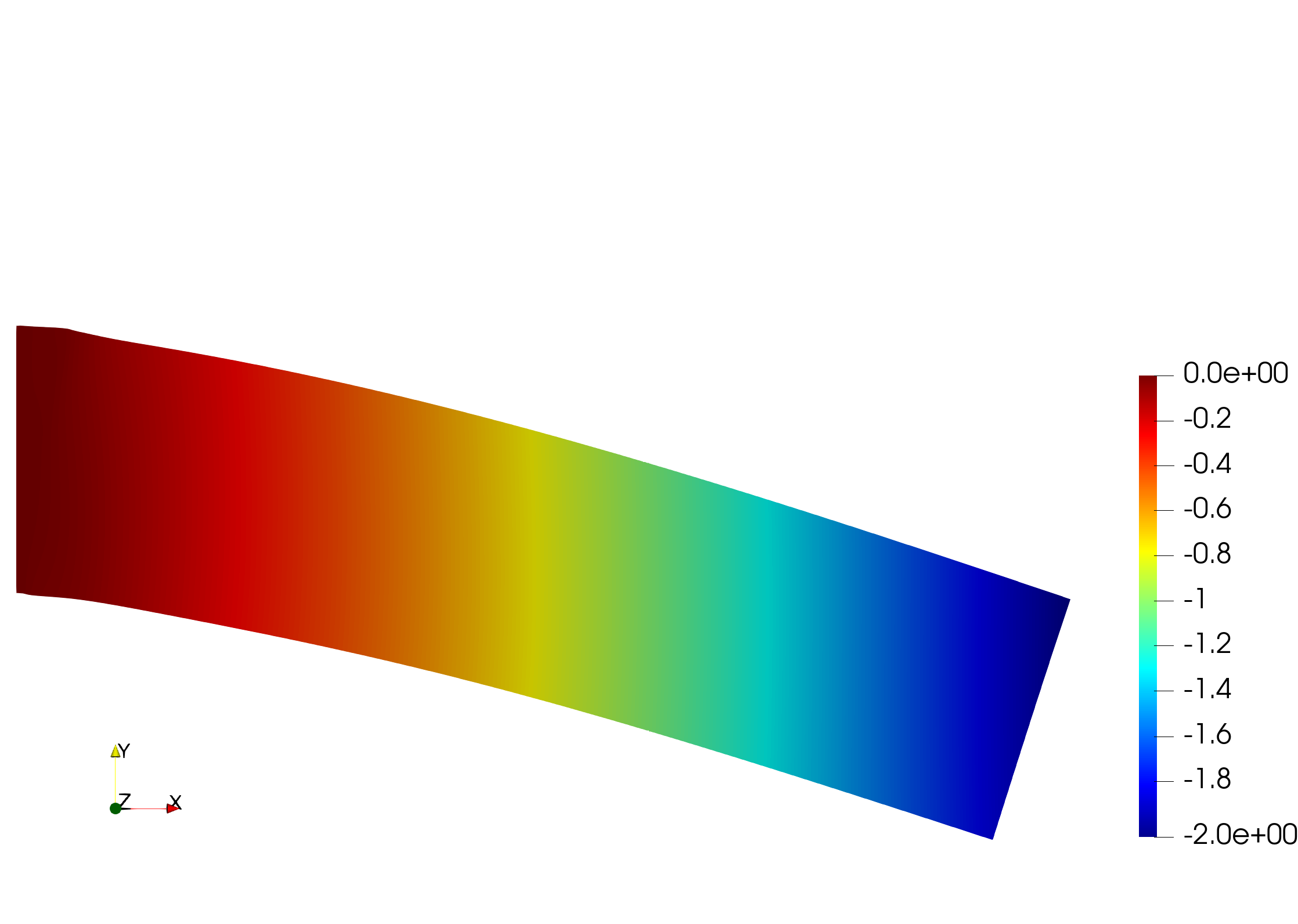}
         \label{fig:L3M2}
     }
     \subfloat[111$\times$30 grid]{
         \includegraphics[trim={0cm 0cm 1cm 20cm},clip,width=0.33\textwidth]{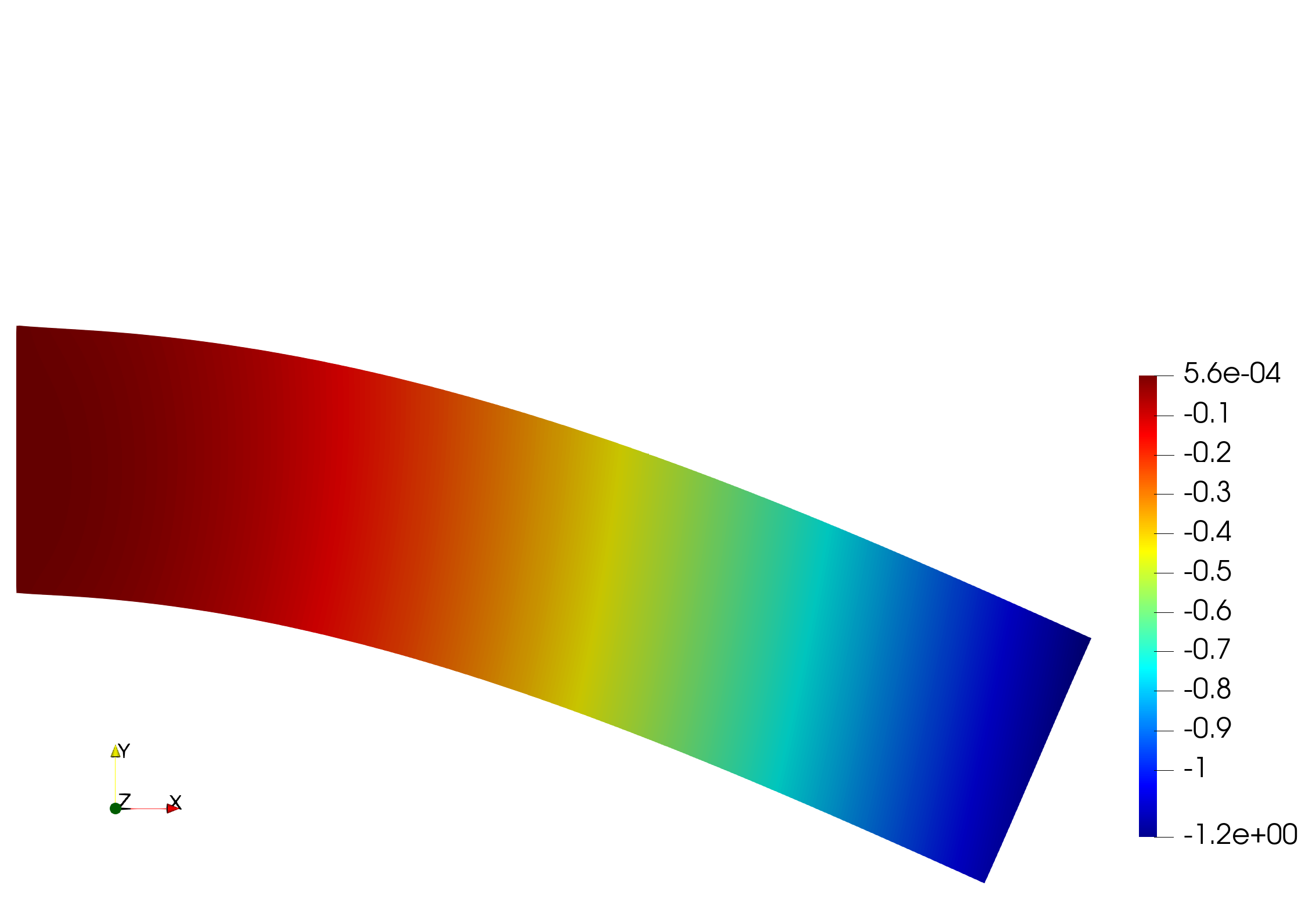}
         \label{fig:L3M3}
     }
    \caption{Effect of grid refinement on strain localization instability. Scaling factors of 0.05, 0.6 and 1 are used to plot the deformed shape in \psubref{fig:L3M1}, \psubref{fig:L3M2} and \psubref{fig:L3M3}, respectively.}
    \label{refinement_ext}
\end{figure}

The three previous sections established that instabilities might occur for AD-based formulations. Therefore, it is interesting to study how this instability can be remedied besides applying the load gradually in multiple steps. As analyzed in \sref{grad_cal}, the instability roots from possible strain localization that happens between two nodes. It is therefore intuitive to refine the grid for NN model training. We doubled the node count from 3700 to 7436 to form a 44-by-13-by-13 grid to solve the hyperelastic beam problem subjected to a load $t=-15$. Since the AD-based MLP-DEM method could not converge for this grid size, we crated another grid with 21708 nodes, forming a 67-by-18-by-18 layout. The results are summarized in \tref{tab:refine}. For all cases, finite element meshes with the same grid layouts were created for the FEM comparison. In addition to testing our implementations, the MLP-DEM model implemented by Nguyen-Thanh et al. \cite{nguyen2020deep} was also tested to see if grid refinement has similar effects on the results. \sref{external_example} shows that a downward traction of -5 induced instability on a 37-by-10 grid. We performed uniform grid refinement twice and presented the contour plots of the Y displacements in \fref{refinement_ext}.

\begin{table}[h!]
    \caption{Effects of grid refinement, Neo-Hookean model, $t=-15$ }
    \small
    \centering
    \begin{tabular}{ccccc}
     Method & \vline & 37$\times$10$\times$10  & 44$\times$13$\times$13 & 67$\times$18$\times$18  \\
    \hline
     & \vline & \multicolumn{3}{c}{Mean percent difference (\%)} \\
    GCN-DEM (AD) & \vline  & \textcolor{red}{254.30} & 2.28 & / \\
    MLP-DEM (AD) & \vline  & \textcolor{red}{361.10} & \textcolor{red}{212.90} & 4.26\\
    GCN-DEM (SF) & \vline  & 2.76 & 2.28 & /\\
    MLP-DEM (SF) & \vline  & 1.13 & 3.90 & /\\
    \hline
     & \vline & \multicolumn{3}{c}{Final loss function value} \\
    GCN-DEM (AD) & \vline  & \textcolor{red}{-186.68} & -23.07 & /\\
    MLP-DEM (AD) & \vline  & \textcolor{red}{-177.65} & \textcolor{red}{-111.91} & -22.81\\
    GCN-DEM (SF) & \vline  & -22.71 & -22.72 & /\\
    MLP-DEM (SF) & \vline  & -22.69 & -22.64 & /\\
    \hline
     & \vline & \multicolumn{3}{c}{Train time [s]} \\
    GCN-DEM (AD) & \vline  & \textcolor{red}{130.10} & 69.01 & /\\
    MLP-DEM (AD) & \vline  & \textcolor{red}{133.10} & \textcolor{red}{221.50} & 302.50\\
    GCN-DEM (SF) & \vline  & 74.19 & 103.70 & /\\
    MLP-DEM (SF) & \vline  & 136.90 & 230.50 & /\\
    \end{tabular}
    \label{tab:refine}
\end{table}

From \tref{tab:refine}, we see that GCN-DEM recovered from instability upon doubling the node count, achieving an accuracy similar to that obtained by the SF-based GCN-DEM in the original coarse grid. MLP-DEM did not recover from instability upon doubling the node count and instead required an even finer mesh to remain stable. We also note that SF-based simulations remained stable in all cases and achieved similar accuracy compared to their AD-based counterparts. A similar trend is also observed in \fref{refinement_ext}, where we see that the first uniform refinement (\fref{fig:L3M2}) reduced the amount of strain localization while the second uniform refinement (\fref{fig:L3M3}) produced reasonable displacement values. From the results, we conclude that AD-based gradient computation should be used in conjunction with a fine grid during training to avoid strain localization instability. If a coarse grid is desired, then AD-based gradient computation may not be suitable, and SF-based gradient computation is recommended for increased robustness.

\section{Conclusions and future work}
\label{sec:conc}
In this work, we developed a DEM model based on GCN, which is different from the typical DEM models that use an MLP network. In addition, we implemented and tested two different spatial gradient computation techniques, one using AD and the other one using SF. Numerical tests were performed using two material models, linear elastic and hyperelastic, to test the model performance in linear and nonlinear settings. Special attention is paid to studying the stability of the algorithm changes with increasing load magnitude, and how different spatial gradient calculation methods affect the stability.

The current work is novel in two main aspects: to the best of the authors' knowledge, this is the first time in the literature that a graph convolutional network is employed in the deep energy method, and we present a comparison of accuracy and computational time with the traditional DEM models based on multilayer perceptron networks. In addition, we performed a critical evaluation of two spatial gradient computation methods to investigate their effects on solution accuracy, solution time and most importantly, solution stability. Through two 3D examples using linear elastic and hyperelastic material models, we show that our GCN-DEM model shows higher computational efficiency than the MLP-DEM model while maintaining similar, if not better, solution accuracy. This comparison highlights the important role of network type in the energy-based models, sheds light on the future deployment of different types of NNs in the deep energy framework, and shows a promising direction to pursue to improve the quality of deep energy-based methods to solve PDEs. 

Another key finding of this current study is recognizing that the AD-based spatial gradient computation can lead to instability in the DEM framework due to its inability to detect erroneous strain localization. This result means that the AD-based DEM framework might fail to converge to the actual solution if the applied load in a single load step is large or when a coarse mesh is used. On the contrary, SF-based gradient computation demonstrated superior stability, especially in large deformations, while delivering a solution accuracy similar to that of AD-based DEM (when the AD-based one is stable). This robustness is highlighted by the fact that the SF-based DEM remained stable and gave accurate solutions in all examples presented in this work. Although the AD-based DEM method recovered from instability upon mesh refinement, we argue that instability would return for a larger load, and using a fine mesh should \emph{not} be required for a method to remain stable. The superior stability of the SF-based approach renders it suitable in simulations involving large deformations, which is typical for rubber-like hyperelastic materials. The examples in this work demonstrated the ability to apply the full magnitude of an external load in a single load step while maintaining high solution accuracy, something that FEM struggles to do. This feature renders the method computationally efficient for very nonlinear materials, where FEM typically requires breaking down the load into many smaller load steps to achieve convergence. 

The ability of the SF-based DEM model to solve hyperelastic material problems in a single load step renders it very attractive in the topology optimization of nonlinear hyperelastic materials. For classical FEM-based methods, multiple Newton-Raphson iterations are needed in each design iteration to achieve global force convergence, and the process is repeated for many topology optimization iterations. Using the DEM framework to substitute nonlinear FE simulations similar to our previous work \cite{he2022deep} might speed up the overall solution process and will be our future work. In addition, we also plan to extend the SF-based GCN-DEM network to unstructured tetrahedral meshes to perform simulations in irregular domains to harness the full power of the graph convolution network.

\section*{Replication of results}
The data and source code that support the findings of this study can be found at: \url{https://github.com/Jasiuk-Research-Group}. \textcolor{red}{Note to editor and reviewers: the link above will be made public upon the publication of this manuscript. During the review period, the data and source code can be made available upon request to the corresponding author.}

\section*{Conflict of interest}
The authors declare that they have no conflict of interest.


\section*{CRediT author contributions}
\textbf{Junyan He}: Conceptualization, Methodology, Software, Formal analysis, Investigation, Data Curation, Writing - Original Draft.
\textbf{Diab Abueidda}: Conceptualization, Supervision, Writing - Review \& Editing.
\textbf{Seid Koric}: Supervision, Writing - Review \& Editing.
 \textbf{Iwona Jasiuk}: Supervision, Resources, Writing - Review \& Editing, Funding Acquisition.

\bibliographystyle{unsrtnat}
\setlength{\bibsep}{0.0pt}
{\scriptsize \bibliography{References.bib} }
\end{document}